\documentclass[twocolumn,twocolappendix]{aastex631}

\usepackage{graphicx}	
\usepackage{amsmath}	
\usepackage{amssymb}	

\usepackage{natbib}
\usepackage{enumitem}
\usepackage[flushleft]{threeparttable}
\usepackage{colortbl}
\usepackage[normalem]{ulem}
\usepackage{color}
\usepackage{bm} 
\usepackage{gensymb}
\definecolor{darkblue}{rgb}{0.0,0.0,0.8}
\definecolor{darkred}{rgb}{0.75,0.,0.25}
\definecolor{darkorange}{rgb}{1,0.3,0.0}
\definecolor{darkgreen}{rgb}{0.0,0.6,0.0}
\definecolor{darkpurple}{rgb}{0.8,0.,0.9}
\definecolor{brown}{rgb}{0.65,.16,0.16}
\definecolor{grey}{rgb}{0.4,0.5,0.6}
\definecolor{white}{rgb}{1,1,1}
\definecolor{trolleygrey}{rgb}{0.5, 0.5, 0.5}
\definecolor{lavender}{rgb}{0.835,0.812,0.969}
\definecolor{pastelorange}{rgb}{0.99,0.92,0.82}
\definecolor{pastelblue}{rgb}{0.85,0.93,0.99}

\newcommand{\darkg}[1]{\textcolor{trolleygrey}{#1}}

\renewcommand{\d}{{\rm d}}
\newcommand{\mpo}{{\sc MAMPOSSt-PM}}
\newcommand{\jampy}{{\sc JamPy}}
\newcommand{\scf}{{\sc scalefree}}

\newcommand{\gaia}{{\sl Gaia}}
\newcommand{\hst}{{\sl HST}}
\newcommand{\jwst}{{\sl JWST}}

\newcommand{\msun}{\rm M_\odot}
\newcommand{\kms}{\rm km\,s^{-1}}
\newcommand{\masyr}{\rm mas\,yr^{-1}}

\begin{document}

\title{HSTPROMO Internal Proper Motion Kinematics of Dwarf Spheroidal Galaxies:\\ I.~Velocity Anisotropy and Dark Matter Cusp Slope of Draco}

\correspondingauthor{Eduardo Vitral}
\email{evitral@stsci.edu}

\author[0000-0002-2732-9717]{Eduardo Vitral}
\affiliation{Space Telescope Science Institute, 3700 San Martin Drive, Baltimore, MD 21218, USA}
\affiliation{Institut d’Astrophysique de Paris, CNRS, Sorbonne Université, 98bis Boulevard Arago, F-75014, Paris, France}

\author[0000-0001-7827-7825]{Roeland P. van der Marel}
\affiliation{Space Telescope Science Institute, 3700 San Martin Drive, Baltimore, MD 21218, USA}
\affiliation{Center for Astrophysical Sciences, The William H. Miller III Department of Physics \& Astronomy, Johns Hopkins University, Baltimore, MD 21218, USA}

\author[0000-0001-8368-0221]{Sangmo Tony Sohn}
\affiliation{Space Telescope Science Institute, 3700 San Martin Drive, Baltimore, MD 21218, USA}

\author[0000-0001-9673-7397]{Mattia Libralato}
\affiliation{AURA for the European Space Agency (ESA), Space Telescope Science Institute, 3700 San Martin Drive, Baltimore, MD 21218, USA}
\affiliation{INAF - Osservatorio Astronomico di Padova, Vicolo dell'Osservatorio 5, Padova I-35122, Italy}

\author[0000-0003-4922-5131]{Andrés del Pino}
\affiliation{Centro de Estudios de Física del Cosmos de Aragón (CEFCA), Unidad Asociada al CSIC, Plaza San Juan 1, E-44001, Teruel, Spain}

\author[0000-0002-1343-134X]{Laura L. Watkins}
\affiliation{AURA for the European Space Agency (ESA), Space Telescope Science Institute, 3700 San Martin Drive, Baltimore, MD 21218, USA}

\author[0000-0003-3858-637X]{Andrea Bellini}
\affiliation{Space Telescope Science Institute, 3700 San Martin Drive, Baltimore, MD 21218, USA}

\author[0000-0003-2496-1925]{Matthew G. Walker}
\affiliation{McWilliams Center for Cosmology, Carnegie Mellon University, 5000 Forbes Avenue, Pittsburgh, PA 15213, USA}

\author[0000-0003-0715-2173]{Gurtina Besla}
\affiliation{Steward Observatory, University of Arizona, 933 North Cherry Avenue, Tucson, AZ 85721, U.S.A.}

\author[0000-0002-9197-9300]{Marcel S. Pawlowski}
\affiliation{Leibniz-Institute for Astrophysics Potsdam (AIP), An der Sternwarte 16, 14482 Potsdam, Germany}

\author[0000-0001-8956-5953]{Gary A. Mamon}
\affiliation{Institut d’Astrophysique de Paris, CNRS, Sorbonne Université, 98bis Boulevard Arago, F-75014, Paris, France}





\begin{abstract}
We analyze four epochs of \hst\ imaging over 18 years for the Draco dwarf spheroidal galaxy. We measure precise proper motions (PMs) for hundreds of stars and combine these with existing line-of-sight (LOS) velocities. This provides the first radially-resolved 3D velocity dispersion profiles for any dwarf galaxy. These constrain the intrinsic velocity anisotropy and resolve the mass-anisotropy degeneracy. We solve the Jeans equations in oblate axisymmetric geometry to infer the mass profile. We find the velocity dispersion to be radially anisotropic along the symmetry axis and tangentially anisotropic in the equatorial plane, with a globally-averaged value $\overline{\beta_{\rm B}}=-0.20^{+ 0.28}_{- 0.53}$, (where $1 - \beta_{\rm B} \equiv \langle v_{\rm tan}^2 \rangle / \langle v_{\rm rad}^2 \rangle$ in 3D). The logarithmic dark matter (DM) density slope over the observed radial range, $\Gamma_{\rm dark}$, is $-0.83^{+ 0.32}_{- 0.37}$, consistent with the inner cusp predicted in $\Lambda$CDM cosmology. As expected given Draco's low mass and ancient star formation history, it does not appear to have been dissolved by baryonic processes. We rule out cores larger than 487, 717, 942~pc at respective 1-, 2-, 3-$\sigma$ confidence, thus imposing important constraints on the self-interacting DM cross-section. Spherical models yield biased estimates for both the velocity anisotropy and the inferred slope. The circular velocity at our outermost data point (900~pc) is $24.19^{+ 6.31}_{- 2.97}~\kms$. We infer a dynamical distance of $75.37^{+ 4.73}_{- 4.00}$~kpc, and show that Draco has a modest LOS rotation, with $\left<v / \sigma \right> = 0.22 \pm 0.09$. Our results provide a new stringent test of the so-called `cusp-core' problem that can be readily extended to other dwarfs.
\end{abstract}



\section{Introduction} \label{sec: intro}

Decades of astrophysical evidence support the notion that most of the matter in the Universe is dark. However, the nature of this dark matter (DM) remains a mystery. The most likely candidate is some form of cold DM (CDM), consisting of collisionless particles that cannot (yet) be detected directly, but that interact through gravity.

Some of the best systems to study DM are the ``classical'' dwarf spheroidal galaxies (dSphs) in the Milky Way (MW). They are strongly DM dominated \citep{Pryor&Kormendy&Kormendy90}, and have a large number of bright stars that can be resolved due to their proximity. The stars' motions contain information about the gravitational potential in which they move, and thus a large observational effort has been invested in obtaining their line-of-sight (LOS) velocities ($v_{\rm LOS}$, e.g. \citealt{Tolstoy+04,Walker+07,Gilmore+22}). Results from analyzing these data have been inconclusive about some CDM predictions. A conspicuous example of this is the so-called ``cusp-core problem'': the tension around the predicted and observed DM mass-density profiles of galaxies. CDM halos in collisionless cosmological $N$-body simulations follow a nearly universal mass-density profile that increases and diverges toward the center, forming a `cusp' \citep*{Navarro+97}. In contrast, observations of some dSphs
favor shallower density profile slopes, consistent with a constant density `core' at the center \citep[e.g.][]{Battaglia+08,Walker&Penarrubia&Penarrubia11,Amorisco&Evans12,Brownsberger&Randall&Randall21}.

Various solutions have been proposed to explain this and other discrepancies. Some propose fundamental changes in the nature of DM, such as warm DM (WDM), e.g. sterile neutrinos and gravitinos, that predict lower central DM densities and cored profiles \citep{Dalcanton&Hogan01}, or self-interacting DM (SIDM) for which DM particles in the central region thermalize via collisions and thereby form a cored profile \citep[e.g.][]{Sameie+20}. Others include the impact of baryons, which may transform cusps into cores by transferring energy and mass to the outer parts of the halos, e.g. via supernova feedback \citep{Read&Gilmore05,Pontzen&Governato12,Brooks&Zolotov&Zolotov14}, or star formation events \citep{Read+18}.
Recent studies have also found that the orientation of a galaxy with respect to the viewer has a large impact on the derived velocity dispersion, resulting in a range of density slopes fitting the data \citep{Genina+18}. 

Significant uncertainties are introduced by the fact that most observational studies are based solely on $v_{\rm LOS}$ measurements, which constrain only one component of motion. Consequently, interpretations rely on substantial assumptions, in particular that $v_{\rm LOS}$ is representative of the three-dimensional (3D) velocities.\footnote{By this statement, we mean that $v_{\rm LOS}$ is sometimes used to infer mass and/or anisotropy properties that are not uniquely constrained solely by the second-order moments of this single dimension.} These assumptions have been challenged by alternatives implying that the inferred, excessive dynamical mass-to-light ratios could be due to e.g. modified gravity \citep{McGaugh&Wolf&Wolf10} or out-of-equilibrium dynamics caused by tidal interaction with the MW \citep{Klessen&Kroupa&Kroupa98, Hammer+18}, although the latter is hard to explain for satellites on orbits having higher pericenter values reported by \cite{Li+21}, \cite{Battaglia+22} and \cite{Pace+22} from \gaia-based systemic proper motions. 

Multiple techniques have been used to model $v_{\rm LOS}$ dispersion ($\sigma_{\rm LOS}$) profiles and, thus, constrain mass density profiles of dSphs. Examples include Jeans models \citep{Walker+09, Zhu+16, Read&Steger17}, distribution function (DF) fitting \citep{Wilkinson+02, Vasiliev19agama}, and Schwarzschild orbit superposition modeling \citep{Breddels+13,Kowalczyk+19}, each with their own strengths and weaknesses. However, all modeling techniques face the same problem: When only $v_{\rm LOS}$ are used, there is a strong degeneracy between the mass density profile $\rho(r)$ and the velocity anisotropy profile $\beta(r)$, which quantifies differences in velocity dispersions in orthogonal directions \citep{Binney&Tremaine87,Binney&Mamon82}. Some models mitigate this degeneracy by restricting parameter space or using higher-order moments \citep{Vasiliev19agama,Genina+20,Read+21}, but having only the LOS component of motion fundamentally limits what can be achieved.

The key to progress is to measure the internal proper motion (PM) kinematics of stars. The radial and tangential PM components directly measure the projected velocity dispersion anisotropy, which, under assumptions of inclination and intrinsic shape, uniquely determines $\beta(r)$ without requiring any dynamical modeling (e.g. \citealt{vanderMarel&Anderson&Anderson10}). 
This makes PMs crucial for dynamical modeling of dSphs, with models making use of PMs performing consistently better than those based solely on $v_{\rm LOS}$ \citep{Read+21}.
Different techniques can be used to measure internal PM kinematics, but all of them require combining two or more epochs of observations to determine PMs of individual stars. At typical distances of MW dSphs, the only feasible instruments currently available for measuring individual PMs are \gaia, the \textit{Hubble Space Telescope} (\hst), and \jwst. 

\gaia\ has been tremendously successful in revolutionizing our view of the MW and its satellites, but the relatively shallow limiting magnitude (G$\sim$21 mag) and its large PM uncertainties for typical dSphs stars in the MW halo \citep[e.g.][]{Pace+22,Vitral21} hinder its use for a direct measurement of \textit{internal} PM dispersions \citep{MartinezGarcia+21}. An alternative is to combine \gaia\ astrometry with that from another instrument (e.g. \hst) to achieve longer time baselines, and thus lower PM uncertainties. This procedure has been applied in \cite{Massari+17,Massari+20} and \cite{delPino+22}. However, even with the \gaia-end-of-mission PM uncertainties reduced by a factor of 3, the number of stars available to measure PM dispersion profiles will always be confined to those near the tip of the red giant branch due to the limiting magnitude of \gaia. This is insufficient to discriminate between cusp and core models \citep{Strigari+18,Guerra+23}.

While the \jwst\ time baseline is still too short (due to its recent launch) for a robust \jwst\ vs.~\jwst\ PM computation, comparing positions of stars in images obtained with the same detectors onboard \hst\ over time is the best means to obtain precise PMs of thousands of individual stars.
\hst\ is exquisitely well suited for astrometric and PM science, due to is stability, high spatial resolution, and well-determined point spread functions (PSFs) and geometric distortions. By combining two or more epochs of space-based imaging, it is possible to measure precise internal PMs in nearby stellar systems (e.g. \citealt{Libralato+22}). 

In the current work, we combine 18 years of \hst\ data, mostly obtained in the context of our HSTPROMO (High-Resolution Space Telescope PROper MOtion) Collaboration,\footnote{\url{https://www.stsci.edu/~marel/hstpromo.html}} to measure PMs of hundreds of stars in the Draco dSph. With this, we measure its internal PM dispersion profile for the first time and thus provide unprecedent constraints on its DM density slope. We describe the datasets we used in Section~\ref{sec: data}, we explain the methods used to analyze the data in Section~\ref{sec: methods}, and we present our results in Section~\ref{sec: results}. We comment on the robustness of our findings in Section~\ref{sec: robust}, and discuss and conclude our work in Sections~\ref{sec: discussion} and \ref{sec: conclusion}, respectively.
Throughout the paper, we use lower case $r$ to denote 3D distances, and upper case $R$ to denote 2D projected distances.

\section{Draco Data and General Characteristics} \label{sec: data}

The Draco dSph is an excellent candidate to test the predictions of CDM scenarios. Its star formation shut down long ago ($\sim$10 Gyr; \citealt{Aparicio+01}), making it a prime candidate for hosting a `pristine' DM cusp, unaffected by baryonic processes \citep{Read+18}. Interestingly, the stellar mass of Draco is clearly below the limit where stellar feedback, as implemented in current cosmological simulations, should still produce a core \citep{Fitts+17}.\footnote{\cite{Fitts+17} place this limit at $M_{\star} = 2 \times 10^{6}~\msun$.} Furthermore, Draco is one of the most DM dominated satellites of the MW \citep{Kleyna+02}, and seemingly unaffected by Galactic tides that might heat its velocity dispersion profile \citep{Odenkirchen+01,Segall+07}.

Recent efforts to infer the DM density of Draco from Jeans modeling of LOS velocities have yielded similar results: \cite{Read+18} fit rotation-less spherical Jeans models combined with higher-order LOS moments and report a DM density slope at 150~pc of $-0.95_{-0.46}^{+0.50}$ (95\% intervals); \cite{Hayashi+20} applied rotation-less axisymmetric models to LOS data, and report a cusp with `high probability' and a formal measurement of the asymptotic DM slope of $-1.03_{-0.15}^{+0.14}$ (68\% intervals).
Meanwhile, when formulating dynamical mass estimators based on PM dispersions, \cite{Lazar&Bullock20} found the then available data to be insufficient for the purpose of constraining the asymptotic DM slope.
Below, we describe the main characteristics of the new datasets we employ, and how those are able to grasp the dynamical status of Draco in more detail and with better accuracy.

\subsection{Projected Density} \label{ssec: draco}

\subsubsection{Center} \label{sssec: center}

The quoted center for Draco in the \cite{McConnachie12} catalog is the one from \cite{Wilson55}, when the dSph was discovered. After that, more detailed sky surveys have allowed further refinement of this measurement. In particular, \cite{Odenkirchen+01} and \cite{Martin+08} used SDSS data to estimate the values quoted in Table~\ref{tab: overview}, and \cite{Vitral21} computed its center from \gaia\ EDR3 data assuming a Plummer \citep{Plummer1911} spherical model. In this paper, we compute it again from \gaia\ EDR3, but using a more refined version of the \cite{Vitral21} algorithm, which allows for an elliptical Plummer distribution (see Appendix~\ref{app: dens-prof} for analytical expressions of density profiles). The overall parameterizations are thus similar to the ones reported in \cite{Vitral21}, with the exception of the elliptical Plummer, which adds two extra free parameters to the fit: (i) a projected angle $\theta$ in the sky and (ii) the ellipticity implied by the minor-axis scale length of the projected ellipse.

Our fit was performed by a Monte Carlo Markov Chain (MCMC) routine that uses the software \textsc{emcee} \citep{emcee}. We selected the most probable values from the joint MCMC posterior chain as the parameters, and assigned uncertainties based on its difference to the $16^{\rm th}-84^{\rm th}$ percentiles of the respective posterior distribution. Our best fit center, projected angle and projected ellipticity are listed in Table~\ref{tab: overview}. Overall, the fits agree very well with the estimates from \cite{Wilson55,Odenkirchen+01} and \cite{Martin+08}.

\begin{deluxetable*}{lllllrrrr}
\tablecaption{Overview of Draco parameters.}
\label{tab: overview}
\tablewidth{750pt}
\renewcommand{\arraystretch}{1.1}
\tabcolsep=3pt
\tabletypesize{\scriptsize}
\tablehead{
\colhead{Reference} &
\colhead{Data} &
\colhead{$\alpha_{0}$} &
\colhead{$\delta_{0}$} &
\colhead{$\theta$} &
\colhead{$\epsilon$} &
\colhead{$r_{h}$} &
\colhead{$\mu_{\rm LOS}$} & 
\colhead{$\left< v_{\rm LOS}/\sigma_{\rm LOS} \right>$} \\
\colhead{} &
\colhead{} &
\colhead{[hh~mm~ss]} &
\colhead{[dd~mm~ss]} &
\colhead{[deg]} &
\colhead{} &
\colhead{[$'$]} &
\colhead{$\left[ \kms \right]$} &
\colhead{} \\
\colhead{(1)} &
\colhead{(2)} &
\colhead{(3)} &
\colhead{(4)} &
\colhead{(5)} &
\colhead{(6)} &
\colhead{(7)} &
\colhead{(8)} &
\colhead{(9)}
}
\startdata
This work & \gaia\ EDR3 & $17 \ 20 \ 16.42^{+0.02}_{-0.01}$  & $+57 \ 55 \ 06.60^{+0.07}_{-0.32}$  & $89^{+10}_{-6}$ & $0.25^{+0.03}_{-0.03}$& $10.4^{+0.3}_{-0.4}$ & $-291.73^{+0.48}_{-0.48}$ & $0.22^{+0.09}_{-0.09}$ \\
\cite{Martin+08} & SDSS & $17 \ 20 \ 14.4^{+0.6}_{-0.6}$  & $+57 \ 57 \ 54^{+8}_{-8}$  & $89^{+2}_{-2}$ & $0.31^{+0.02}_{-0.02}$ & $10.0^{+0.3}_{-0.2}$ & -- & -- \\
\cite{Odenkirchen+01} & SDSS & $17 \ 20 \ 13.2^{+1.4}_{-1.4}$ & $+57 \ 54 \ 60^{+0.07}_{-0.07}$ & $89^{+3}_{-3}$ & $0.28^{+0.01}_{-0.01}$ & $12.4^{+1.6}_{-1.2}$ & -- & -- \\
\cite{Wilson55} & Palomar & $17 \ 20 \ 12.4$ & $+57 \ 54 \ 55$ & --  & --  & --  & -- & --
\enddata
\begin{tablenotes}
\scriptsize
\item \textsc{Notes} --  
Columns are 
\textbf{(1)} Reference where the values are reported; 
\textbf{(2)} data source of respective estimates (columns 3--7); 
\textbf{(3)} right ascension of Draco center; 
\textbf{(4)} declination of Draco center; 
\textbf{(5)} projected angle in the sky, from North to East; 
\textbf{(6)} projected ellipticity in the sky, defined as $1 - b/a$, with $a$ and $b$ the major and minor axes of the projected ellipse, respectively; 
\textbf{(7)} 3D half-number radius of a Plummer model fit for this work, of an exponential model fit for \cite{Martin+08} and of a S\'ersic model fit for \cite{Odenkirchen+01} (using the 3D deprojection method from \citealt{VitralMamon21}); 
\textbf{(8)} bulk line-of-sight velocity and
\textbf{(9)} mean rotation fraction in the line-of-sight.
\end{tablenotes}
\end{deluxetable*}

\subsubsection{Surface Density Profile} \label{sssec: surf-dens}

\begin{figure}
\centering
\includegraphics[width=\hsize]{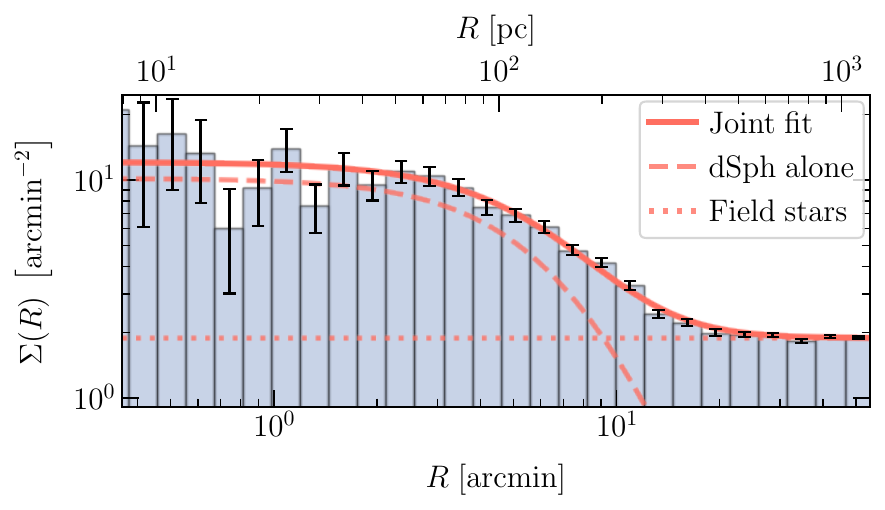}
\caption{\textit{Surface density:} Goodness of fit of a Plummer \cite{Plummer1911} spherical model to \gaia\ EDR3 data of Draco, with significant field-star interlopers. The surface density fits follow the formalism detailed in \cite{Vitral21}, which assumes a constant contribution of field stars.}
\label{fig: surf-dens}
\end{figure}

Further on, we will set up not only axisymmetric Jeans models, but also spherical models to fit our dataset (see Section~\ref{sec: methods}). For that purpose, it is of interest to know the best scale radius of the observed data, assuming a spherical density profile, so that we can set reasonable priors to our models. Following \cite{Massari+20,Hayashi+20}, we assume a Plummer model. We derive the Plummer scale radius of Draco using \gaia\ EDR3 data, using the same formalism as in \cite{Vitral21}, and with the $(\alpha_{0}, \delta_{0})$ centers calculated in Section~\ref{sssec: center}.

Figure~\ref{fig: surf-dens} displays the goodness-of-fit of our spherical Plummer profile to the \gaia\ EDR3 data. This satisfactory agreement yields a 3D half-number radius of $10.4^{+0.3}_{-0.4}$ arcmin, which lies between the values of $10.0^{+0.3}_{-0.2}$ and $12.4^{+1.6}_{-1.2}$ arcmin, estimated by \cite{Martin+08} and \cite{Odenkirchen+01} with SDSS data, respectively, for an exponential model, and a S\'ersic model.

\subsubsection{Inclination} \label{sssec: inc}

Due to the elliptical projected shape of Draco, we choose to model it as an oblate spheroid with a flattening parameter (i.e. intrinsic axial ratio) $q$. This relates to the projected axial ratio of Draco, $q_{\rm p}$, through the equation  \citep{Binney&Tremaine87},
\begin{equation} \label{eq: q-i-relation}
    q_{\rm p}^{2} = \cos^{2}{i} + q^{2} \sin^{2}{i} \ ,
\end{equation}
where $i$ is the inclination of the spheroid (see Section~\ref{sapp: formalism-scf} below, where an edge-on model is defined to have $i = 90\degree$).
We derive $q_{\rm p}$ from the ellipticity value in Table~\ref{tab: overview}, which yields $q_{\rm p} = 0.745 \pm 0.051 $. The intrinsic axial ratio $q$ is not known, but its probability distribution can be assumed to follow the general flattening probability distribution of oblate elliptical galaxies in the nearby Universe. This is given by equation~(4) from \cite*{Lambas+92},
\begin{equation} \label{eq: lambas-q}
    \psi_{\rm obl}(q) = \frac{2}{\pi} \sqrt{1 - q^{2}} \left( 0.344 + 21.272 \, q^{2} -29.24 \, q^{4} \right) \ ,
\end{equation}
with $q \in [0, 0.9]$. This can be used to obtain a probability distribution function (PDF) for the inclination of Draco, as follows: 

\begin{enumerate}
    \item We draw $N$ values\footnote{We use $N = 10^{7}$.} of flattening from the PDF in \cite{Lambas+92}, which we label $q_{\rm all}$.
    \item We draw $N$ inclination values according to $i_{\rm all} = \arccos{(\rm U)}$, where U is the uniform distribution within the $[0, \, 1]$ interval. This ensures that the inclinations are sampled uniformly on the surface of a unit sphere from face-on ($i = 0\degree$) to edge-on ($i = 90\degree$) cases.
    \item For each of those inclinations, we compute the respective projected flattening from Eq.~(\ref{eq: q-i-relation}), using the $q_{\rm all}$ values previously drawn, and we label these as $q_{\rm p, \, all}$.
    \item From those ($q_{\rm p, \, all}, \, i_{\rm all}$) pairs, we keep the ones that satisfy $|q_{\rm p, \, all} - q_{\rm p, \, Draco}| < 0.01$.
\end{enumerate}

\begin{figure}
\centering
\includegraphics[width=\hsize]{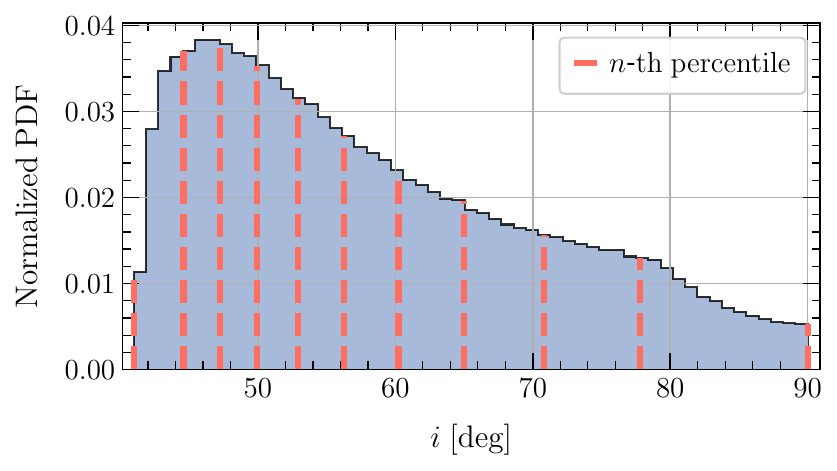}
\caption{\textit{Inclinations:} Probability distribution function of inclinations for Draco, computed according to the methodology described in Section~\ref{sssec: inc}. The \textit{dashed red} lines indicate the $n$-th percentiles of the distribution, with $n$ spaced from zero to a hundred, on intervals of ten.}
\label{fig: inc-dist}
\end{figure}

The remaining pairs from the last step above yield the projection of the inclination PDF onto the observed projected axial ratio of Draco, which is depicted in Figure~\ref{fig: inc-dist}. The resulting $[16, \, 50, \, 84]$ percentiles of the inclination and flattening final distributions are $[46.1\degree, \, 56.3\degree, \, 73.4\degree]$ and $[0.38, \, 0.60, \, 0.72]$, respectively.

\subsection{Line-of-Sight Velocities} \label{ssec: los}

Draco has been the subject of many observational campaigns to obtain LOS velocities of samples
of individual stars \citep[e.g.][]{Armandroff+95,Wilkinson+04,Walker+15}. Recently, \cite{Walker+23} provided the most complete catalog of dwarf galaxy LOS kinematics, including also metallicities and stellar parameters. Here, we make use of this catalog to complement our PM dataset. In this Section, we study some of the main aspects of this LOS dataset, including its interloper contribution, the implied galaxy rotation, and the influence of binaries on the inferred kinematics.

\subsubsection{Interloper cleaning} \label{sssec: los-interloper}

To best interpret our results based on LOS data, we need to remove interlopers (essentially, stars in the foreground and background). Hence, we perform a multi-dimensional mixture model to assign membership probabilities to each star in our subset, and then select it (or not) based on a threshold probability.

We first narrow our study to catalog stars that satisfy \texttt{good$\_$obs $==$ 1}, as suggested in \citet[][section~5]{Walker+23}. This essentially removes stars having high $v_{\rm LOS}$ uncertainties. Next, we select the parametrizations that we use to model the joint PDF of Draco stars (tracers) and interlopers, in each dimension of the data (i.e. $v_{\rm LOS}$, $T_{\rm eff}$, $\log{g}$,\footnote{Throughout this work, we denote the logarithm on base 10 as $\log$, and the logarithm in the natural base as $\ln$.} [Fe/H], [Mg/Fe]). 
The tracer PDF of $v_{\rm LOS}$, $\log{g}$, [Fe/H] and [Mg/Fe] were modeled as a Gaussian, while the tracer PDF of $T_{\rm eff}$ was modeled as a double Gaussian.
The interloper PDF of $\log{g}$, [Fe/H] and [Mg/Fe] were modeled as a triple Gaussian, while the interloper PDF of $v_{\rm LOS}$ and $T_{\rm eff}$ were modeled as a log-Gauss and double Gaussian distributions, respectively.
These choices of PDFs were done so as to maximize the goodness-of-fit.

\begin{figure}
\centering
\includegraphics[width=\hsize]{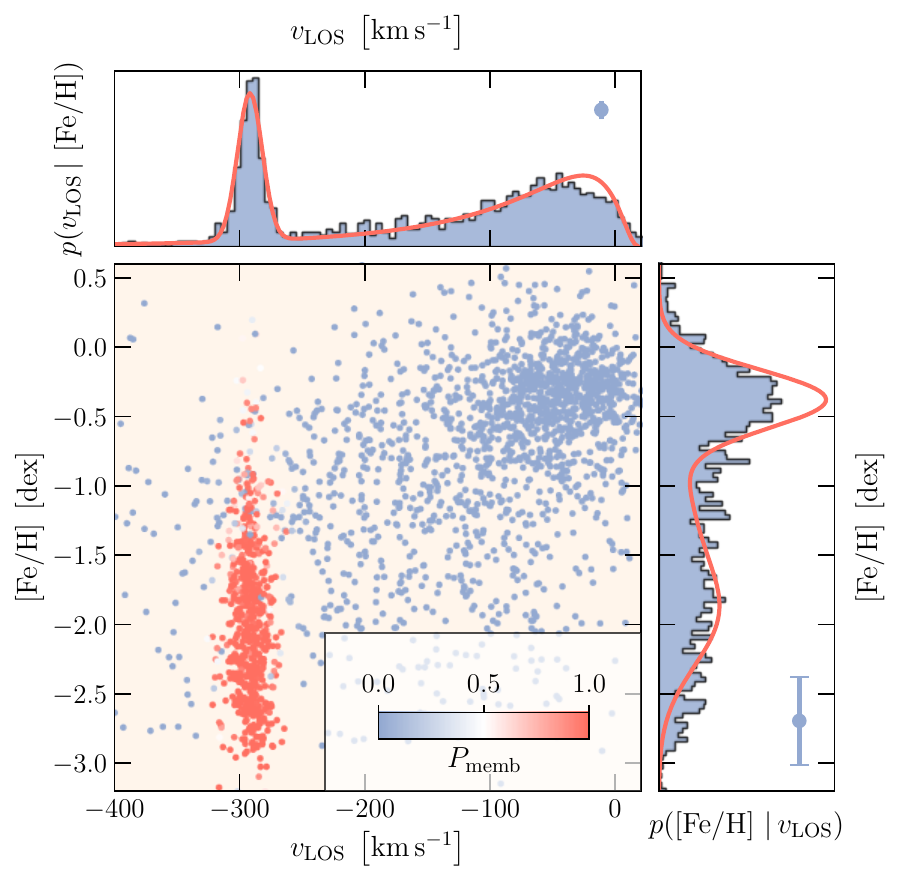}
\caption{\textit{Interloper cleaning:} Goodness of fit of our multi-dimensional mixture model, projected on the $v_{\rm LOS}$ and [Fe/H] dimensions. The central panel shows the stars from the \cite{Walker+23} catalog on the sky region of Draco, in the $v_{\rm LOS}$ -- [Fe/H] plane, color-coded by their Draco membership probability. The upper and right panels display the histogram of this plane projected on the $v_{\rm LOS}$ and [Fe/H] dimensions, respectively. On the corner of each side panel, we represent the median uncertainty on the respective dimension by an error bar. Stars with $P_{\rm memb} > 0.99$ were selected for inclusion in our analysis.}
\label{fig: vlos-feh}
\end{figure}

We fitted this multi-dimensional distribution through an MCMC routine on discrete data and considered the region between the $16^{\rm th}-84^{\rm th}$ percentiles of each posterior distribution as the uncertainty on our fits, and the most probable values from the joint MCMC posterior chain as the best parameters. Figure~\ref{fig: vlos-feh} showcases the goodness of our fit, projected on the  $v_{\rm LOS} - $[Fe/H] dimensions. From our fits, we assigned as Draco members the stars having a membership probability higher than 99\%, which removes most of the stars beyond a little more than 3-$\sigma_{\rm LOS}$ from the bulk $v_{\rm LOS}$. This final subset was composed of 435 stars with $v_{\rm LOS}$ data and uncertainties smaller than the value of $\sigma_{\rm LOS}$. Our measured value for the bulk LOS motion, $\mu_{\rm LOS}$,\footnote{We label the bulk LOS motion of Draco as $\mu_{\rm LOS}$, and the first order moment over the major axis, which relates to rotation, as $\left<v_{\rm LOS}\right>$.} is presented in Table~\ref{tab: overview}. For this subset, we chose not to correct for perspective effects caused by Draco's bulk motion, since those have negligible effects.
Indeed, the RMS correction for the sample stars implied by eq.~(13) of \cite{vanderMarel+02} is only $0.2~\kms$, while the rotation and velocity dispersion profiles inferred further in this work change by at most $0.1~\kms$, which is well below their respective measurement uncertainties.

\subsubsection{Rotation} \label{sssec: rotation}

Like most galaxies \citep[e.g.][]{MartinezGarcia+23}, it is possible that Draco possesses detectable mean rotation. We estimate here the rotation fraction of Draco, $\left< v/\sigma \right>$, from its $v_{\rm LOS}$.

\begin{figure}
\centering
\includegraphics[width=\hsize]{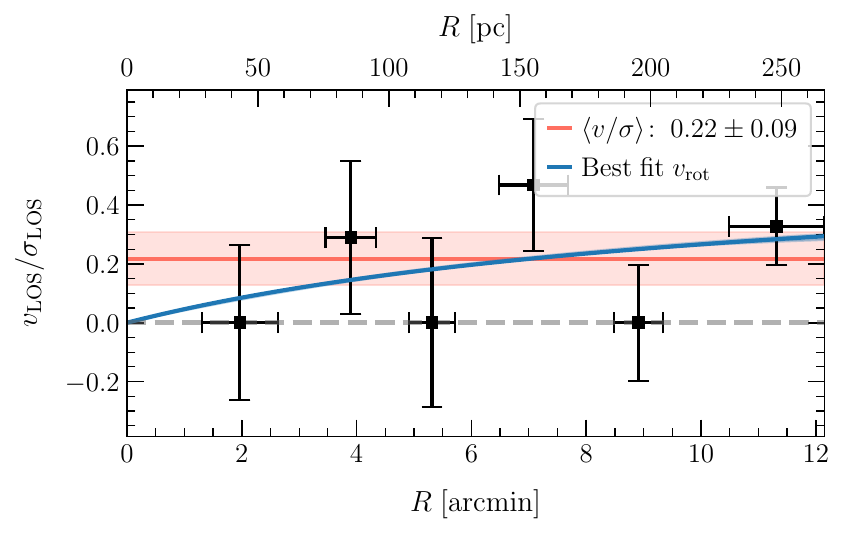}
\caption{\textit{Rotation:} Rotation profile of Draco, as a function of its projected radius, computed with LOS data. The black dots and error bars show the $v/\sigma$ ratio per bin, while the red and blue curves (and respective uncertainty regions) show, respectively, the overall mean through the Draco field, and the best fit of Eq.~(\ref{eq: rotation-prof}). We add a dashed-gray line to represent the case of no rotation. From this plot, we conclude that Draco has a modest amount of rotation.}
\label{fig: vlos-rotation}
\end{figure}

We partition the LOS data into six concentric annuli on the sky, all of which having the nearly the same amount $N\sim50$ of stars. We then perform a sinusoidal fit with free amplitude for each partition (i.e. to the respective $v_{\rm LOS}$ vs. projected angle quantities), and free mean velocity and phase. The mean and phase\footnote{The phase is taken with respect to the position angle of the minor axis (see column~5 from Table~\ref{tab: overview}). We found that while the rotation curve was robustly constrained by the data, the exact angle of the rotation axis was not. The fits showed angle variations between annuli, and the best-fit angle also depended on the exact choice of annuli, both in excess of the formal uncertainties. While a kinematic axis intermediate between the major and minor photometric axis appeared formally preferred by the fits, we concluded after experimentation that an oblate model with rotation around the photometric minor axis was acceptable. In any case, the data do show a preference for one spin sign (i.e. receding relative velocities on the Western longitudes) rather than another.}
are forced to be the same for all annuli (thus, in total, 8 free parameters). The measured amplitude and $\sigma_{\rm LOS}$ per annulus allow us to construct the rotation profile displayed in Figure~\ref{fig: vlos-rotation}, which has a mean $\left< v_{\rm LOS}/\sigma_{\rm LOS} \right> = 0.22 \pm 0.09$ averaged over all radii  (listed in Table~\ref{tab: overview}). For visualization, the blue line in the Figure displays the best fit using a parametrization of the form
\begin{equation} \label{eq: rotation-prof}
    \left(\frac{v}{\sigma}\right)_{\rm rot} = \left(\frac{v}{\sigma}\right)_{0} \, \frac{(R/R_{0})}{[1 + (R/R_{0})]^{1 + \zeta}} \ ,
\end{equation}
which increases linearly at small projected radii, and falls as a power-law at higher projected radii.

Our results lie between those found by \cite{Hargreaves+96} and \cite{Kleyna+02}, who used less complete datasets than ours. They found a rotation amplitude of $0.7 \ \kms$ around Draco's minor axis and of $6 \ \kms$ at 30 arcmin, respectively. Meanwhile, our $(v/\sigma)_{\rm rot}$ fit predicts $v/\sigma = 0.4$ at 30 arcmin, which translates to a rotation amplitude, $v_{\rm rot}$, of less than $4 \ \kms$ at this projected radius.
Our fit also agrees well with \cite{MartinezGarcia+21}, where internal rotation was confirmed using \gaia\ EDR3 PMs.
From those results, we conclude that although Draco has some mean rotation, it is small when compared to the overall velocity dispersion, especially at inner radii where most of the data are concentrated (see Figure~\ref{fig: surf-dens} for comparison). 

\subsubsection{Binaries} \label{sssec: binaries-los}

Per construction, measurements of $v_{\rm LOS}$ from spectral lines are subject to Doppler shifts from unresolved binary motion. As a consequence, single epoch LOS measurements can carry an overestimated $\sigma_{\rm LOS}$ and thus an overestimation of the system's total mass due to binary motion. Meanwhile, given the multi-epoch requirement of PM measurements, those end up averaging the motion of unresolved binaries to zero, such that the mentioned overestimation becomes negligible. For example, while \cite{Bianchini+16} showed that globular star clusters with unresolved binary fractions up to $50\%$ should introduce changes $< 6\%$ on the PM velocity dispersion, \cite{Pianta+22} recently performed simulations of dSphs to argue that one could reach much higher changes when using only LOS data.

\begin{figure}
\centering
\includegraphics[width=\hsize]{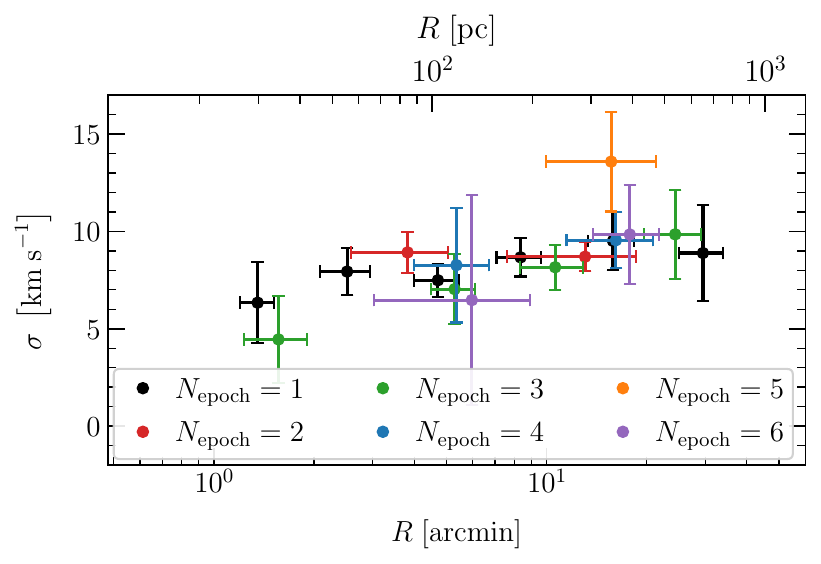}
\caption{\textit{Impact of unresolved binaries:} Multi-epoch velocity dispersion profiles (as a function of projected radii) for groups of LOS data constructed from a different number of epochs. The number of stars having multi-epoch observations is scarcer, and thus we have fewer radial bins for those. The excellent agreement within 1-$\sigma$ between all $N_{\rm epoch}$, and the fact that the $\sigma_{\rm LOS}$ profiles associated with higher $N_{\rm epoch}$ are not considerably smaller than the $N_{\rm epoch} = 1$ subset give us good confidence that unresolved binaries do not affect our mass estimates beyond the statistical uncertainties.}
\label{fig: multi-epoch-disp}
\end{figure}

Such an undesirable effect can be almost completely corrected by obtaining multi-epoch LOS observations, as recently argued by \cite{Wang+23}. Given Draco's high binary fraction of $50\%$ \citep{Spencer+18}, we perform here a multi-epoch test to gauge the influence of unresolved binaries on our cleaned Draco LOS dataset. To do so, we plot the radial profile of $\sigma_{\rm LOS}$ for groups of LOS data constructed from a different number of epochs. If unresolved binaries are to affect our LOS data as proposed by \cite{Pianta+22}, then one should expect multi-epoch velocity dispersions to be considerably smaller than the ones computed from single epoch exposures.

Figure~\ref{fig: multi-epoch-disp} displays our multi-epoch comparison, where velocity dispersion profiles are computed according to \citet[][appendix~A]{vanderMarel&Anderson&Anderson10} and \citet[][section~3.2.1]{Vitral+23}. The number of stars having multi-epoch observations is scarcer, and thus we have fewer radial bins for those. In any case, all our multi-epoch radial dispersion profiles agree within 1-$\sigma$ to the single-epoch measurement. Hence, Figure~\ref{fig: multi-epoch-disp} reassures us that for the cleaned LOS subset of \cite{Walker+23}, the effects of unresolved binaries in the velocity dispersion of Draco are within the statistical uncertainties. 
Finally, we revisit this conclusion in Section~\ref{ssec: binaries}, where we compare our results for subsets with and without LOS data.

\begin{figure}
\centering
\includegraphics[width=0.9\hsize]{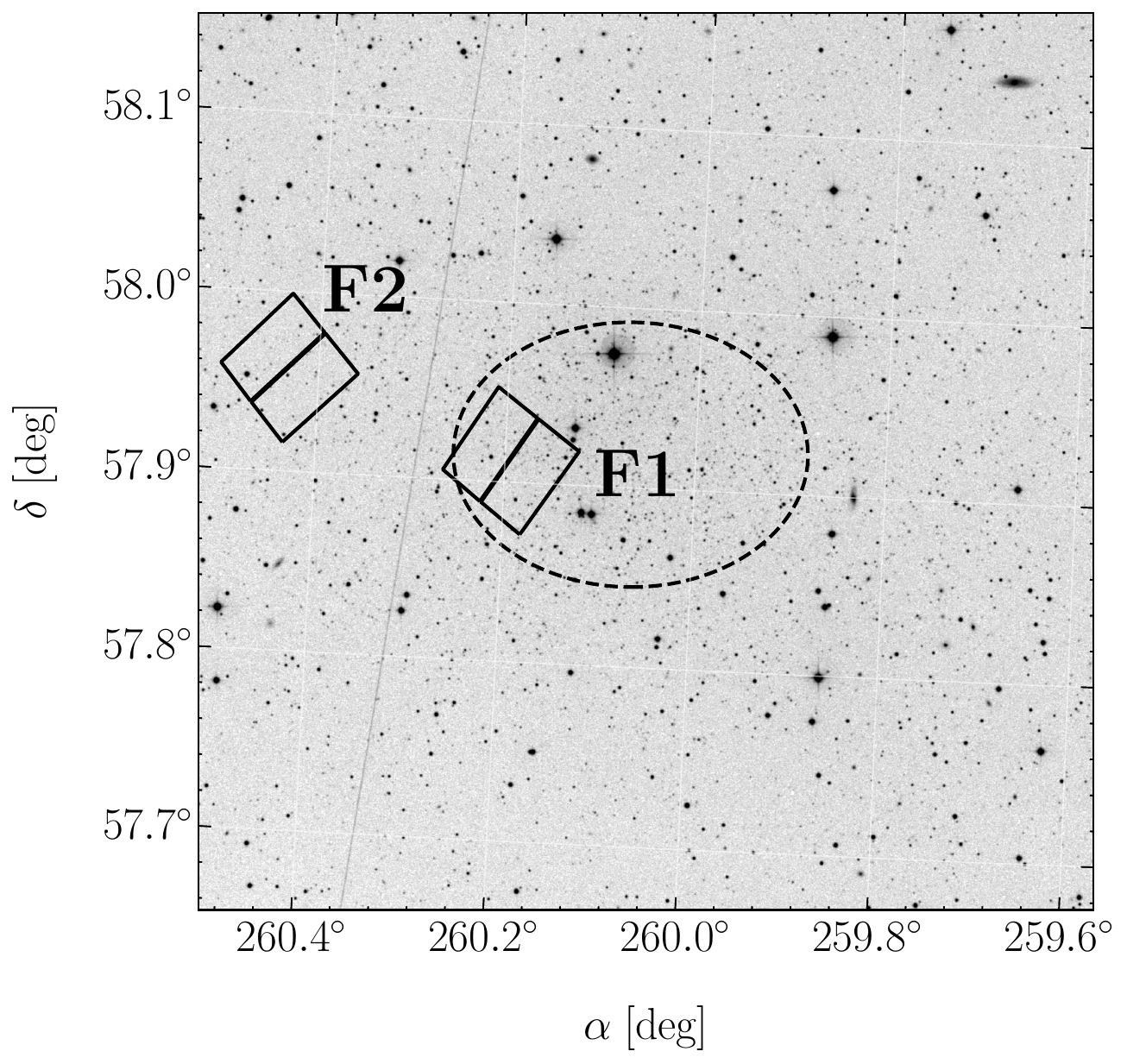}
\caption{\textit{Observed fields:} \hst\ target fields (background image is from the STScI Digitized Sky Survey, see acknowledgments), with a black ellipse showing Draco’s half-number radius.}
\label{fig: observed-fields-dss}
\end{figure}

\subsection{Proper motions} \label{ssec: pms}

\subsubsection{Observations and Astrometric Catalogs} \label{sssec: obs}

For our new PM measurements of Draco stars, we used multi-epoch \hst\ ACS/WFC imaging data. Descriptions about field locations and observations during earlier epochs for our target fields F1, and F2 are provided in \citet{Sohn+17}. The field locations are also shown in Figure~\ref{fig: observed-fields-dss}. 
In summary, the F1 field had three epochs of imaging data obtained in 2004, 2006, and 2013, while F2 had two epochs of imaging data obtained in 2004 and 2012. All fields were observed once again on October–November 2022 through our \hst\ program GO-16737 (\citealt{2021hst..prop16737S}) using the same filter (F606W), telescope pointing, and orientation as in the previous epochs.\footnote{We also observed another field F3, but found the resulting PMs of insufficient accuracy for the present purpose, due to the use of different \hst\ filters per epoch.} In this latest epoch, we obtained 15 individual exposures with each exposure lasting 430 seconds for each field.

The data analysis largely followed the procedures described in \citet{Bellini+2018} and \citet{Libralato+2018}. Here, we provide only a high-level outline of the PM derivation process and refer the reader to those papers for more details about the methodology. We downloaded the flat-fielded {\tt \_flt.fits} images of all target fields for each epoch from the Mikulski Archive for Space Telescopes (MAST) and processed them using the {\tt hst1pass} program \citep{Anderson2022} to derive a position and a flux for each star in each exposure. Instead of working on the {\tt \_flc.fits} images that are corrected for charge transfer efficiency (CTE) losses, we utilized the table-based CTE correction option in {\tt hst1pass}, which is an improved version of the ones used in previous works (\citealt{Anderson2022}, Anderson in prep.). We applied corrections to the positions using the ACS/WFC geometric distortions based on \citet{Kozhurina-Platais+2015}; these were further extended to include time-dependent distortion variations beyond 2020 (V. Kozhurina-Platais, private communication). 

For each field, we constructed a ``master frame'' using the average positions of stars from the repeated first-epoch exposures. The ($X, Y$) axes of these master frames were aligned with ($\alpha, \delta$) by registering the stellar positions to the \gaia\ DR3 astrometric system. We aligned the positions of stars from the other epochs to these master frames using a six-parameter linear transformation, and determined average positions for each epoch. By construction, this procedure aligns the star fields between different epochs leading to zero PM on average for the Draco dSph stars themselves. This does not affect our results since we are mostly interested in measuring the internal velocity dispersion on the plane of the sky (see Section~\ref{sssec: loc-corr-implementation} below for more discussion of this topic). Uncertainties on the average positions were determined from the repeated measurements as the root mean square divided by the square-root of the number of exposures. In the end, for each field per epoch, we prepared a catalog that includes positions of stars measured as described above as well as average instrumental\footnote{The instrumental magnitude in a given filter is defined here as ${\rm mag} = - 2.5 \log c$, where $c$ is the number of photon counts per exposure for a source.} F606W and F814W magnitudes (from the 2012--2013-epoch data) output by {\tt hst1pass}. 

\subsubsection{Photometric cleaning} \label{sssec: phot}

Once our observations are reduced and we have the master frame ($X, Y$) positions of sources at each epoch, for each field, we first perform a photometric cleaning of the data. The goal of this step is mainly (i) to remove interlopers, (ii) to remove background galaxies, and (iii) to remove stars associated with poor photometry that might bias our PM analysis. Points (i) and (ii -- iii) are performed independently, and we further select stars that simultaneously survived both sets of cleaning cuts.

\begin{figure}
\centering
\includegraphics[width=\hsize]{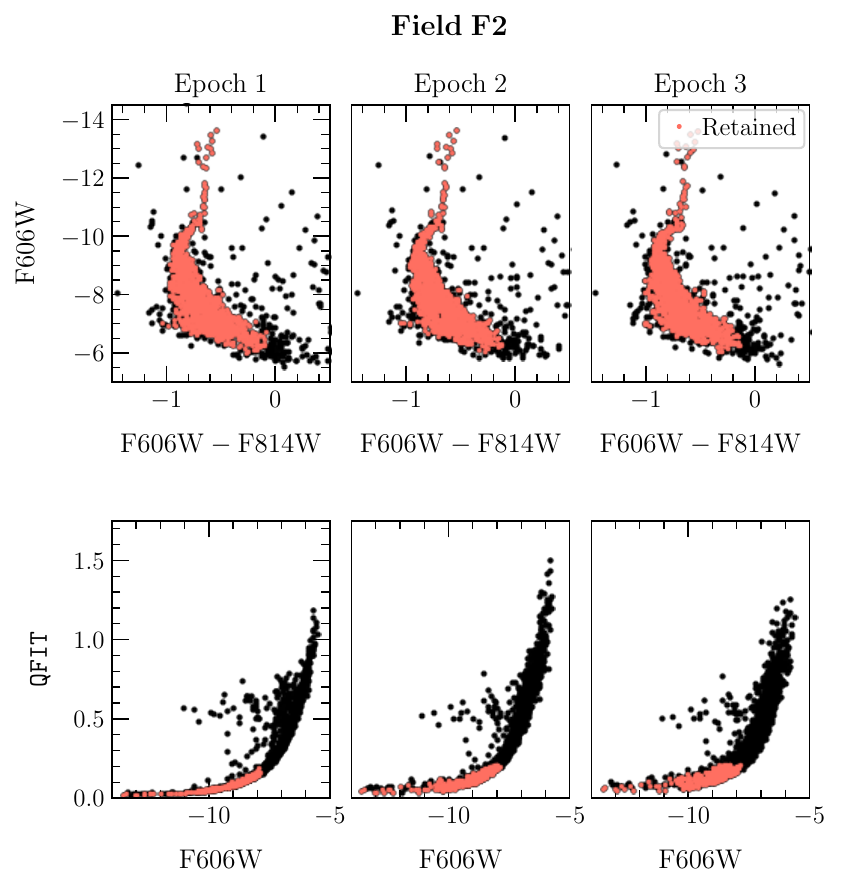}
\caption{\textit{Photometric cleaning:} The \textbf{upper panels} show the CMDs of our initial interloper cleaning of Field~2 (using instrumental magnitudes), with the original stars in the subset in \textit{black}, and the retained ones in \textit{orange}. The \textbf{lower panels} show the \texttt{QFIT}-based cleaning of Field~2, with the original stars in the subset in \textit{black}, and the retained ones in \textit{orange}. The plots attest to the effectiveness of our friends-of-friends photometric and interloper cleaning.}
\label{fig: phot-clean}
\end{figure}

Point (i) is accomplished by performing a cleaning on the color-magnitude diagram (CMD) of each field, at each epoch. We use a friends-of-friends procedure where we assign as an interloper a star whose distances to other stars in the CMD\footnote{The distance is defined in F606W vs.~(F606W -- F814W) space using percentiles to normalize each dimension, similarly to what is explained in section~3.3 of \citet[][eqs.~2 -- 3]{Vitral+22}.} is greater than typical distances in the subset. To do so, we define, after inspection of the CMD-distance distribution at each epoch, fiducial distance-thresholds to use in this cleaning. Since this step is likely to remove bright stars on the tip of the red giant branch and the horizontal branch (this is because they do not have a high number of neighbors), we reintroduce them to the cleaned subset. They are likely dSph members and would be filtered in further steps if they are not. As an example, Figure~\ref{fig: phot-clean} (upper panels) display the results of this CMD cleaning for the three epochs of Field 2.

Next, to address point (ii), we remove sources likely to be background galaxies, which lie on the upper side of the \texttt{QFIT} -- F606W diagram,\footnote{The \texttt{QFIT} parameter is a combined
measure of goodness of fit and S/N (see \citealt{Anderson+06,Libralato+14} for details).} departing from the bulk set of stars. This step is performed with a similar friends-of-friends analysis as for point (i), with different distance thresholds per field and per magnitude range. Finally, we proceed to point (iii) by removing stars that satisfy \texttt{QFIT}$\geq 0.2$, as they are associated with poor PSF fits. The final \texttt{QFIT} cleaning of our subset is displayed in the lower panels of Figure~\ref{fig: phot-clean}.

\subsubsection{Local corrections} \label{sssec: loc-corr-implementation}

After having a photometrically-cleaned subset which is also devoid, at least to a large extent, of interlopers, we proceed to compute the PM of each star in our subset. Essentially, the raw PMs are computed by a least-squares line fit of the master frame ($X, Y$) positions as a function of the epoch time. We use the \textsc{numpy.polyfit} routine from \textsc{Python}, assuming the ($X, Y$) uncertainties calculated in Section~\ref{sssec: obs}, and no $\chi^{2}$ re-scaling.\footnote{Essentially, the $\chi^{2}$ quantity is defined as $\chi^{2} = \sum^{N}_{i} \frac{(y_i - f(x_i))^{2}}{(N - N_{\rm free})}$, with $N_{\rm free} = 2$ (i.e. a line) and $N$ being the number of epochs used in the fit.} We store the $\chi^{2}$ of the fit for later data-cleaning.

\begin{figure}
\centering
\includegraphics[width=\hsize]{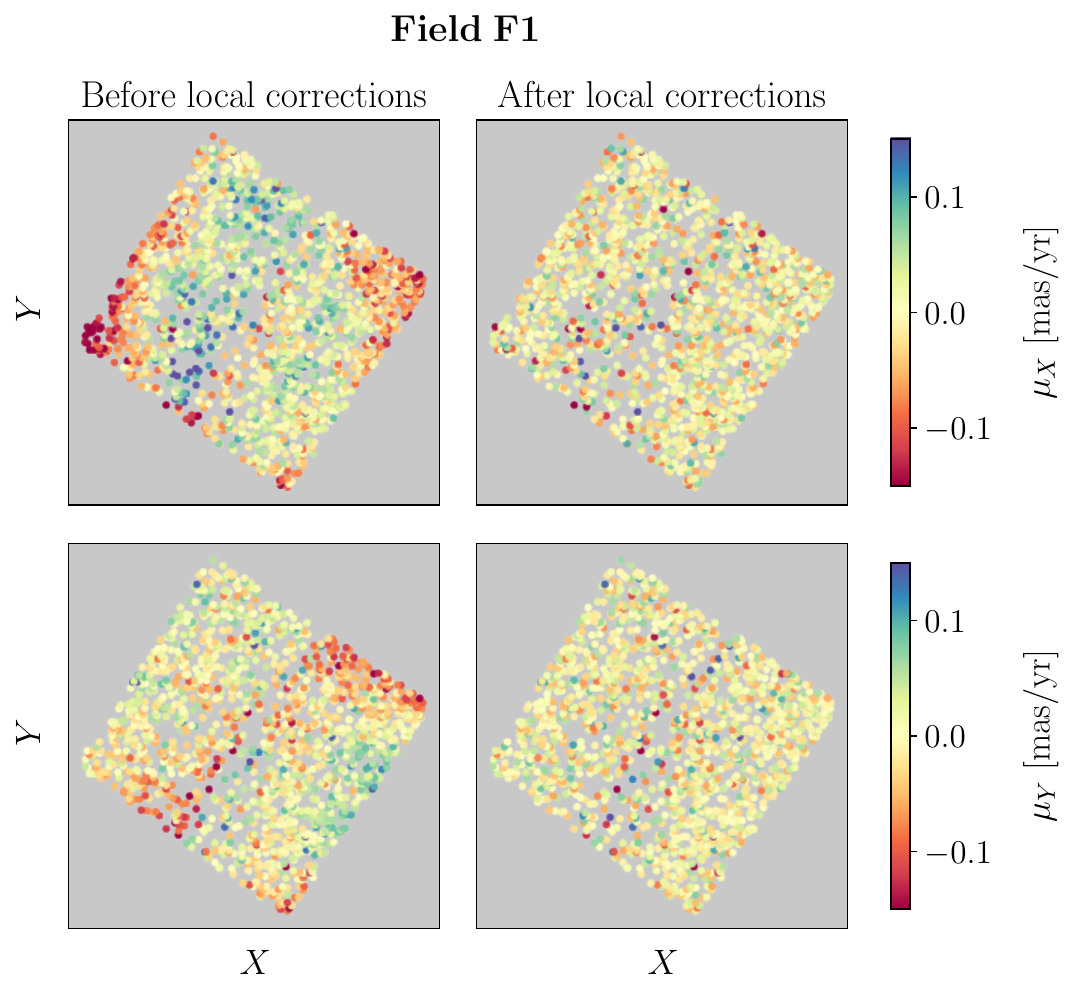}
\caption{\textit{Local corrections:} Effect of local proper motion (PM) corrections on our dataset (here, shown for Field~1). The panels show the master frame ($X, Y$) positions of the sources, color-coded according to the PM in each direction (for reference, Draco's typical PM dispersion peaks around $0.03 \ \masyr$). The \textbf{left} panels show clearly artificial PM shifts related to uncorrected residual CTE and 
geometric distortion effects, especially around the detector edges and boundaries. The \textbf{right} panels show our  corrected sample, which has a generally homogeneous PM distribution throughout the whole field, centered around null PMs.}
\label{fig: local-corrections}
\end{figure}

The raw PMs may contain low-level systematic effects related to the CTE issues of \hst's degrading charge-coupled devices (CCD), as well as from subtle variations in  geometrical distortion between epochs. As a result, some regions of the observed fields may present systematically higher/lower PMs. This problem has been previously reported, for instance, in \citet{Bellini+14} and \citet{Libralato+22}. We display this for our Field~1 in the left panels of Figure~\ref{fig: local-corrections}. The best procedure to correct for these effects is to perform a local PM correction that shifts those regions back to the bulk PM of the field. In practice, we follow the procedures laid out in previous works \citep[e.g.][]{Bellini+14,Libralato+22} that have constructed \hst\ PMs by looping over each star and removing the median PM of a local net of the ten closest\footnote{Here, `closest' refers to ($X, Y$) spatial positions. We verified that the systematics observed in Figure~\ref{fig: local-corrections} pertained mostly to geometrical distortions (rather then CTE), where distances in magnitude space are not relevant and could instead bring farther away stars into the local net.} stars. This process adds an extra layer of uncertainties (basically the uncertainty on the median,\footnote{The uncertainty on the median for a Gaussian distribution is given by $\epsilon_{\rm median} = \sqrt{\frac{\pi}{2} \frac{n}{(n-1)}} \, \epsilon_{\rm mean} = \sqrt{\frac{\pi}{2} \frac{n}{(n-1)}} \, \frac{\sigma}{\sqrt{n}}$ \citep{Kenney&KeepingPT1}, where $\sigma$ and $n$ are the distribution standard deviation and number of samples, which we fix to ten.} which we add quadratically to the original PM uncertainty), but successfully renders the PM dataset more homogeneous. The right panels of Figure~\ref{fig: local-corrections} show that is successfully removes most of the systematic effects. 

This local correction step removes not only streaming artifacts from the data, but also any variations in mean streaming intrinsic to the galaxy. However, it preserves the local velocity dispersion, which is most critical to perform mass-modeling.
We verified with axisymmetric rotating mock datasets that this step does not significantly change the second order velocity moment of the data (changes remain smaller than $\sim5\%$ for all possible inclinations). Moreoever, our axisymmetric model fitting in Section~\ref{ssec: axi-good-fit} below explicitly accounts for the fact that any mean streaming in the PM directions is not observationally constrained.

\subsubsection{Sky coordinates} \label{sssec: sky-coord}

As explained in Section~\ref{sssec: obs}, our master frame ($X, Y$) positions are already aligned, per construction, with sky coordinates, by using \gaia\ reference frames. This means that our PMs computed in ($X, Y$) directions are straightforwardly converted as $\mu_{\alpha, *} = - \mu_{X}$ and $\mu_{\delta} = \mu_{Y}$, where we denote $\mu_{\alpha, *} \equiv \mu_{\alpha} \cos{\delta}$.

To convert ($X, Y$) positions to ($\alpha, \delta$), we first perform a naive translation/rotation such that the coordinates match approximately the true ones. Next, we select brighter sources and associate them with \gaia\ EDR3 catalog sources. We perform a final translation/rotation to minimize the logarithm of the sum of distances between those matches, and apply the respective conversion parameters to our dataset. We verified that our matches are performed correctly by comparison to the Draco subset from \cite{delPino+22}.

\subsubsection{Outliers and underestimated errors} \label{sssec: under-err} \label{sssec: literature-comp}

After the conversion to sky coordinates, our PM subset is nearly ready for use. However, there might still be hidden interlopers in the data with unusually high PMs, or stars with underestimated errors that might bias our results. 

A rapid test to probe the number of such stars is to fit the PM distribution with a Gaussian (in both radial and tangential directions), and to compare the fraction of stars beyond 3-$\sigma$ to the fraction $\sim0.27\%$ predicted in this Gaussian. When performing this exercise, we observe that the fraction of stars in the wings of the distribution increases as we consider stars with higher PM uncertainties. This not only shows that we could be encompassing interlopers, but also points to the possibility of underestimated errors towards fainter stars.

\begin{figure}
\centering
\includegraphics[width=0.95\hsize]{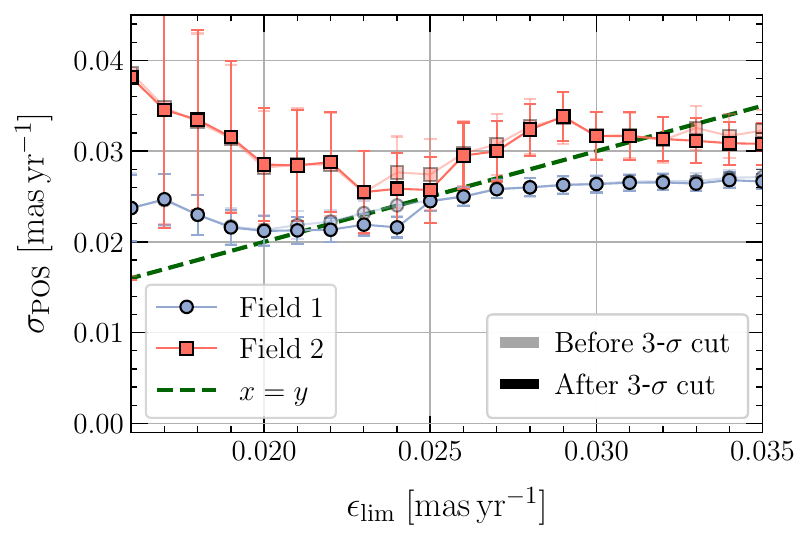}
\caption{\textit{Underestimated errors:} Measured proper motion (PM) dispersion, $\sigma_{\rm POS}$, as a function of the maximum PM uncertainty in the dataset, $\epsilon_{\rm lim}$. \textbf{Blue} curves relate to Field~1, while the \textbf{red} ones relate to Field~2. \textit{Opaque} curves relate to the PM subsets without the 3-$\sigma$ cleaning explained in Section~\ref{sssec: under-err}, while the \textit{solid} curves show results for the PM subsets with such cut. For reference, we display a $x = y$ line in dashed \textbf{green}. This plot shows that it is important to impose a cut in PM errors at $\epsilon_{\rm lim} \sim 0.024~\masyr$. Inclusion of stars with PM uncertainties in excess of the intrinsic galaxy dispersion yields an overestimated value for the galaxy dispersion.}
\label{fig: 3sig-clean}
\end{figure}

\begin{figure*}
\centering
\vspace{1cm}
\includegraphics[width=0.7\hsize]{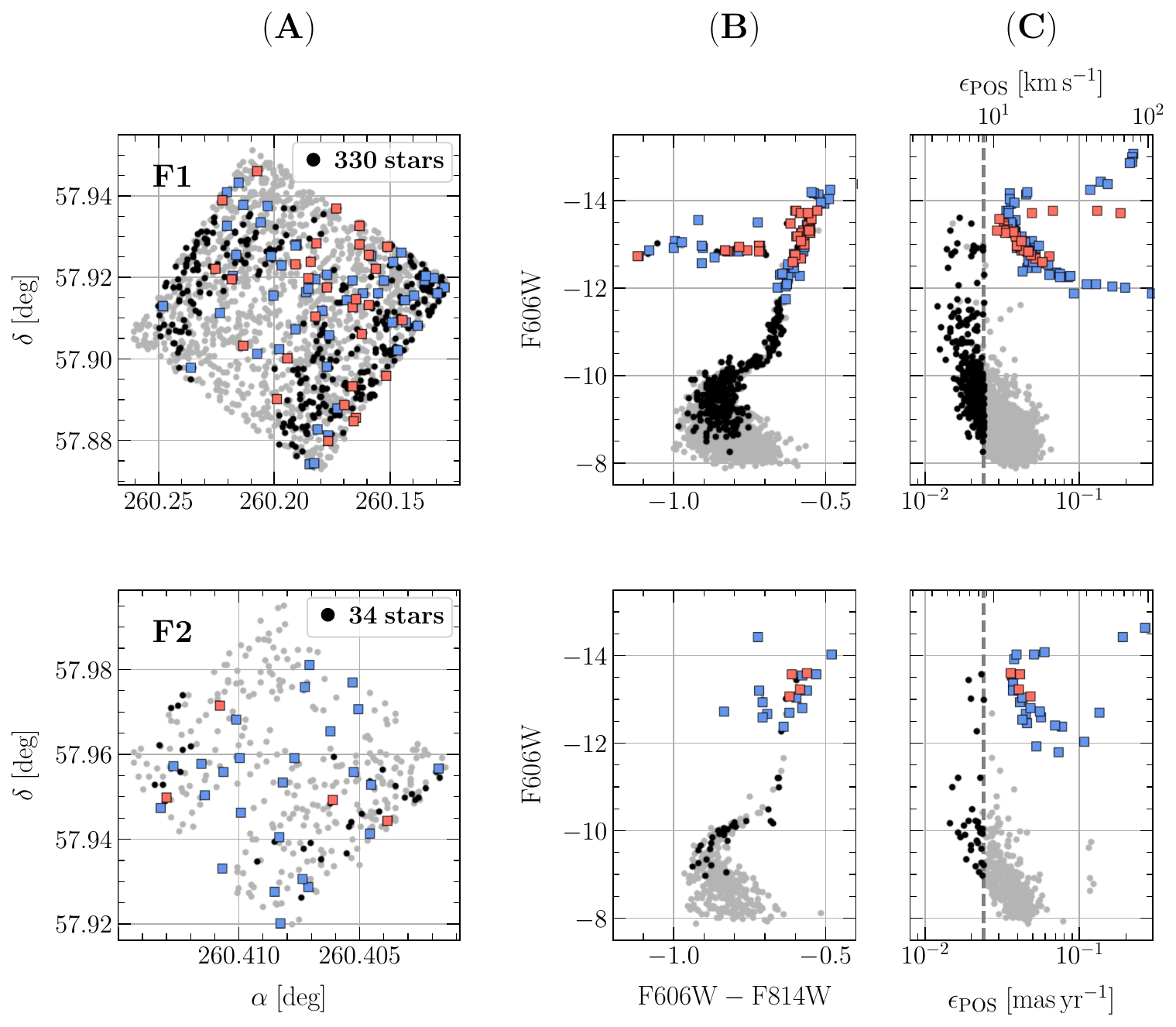}
\caption{\textit{Data overview:} \textbf{(A)} Zoom-in of studied fields. \textit{Black dots} indicate stars that have PM accuracies good enough to fulfill our scientific goals (i.e. $\epsilon_{\rm POS} \lesssim \sigma_{\rm POS}$) and that were used in our dynamical modeling, while the \textit{gray dots} indicate stars that do not, or that failed our multiple steps of data cleaning. The \hst$+$\gaia\ sample compiled with \textsc{GaiaHub} \citep{delPino+22} is marked in \textit{blue}, while the \textit{orange} stars depict the dataset used by \cite{Massari+20}. \textbf{(B)} CMDs based on \hst\ data (using instrumental magnitudes), using the same symbols as (A). \textbf{(C)} F606W magnitude as a function of the 1D PM error (calculated as in eq.~[B2] from \citealt{Lindegren+18}), using the same symbols as (A), with a dashed line representing the $\epsilon_{\rm POS}$ threshold we use. Our dataset comprises hundreds of stars with $\epsilon_{\rm POS} \lesssim \sigma_{\rm POS}$, compared to zero such stars in previous PM analyses of Draco.}
\label{fig: draco-fields}
\end{figure*}

\begin{figure*}
\centering
\includegraphics[width=0.9\hsize]{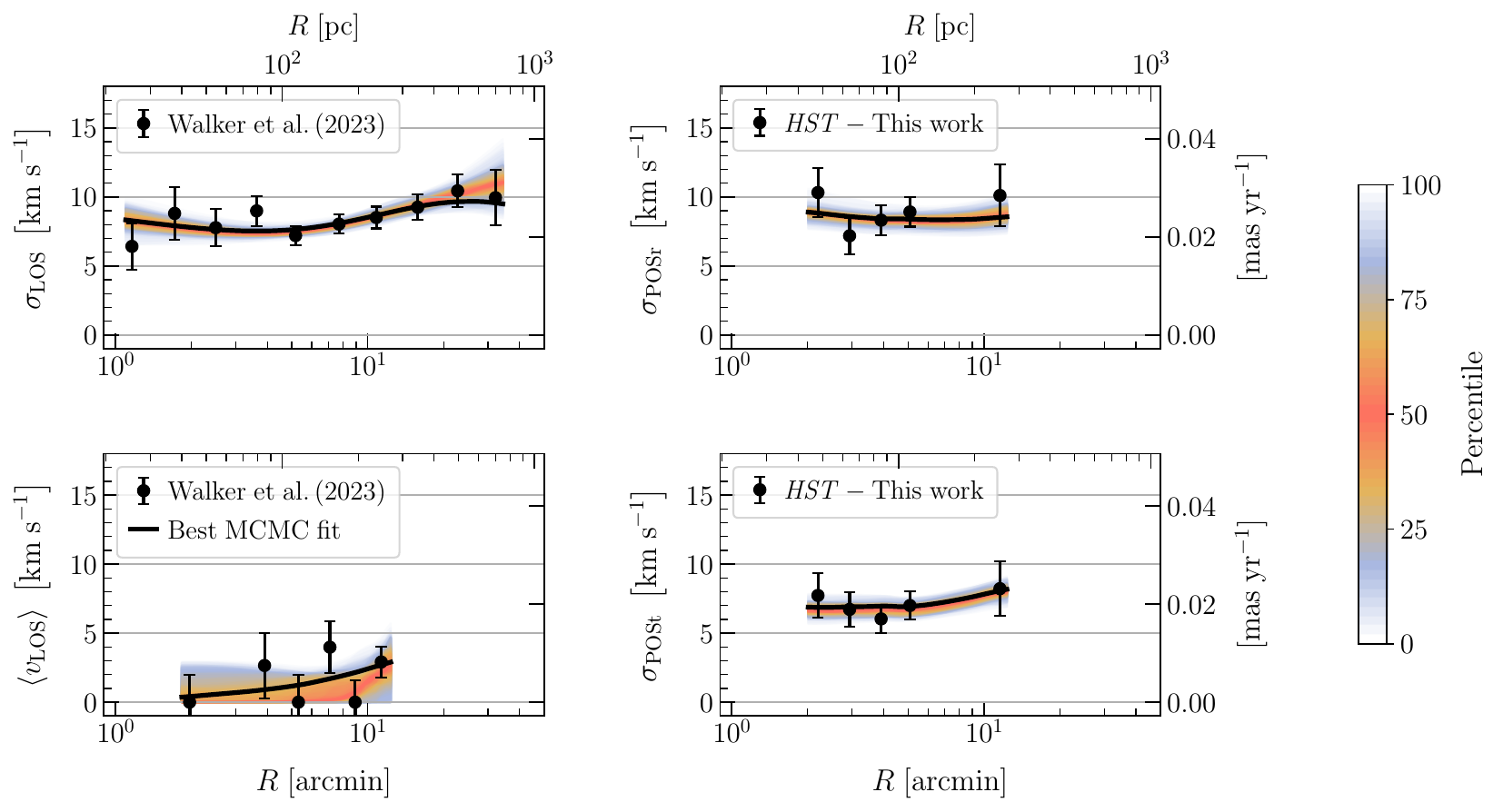}
\caption{\textit{Observations and model comparison:} The quantities showcased are --  
\textit{upper left}: velocity dispersion in the line-of-sight;
\textit{upper right}: plane-of-sky velocity dispersion in radial direction;
\textit{lower left}: line-of-sight rotation amplitude;
\textit{lower right}: plane-of-sky velocity dispersion in tangential direction.
The black circles and error bars represent the data, computed from the catalog of \cite{Walker+23} for the leftmost panels, and from our \hst\ program in the rightmost panels. Model predictions and the adopted galaxy distance (which we use to convert $\masyr$ to $\kms$ in the rightmost panel) are from our axisymmetric \jampy\ MCMC fits for $i = 57.1\degree$ as discussed in Section~\ref{ssec: axi-good-fit} below. 
Our best fit (as defined in Section~\ref{ssec: jeans-jampy}) is depicted as a black solid line, which we interpolate with respect to projected radius $R$ from the actual data $R$ values (this is done for visualization purposes, since there is also a dependence with the projected angle $\xi$).
The percentile from our MCMC chains are color-coded as in the color bar scheme, in the right.
}
\label{fig: disp-axi}
\end{figure*}

When measuring the velocity dispersion of a subset (as explained in Section~\ref{sssec: PMsigmaprofiles}), we are actually fitting the quadratic sum of the intrinsic dispersion and the errors associated with the tracers. If the errors are underestimated, the intrinsic dispersion will be overestimated. Figure~\ref{fig: 3sig-clean} shows the PM velocity dispersion (i.e. $\sigma_{\rm POS}$, where `POS' stands for `plane-of-sky') of stars with maximum PM uncertainty $\epsilon_{\rm lim}$. The curves show that the fitted $\sigma_{\rm POS}$ starts to increase as we include stars with errors beyond $\epsilon_{\rm lim} \gtrsim 0.024~\masyr$, roughly equal to the intrinsic velocity dispersion of the galaxy. Those are  fainter/high-magnitude stars that  likely have underestimated errors. Hence, for further analysis we removed all stars whose PM uncertainties exceed the threshold of $\epsilon_{\rm lim} = 0.024~\masyr$. In Section~\ref{ssec: error-limit} we further test the impact of this choice.

Given the possible issues related to stars with large PMs, we decided to also impose a 3-$\sigma$ cut on our PM sample. This can jointly remove unwanted interlopers and remaining stars with underestimated errors. Comparison of the solid and opaque lines in Figure~\ref{fig: 3sig-clean} shows that this does not strongly change the inferred $\sigma_{\rm POS}$. Nonetheless, the downside of any velocity cut is that it yields a slight underestimate of the true velocity dispersion (essentially, $\sim99.73\%$ of the true value for a 3-$\sigma$ cut of a Gaussian). To assure that this does not bias our dynamical modeling, which depends in part on comparison of LOS and PM kinematics, we performed the same cut in our LOS dataset (on top of the previous membership probability cut). This did not significantly change the LOS dataset, which already had a cut within a few $\sigma$ due to the larger fraction of interlopers. 

Our final dataset is the most complete and accurate PM catalog of a dSph to date, comprising 364 well measured stars.  
Comparatively, Figure~\ref{fig: draco-fields} shows that it comprises nearly ten times more stars than in \citet[][orange squares]{Massari+20}, twice more stars than in \citet[][blue squares]{delPino+22}, and it reaches much deeper magnitudes than both datasets could ever do given their necessity for \gaia\ measurements. Moreover, the uncertainties in our PM measurements are all below the local PM dispersion (see dashed gray line), compared to no such stars in both previous studies. This improvement is particularly important, because it is difficult to accurately constrain the PM dispersion of a galaxy based on individual PM measurements with uncertainties that do not resolve this dispersion (which is further compounded by known \gaia\ systematics, e.g. \citealt[]{Fardal+21} and \citealt{Vasiliev&Baumgardt&Baumgardt21}).

\subsubsection{Proper Motion Dispersion Profiles} \label{sssec: PMsigmaprofiles}

Having constructed the PM catalog, we proceed to construct velocity dispersion profiles that will be used throughout the next sections. As in Section~\ref{sssec: binaries-los}, all our computations of the dispersion $\sigma$ of a given random variable follow the recipe presented in \cite{vanderMarel&Anderson&Anderson10}, also recently employed in \cite{Vitral+23}. This consists of a maximum likelihood fit of a Gaussian distribution to the data, aiming to recover the respective standard deviation of the fit. The bias and uncertainty of such an estimate \citep[e.g.][]{Kenney&KeepingPT2} are corrected in a Monte Carlo sense, where we analyze numerous pseudo-data sets in the same fashion as the real data (see appendix~A from \citealt{vanderMarel&Anderson&Anderson10} for details).

For spherically symmetric models of Draco, the velocity dispersion profile can be written as a function of the projected distance to the galaxy's center, $R$. We thus create logarithmically-separated data bins in $R$ whenever we need to visualize $\sigma$. In practice, our spherical modeling deals with discrete data (see Section~\ref{ssec: jeans} below), such that the bin choices we use to visualize our results do not actually matter for the fitting procedure. For the axisymmetric case however, $\sigma$ will also depend on the projected angle $\xi$ of the data bin, defined as the angle between a given point and the projected major-axis of the galaxy. Besides, our fitting approach in this case is frequentist (see Section~\ref{ssec: jeans-jampy}), such that the binning process requires more attention. 

Our LOS kinematics are based on fits of all position angles along an annulus, with the rotation amplitude in Figure~\ref{fig: vlos-rotation} pertaining to the value on the kinematic major axis.\footnote{While our further modeling assumes an oblate dSph with maximum rotation on the equatorial plane, the data points used in our fits pertain to the kinematic major axis found in Section~\ref{sssec: rotation}, which did not align exactly with Draco's major projected axis, but was consistent within the uncertainties.} Instead, for the PMs we have measurements only for specific fields (see Figure~\ref{fig: observed-fields-dss}) that span a small range of position angles. Therefore, whenever sampling velocity moments as explained in Section~\ref{ssec: jeans-jampy}, these moments are averaged over all sky angles for the LOS, while we take the mean sky angle of each radial bin for the POS directions (namely, POSr for the \textit{plane-of-sky radial} direction and POSt for the \textit{plane-of-sky tangential} direction). We use the major axis (i.e. $\xi = 0\degree$) for comparison to the rotation amplitude.

The inferred velocity dispersion profiles in the three orthogonal directions are shown in Figure~\ref{fig: disp-axi}, together with the LOS rotation curve, all with similar $x$ and $y$-scales.\footnote{The adopted galaxy distance (used to transform mas/yr to km/s) and the model predictions in this figure will be discussed in Section~\ref{ssec: axi-good-fit} below. The distance $74.98^{+ 3.95}_{- 3.21}$ kpc is from the model fit, and is close  to the RR Lyrae estimate from \citet[][namely $D = 75.8 \pm 5.4$~kpc]{Bonanos+04}.} This provides, for the first time, radially-resolved 3D velocity dispersion profiles for any dwarf galaxy. Focusing on the observations, we note that the radial PM dispersion is considerably higher than the tangential PM dispersion. Averaged over all radii probed, $\left< \sigma_{\rm POSt} \right> / \left< \sigma_{\rm POSr} \right> = 0.80 \pm 0.08$. The ratio of the LOS dispersion to the PM dispersion is somewhat closer to unity, $\left< \sigma_{\rm LOS} \right> / \left< \sigma_{\rm POS} \right> = 1.08 \pm 0.09$, where $\sigma_{\rm POS}$ represents an average over both PM directions. The first ratio is independent of galaxy distance, while the second is inversely proportional to it. 

The tight observational constraints on ratios like these enable dynamical models of the kinds discussed in Section~\ref{sec: methods} below to strongly constrain the structure of Draco. To understand why, consider first the ratio $\sigma_{\rm POSt} / \sigma_{\rm POSr}$, which is a measure of the projected velocity dispersion anisotropy in the plane of the sky. In spherical geometry, 
\citet{Leonard&Merritt89} and \citet{vanderMarel&Anderson&Anderson10} both showed that there is a direct relation between this projected anisotropy and the  intrinsic three-dimensional velocity dispersion anisotropy. In Appendix~\ref{sapp: beta-dependence-scf} we use scale-free dynamical models of the type discussed in \cite*{deBruijne+96} to show that the same is expected to hold in axisymmetric geometry. The details of the relation depend on quantities that are constrained by observational data, such as the projected axial ratio of the system, the position angle of the tracers on the sky, the radial profiles of the luminous and dark matter densities, the viewing inclination of the galaxy,  etc. But in essence, $\sigma_{\rm POSt} / \sigma_{\rm POSr}$ is a diluted measure (i.e. brought closer to unity due to projection effects) of the intrinsic 3D ratio $\sigma_{\rm tan} / \sigma_r$ (see Figure~\ref{fig: b3-app} in the appendix). Thus, the observed 
$\left< \sigma_{\rm POSt} \right> / \left< \sigma_{\rm POSr} \right>$ implies the presence of radial velocity dispersion anisotropy in Draco. With suitable dynamical modeling, quantitative constraints are obtained on the shape of the 3D velocity dispersion tensor. This then breaks the mass-anisotropy degeneracy that plagues modeling of LOS velocities alone \citep{Binney&Mamon82}, so that the DM density profile can be determined. And with the 3D anisotropy known, the ratio $\sigma_{\rm LOS} / \sigma_{\rm POS}$ allows a kinematic determination of the galaxy distance (as done previously for globular clusters, e.g. \citealt{Watkins+15}). 

\section{Methods} \label{sec: methods}

Various techniques have been used to model the velocity dispersion profiles of dSphs and, thus, constrain their mass density profiles (see introduction). In this work, we employ multiple techniques to exploit our dataset. This helps us to understand any modeling uncertainties, and makes best use of the different codes available in the literature. We summarize them below.

\subsection{Spherical Jeans modeling: \mpo} \label{ssec: jeans}

Although Draco, like many other dSphs, is a flattened system (e.g. Table~\ref{tab: overview} and Figure~\ref{fig: observed-fields-dss}), previous studies have, in general, considered spherical models to fit its internal kinematics \citep[e.g.][]{Read+18,Massari+20}. Hence, it is useful to perform a similar kind of modeling if one wants to better interpret and compare previous results that assumed sphericity.

We perform spherical mass modeling with the Bayesian code \mpo \ (Mamon \& Vitral in prep.), which is an extension of {\sc MAMPOSSt} \citep{Mamon+13} to handle PMs in addition to line-of-sight velocities. \mpo\ is briefly described in section~2 of \cite{VitralMamon21}, and was tested by \cite{Read+21}, who showed that \mpo\ reproduced well the radial profiles of mass density and velocity anisotropy of mock dwarf spheroidal galaxies. \mpo\ is also a faster code than its mass-modeling counterparts (see Table~2 from \citealt{Read+21}), which allows us to probe a wide range of dynamical models in less time, which is useful when defining priors and fitting boundaries (see  Section~\ref{ssec: jeans-jampy} below). 

\subsubsection{General formalism of \mpo} \label{sssec: mpopm}

\mpo\ fits models for the radial profiles of total mass and the velocity anisotropy of the visible stars to the distribution of these stars in projected phase space. The \emph{local} velocity ellipsoid is assumed to be an anisotropic Gaussian, whose axes are aligned with the spherical coordinates. The sizes of the axes are obtained by solving the spherical Jeans equation 
\citep{Binney80}    
\begin{equation}
  \frac{{\rm d}\left (\nu\sigma_r^2\right)}{{\rm d}r} + 2\,\frac{\beta_{\rm B}(r)}{
    r}\,\nu(r)\sigma_r^2(r) = -\nu(r) \frac{G\,M(r)}{r^2} \ ,
\label{eq: jeans}
\end{equation}
assuming a given mass profile $M(r)$ and velocity anisotropy profile $\beta_{\rm B}(r)$, for a previously determined mass density profile 
$\nu(r)$ for the kinematic tracers (here stars). The term $\nu\,\sigma_r^2$ is the dynamical pressure that counteracts gravity.\footnote{The Jeans equation~(\ref{eq: jeans}) is a consequence of the Collisionless Boltzmann Equation, which considers the incompressibility in phase space of the six-dimensional (6D) distribution function (DF). Expressing the DF  in terms of 6D number, mass or luminosity density, implies that the term $\nu$ in the Jeans equation is the number, mass or luminosity density. For the present case of a dSph made of stars, it makes more physical sense to reason with mass density. For such systems, the mass density is proportional to the number density given the lack of substantial mass segregation, so the mass density profile is obtained from deprojecting the observed surface number density profile, and multiplying it by a constant factor.} The \citep{Binney80} velocity anisotropy (`anisotropy' for short) is defined as:
\begin{equation}    \label{eq: binney-beta}
    \beta_{\rm B} \equiv 1 - \displaystyle{\frac{\left< v_{\theta}^{2} \right> + \left< v_{\phi}^{2} \right> }{2 \,\left< v_{r}^{2} \right> }} \ ,
\end{equation}
where the $\left< v_i^2 \right>$ are the second-order velocity moments in spherical polar coordinates.  In both spherical and axisymmetric geometry, the first moments $\left< v_r \right> = \left< v_{\theta} \right> = 0$, so that the corresponding velocity dispersions satisfy $\sigma_r^2 = \left< v_r^2 \right>$ and $\sigma_{\theta}^2 = \left< v_{\theta}^2 \right>$. Also, in spherical geometry $\left< v_{\theta}^2 \right> = \left< v_{\phi}^2 \right>$. The first azimuthal moment $\left< v_{\phi} \right>$ need not generally be zero, so that in general $\left< v_{\phi}^2 \right> = \sigma_{\phi}^2 + \left< v_{\phi} \right>^2$. Our spherical models are constructed to have $\left< v_{\phi} \right> = 0$, but in the axisymmetric models that we present later we do allow for the possibility of mean rotation. 

In \mpo, the likelihood is written
\begin{equation}
    {\cal L} = \prod_i p({\bf v}_\mathbf{i}\,|R_i) \ ,
\end{equation}
where the conditional probability of measuring a velocity ${\bf v}_\mathbf{i}$ at projected radius $R_\mathbf{i}$ is the mean of the local  velocity distribution function, $h({\bf v}\,|\,R,r)$, integrated along the line of sight
\begin{equation}
    p({\bf v}\,|\,R) = \frac{2}{\Sigma(R)}\, \int_R^\infty h({\bf v}\,|\,R,r)\,\nu(r) \,\frac{r}{\sqrt{r^2-R^2}}\,{\rm d}r \ .
    \label{pvofR}
\end{equation}
\mpo\ determines the marginal distributions of the free parameters and their covariances by running the MCMC routine {\sc CosmoMC}\footnote{\url{https://cosmologist.info/cosmomc/}.} \citep{Lewis&Bridle02}.

In practice, we use 6 MCMC chains run in parallel and stop the exploration of parameter space after one of the chains reaches a number of steps $N_{\rm steps} = 10\,000 \,N_{\rm free}$, where $N_{\rm free}$ is the number of free parameters of the model.
We discard the first 3000\,$N_{\rm free}$ steps of each MCMC chain, which are associated with a burn-in phase. From the resulting chain values, we assign uncertainties to our best likelihood parameters using the 16th and 84th percentiles of the respective posterior distribution. If the fit lay below (above) those boundaries, we extended the uncertainty down to (up to) the minimum (maximum) chain value.

\subsubsection{Parametrizations \& Priors of \mpo} \label{sssec: priors}

\mpo\ is a parametric code that fits discrete data. The motivations for our choices of parametrization are described further below. Our choice of priors on the other hand is performed to maximize the entropy of the posterior probability distribution. This can be done by assigning flat priors whenever we assume no previous knowledge on a specific parameter, or Gaussian priors whenever we trust a previous measurement, from a different dataset, with reported mean and uncertainty.

The anisotropic runs of \mpo\ use the generalization (hereafter gOM) of the Osipkov-Merritt model (\citealt{Osipkov79, Merritt85}, Eq.~[\ref{eq: gOM}]) or the generalization (hereafter gTiret) of the (\citealt{Tiret+07}, Eq.~[\ref{eq: gTiret}]) model for the velocity anisotropy profile:
\begin{subequations}
\begin{flalign}
    \beta_{\mathrm{B, \, gOM}}(r) &= \beta_{0} + (\beta_{\infty} - \beta_{0}) \ \displaystyle{\frac{r^2}{r^2 + r_{\beta}^2}} \label{eq: gOM} \ , \\
    \beta_{\mathrm{B, \, gTiret}}(r) &= \beta_{0} + (\beta_{\infty} - \beta_{0}) \ \displaystyle{\frac{r}{r + r_{\beta}}} \label{eq: gTiret} \ ,
\end{flalign}
\end{subequations}
where $r_{\beta}$ is the anisotropy radius. We fit $\beta_{0}$ and $\beta_{\infty}$ using flat priors, from $-1.99$ to $1.99$ to the symmetrized quantity\footnote{$\beta_{\rm sym}$ runs from $-2$ for a model with only circular orbits to $+2$ for a model with only radial orbits, given that $\beta$ ranges between $-\infty$ and $+1$, respectively, for these cases.} $\beta_{\rm sym} = \beta / (1 - \beta/2)$, while  fixing $r_{\beta}$ to the scale radius of the luminous tracers.\footnote{This choice has been show to provide a better fitting convergence in \cite{VitralMamon21}.} Notice that a constant-anisotropy case is obtained by fixing $\beta_{\infty} = \beta_{0}$.

The mass density of the luminous tracers is chosen to be a Plummer model, similarly to what was adopted in \cite{Massari+20} and \citet{Hayashi+20}, and also supported by our previous fits of \gaia\ ERD3 data (see Fig~\ref{fig: surf-dens} in Section~\ref{sssec: surf-dens}). We fit the Plummer $r_{-2}$ radius\footnote{This is defined as the radius where $\frac{{\rm d} \log{\rho}}{{\rm d} \log{r}} = -2$.} with Gaussian priors, using the mean and uncertainty from our fits. The total luminous mass of Draco, $M_{\star}$, was estimated by \cite{Martin+08} from its CMD, by assuming either a \cite{Kroupa+93} or a \cite{Salpeter55} initial mass function (IMF). We estimate the mean and variance of $\log{M_{\star}}$ from both of those values, and use it as a Gaussian prior, which encompasses both estimates within 1-$\sigma$.

We test numerous parametrizations for the DM density profile, including:
\begin{itemize}
    \item A generalized Plummer model, which is a special case of the $\alpha\beta\gamma$ model by \cite{Zhao96}, with $\alpha = 2$ and $\beta = 5$.
    \item The \cite{Kazantzidis+04_densityprofs} model, which is motivated from $N$-body simulations of tidally stripped cuspy DM halos.
    \item The generalized NFW profile, motivated from cosmological simulations by \cite{Navarro+97}.
    \item The \cite{Einasto65} profile, which was used recently by \cite{Jiao+23} to fit the MW DM halo.
\end{itemize}

These density models are all listed in Appendix~\ref{app: dens-prof}, and depend on three quantities: a scale radius,\footnote{While Appendix~\ref{app: dens-prof} and Table~\ref{tab: mass-modeling-sph} display the usual scale radii for those parametrizations, \mpo\ fits the $r_{-2}$ quantity.} a total DM mass ($M_{\rm dark}$, or $M_{-2}$ for the generalized NFW model), and an inner slope $\gamma$ ($n$ index for the Einasto model). We assume flat priors for all these variables:
\begin{itemize}
    \item $\gamma \in [-2, 2]$, which encompasses both cuspy ($\gamma = -1$) and cored ($\gamma = 0$) cases (respectively, $n \in [0.1, 10]$). We allow for positive slopes as to not rule out possible physical mechanisms unforeseen by $\Lambda$CDM.
    \item $\log{(M_{\rm dark} / [\msun])} \in [6, 12]$. \cite{Read+17} extrapolated classical $M_{\star} - M_{200}$ relations to lower-mass dSphs, such that Draco, with a luminous mass $\sim 5 \times 10^{5} \ \msun$, is predicted to have $M_{200} \sim 10^{9} \ \msun$. Hence, our priors largely encompass that range.
    \item $\log{(r_{-2} / [{\rm kpc}])} \in [-1, 1]$. Given the expected $M_{200}$ mass from \cite{Read+17},\footnote{We kindly thank Justin Read for sharing his algorithm to compute the precise $M_{200}$ value expected for Draco.} the concentration relation from \cite{Dutton&Maccio14} yields $r_{-2, \rm NFW} \sim 1$~kpc. Our prior thus encompasses this range within an order of magnitude.
\end{itemize}

Finally, we set Gaussian priors for the bulk $v_{\rm LOS}$ of Draco, using our estimate depicted in Table~\ref{tab: overview}, and we also set Gaussian priors for the distance modulus, defined as $\mu_{0} = 5 \, \log{(D/[{\rm kpc}])} + 10$. The mean and uncertainty on the distance modulus are derived by propagating the value and respective uncertainty on the RR Lyrae estimate from \citet[][]{Bonanos+04}, which yields $\mu_{0} = 19.398 \pm 0.156$.

\subsection{Axisymmetric Jeans modeling: \jampy} \label{ssec: jeans-jampy}

To model our dataset under the assumption of an oblate axisymmetric galaxy, we use the publicly available code \jampy\ \citep{Cappellari08,Cappellari20}, tailored to the analysis of axisymmetric systems. This software was shown to reproduce well the dynamics of mock oblate dSphs with rotation \citep{Sedain&Kacharov23}, and has been  applied in \cite{Zhu+24} to recover DM structural parameters of thousands of galaxies.

\subsubsection{General formalism of \jampy} \label{sssec: jampy}

We use the version of \jampy\ in which the velocity ellipsoid is aligned with spherical coordinates, given that we assume a spherical global potential, which would be only minimally altered by Draco's luminous axisymmetric component. This configuration of \jampy\ considers the Jeans equations for rotating oblate systems (e.g. \citealt*{Bacon+83})
\begin{subequations}
\begin{flalign}
    &\frac{\partial \left(\nu \left<v^{2}_{r}\right>\right)}{\nu \, \partial r} + \frac{(1 + \beta_{\rm J}) \, \left<v^{2}_{r}\right> - \left<v^{2}_{\phi}\right>}{r} = - \frac{\partial \Phi}{\partial r} \ , \\
    &\frac{(1 - \beta_{\rm J})}{\nu} \left[\frac{\partial \left(\nu \left<v^{2}_{r}\right>\right)}{\partial \theta} + \frac{\nu \left<v^{2}_{r}\right>}{\tan{\theta}}\right] - \frac{ \left<v^{2}_{\phi}\right>}{\tan{\theta}} = - \frac{\partial \Phi}{\partial \theta} \ ,
\end{flalign}
\end{subequations}
where $\Phi$ is the gravitational potential, the symbol $\left<.\right>$ indicates the distribution function-averaged quantity, and finally, $\beta_{\rm J}$ is defined as
\begin{equation} \label{eq: beta-jampy}
    \beta_{\rm J} \equiv 1 - \frac{\left<v^{2}_{\theta}\right>}{\left<v^{2}_{r}\right>} = 1 - \frac{\sigma^{2}_{\theta}}{\sigma^{2}_{r}} \ ,
\end{equation}
where the second equality assumes that $\left<v_{\theta}\right> = \left<v_{r}\right> = 0$, as in \mpo. Because of symmetry and continuity, axisymmetric models always have $\left< v_{\phi}\right> = 0$ and $\left< v_{\phi}^2 \right> = \left< v_{\theta}^2\right>$ along the symmetric axis. Hence, along the symmetry axis, $\beta$ as defined by equation~\ref{eq: binney-beta} equals $\beta_{\rm J}$. Models with $\beta_{\rm J} = 0$ yield the same predicted second velocity moments as models in which the distribution function $f(E,Lz)$ does not depend on a third integral. Such models have been widely used for fitting data of axisymmetric systems \citep[e.g.][]{vanderMarel91}. Away from the symmetry axis, models with $\beta_{\rm J} = 0$ do {\it not} have an isotropic velocity dispersion tensor. 

Beyond the non-sphericity, another main difference between \mpo\ and \jampy\ is that the latter allows us to model Draco's rotation, by not imposing $\left<v_{\phi}\right> = 0$ throughout the whole system (this equality was also imposed in many previous analyses such as \citealt{Read+18,Hayashi+20,Massari+20}). The first moment in the $\phi$ direction relates to the second order moment in the radial direction through
\begin{subequations}
\begin{flalign}
    \left<v_{\phi}\right>^{2} &= \left<v^{2}_{\phi}\right> - \sigma^{2}_{\phi} \ , \\
    \sigma^{2}_{\phi} &= (1 - \Omega) \, \left<v_{r}^{2}\right> \ ,
\end{flalign}
\end{subequations}
where the rotation parameter $\Omega$ is introduced. This parameter was named  $\gamma$ in \cite{Cappellari20}, but we change this notation to avoid confusion with the inner slope of the DM mass density, which uses the same symbol.

From those equations, \jampy\ samples projected velocity moments that we use to compute the respective quantities in the LOS and POS directions. To fit our dataset, we employ an MCMC chain using the \textsc{emcee} routine that minimizes the $\chi^{2}$, defined as
\begin{equation} \label{eq: chi2-jampy}
    \chi^{2} = \chi^{2}_{\rm LOS} + \chi^{2}_{\rm POSr} + \chi^{2}_{\rm POSt} + \chi^{2}_{\rm rot} \ ,
\end{equation}
where $\chi^{2}_{x} = \sum (x_{\rm data} - x_{\rm model})^{2}_{i} / \epsilon^{2}_{x, \, i}$. In Eq.~(\ref{eq: chi2-jampy}), the four $\chi^{2}$ terms pertain to the LOS, POSr and POSt velocity dispersions at a given ($R$, $\xi$)\footnote{$R$ is the projected radius, and $\xi$ is the respective position angle in the plane-of-sky.} point, while the last term pertains to the first order moment of the LOS velocity on the major axis.\footnote{Our PM analysis methodology does not allow us to measure any mean streaming (see Section~\ref{sssec: loc-corr-implementation}), so it is not included in the $\chi^2$.}

We then maximize the log-probability of our dataset with the set of parameters $\Theta$, defined as $\ln \Pr\{\Theta\} = - \chi^{2} / 2$, along with respective priors defined further below in Section~\ref{sssec: priors-jampy}.
Our MCMC routine sets a maximum of 10\,000 iterations per fit, which we run in parallel in 64 CPUs. In those configurations, each run takes $\sim 3$~days to complete, and we perform it for a different set of possible inclinations from the PDF derived in Section~\ref{sssec: inc}. We discard the burn-in phase by removing the first 3\,000 steps of the chain and visually checking that the chains remain stable further on.

\subsubsection{Parametrizations \& Priors of \jampy} \label{sssec: priors-jampy}

As mentioned above, the timescales to run converging \jampy\ models are drastically longer than respective \mpo\ runs, which can be explained by both the software languages employed in each code (\textsc{Python} vs. \textsc{Fortran}, respectively), as well as the choice of parametrizations: While \mpo\ uses analytical parametrizations for a set of different models, \jampy\ assumes Multi-Gaussian Expansions (MGE) to model both the potential and the stellar distribution, which allows for more general density profiles at the expense of more time.

Therefore, we use our results from the spherical Jeans modeling to assist our fitting choices with \jampy. 
For instance, since we observed no significant preference for a particular DM density parametrization
in our spherical modeling results (see Section~\ref{ssec: sph-results}), we here decide to use only the generalized Plummer profile, as its analytical expressions are more easily handled when building MGEs in \textsc{Python}. In the absence of external constraints on the geometrical shape of Draco's DM halo, we continue to assume that it is spherical,\footnote{While cosmological simulations tend to favor generally triaxial DM halos \citep[e.g.][]{JS00,Kazantzidis+04_DMshape}, recent observational studies of the MW DM halo support a quasi-spherical potential within the inner $\sim 30$~kpc \citep*{Hattori+21,Wegg+19}.} even when the luminous density is chosen to be axisymmetric (so as to fit the observed projected shape of Draco). Although we use the same priors as \mpo\ for the DM density parameters\footnote{With exception of the DM scale radius, to which we allow a larger prior towards higher radii.} and Draco's distance, we fix the stellar density parameters,\footnote{This means we fix $M_{\star} = 4.7 \times 10^{5}~\msun$, and the major axis of the projected density as the value fitted in Section~\ref{sssec: center}, namely $a_{\rm maj, \star} = 9.1$~arcmin, where $a_{\rm maj}$ is the respective Plummer major axis (see Appendix~\ref{app: dens-prof}).} as we observed no departure from the mean \mpo\ Gaussian priors. In addition, we assume the $\beta_{\rm J}$
parameter to be a constant,\footnote{A similar assumption is also present in \cite{Hayashi+20}, who base their choices on cosmological simulations by \cite{Vera-Ciro+14}. For robustness purposes, we also ran a test with a gOM-like parametrization for $\beta_{\rm J}$ and observed no departure from the constant case.} since we show in Section~\ref{sec: results} that the data does not prefer more general profiles such as the 
Osipkov-Merrit generalization of Eq.~(\ref{eq: gOM}). Indeed, because \mpo\ assumes no rotation and spherical symmetry, such that $\sigma_{\theta} = \sigma_{\phi}$, its velocity anisotropy parameter is equal to \jampy 's $\beta_{\rm J}$.

Finally, we observed that when assuming a spatially constant rotation parameter $\Omega$, we could not fit well enough Draco's observed rotation curve. Hence, we assume the more general behavior
\begin{equation} \label{eq: rotation-jampy}
    \Omega(r) = \Omega_{0} + (\Omega_{\infty} - \Omega_{0}) \frac{1}{1 + (r_{\Omega}/r)^{2}} \ .
\end{equation}

We fit ($\Omega_{0}$, $\Omega_{\infty}$) by assuming flat priors from $-1.99$ to $1.99$ to the symmetrized quantity $\Omega_{\rm sym} = \Omega / (1 - \Omega/2)$, while fixing $r_{\Omega}$ to the luminous scale radius.

As a consistency check that the fitting strategy of the different Jeans modeling algorithms we use do not strongly diverge from each other, we compared two constant-anisotropy runs from \mpo\ and \jampy\ for a spherical geometry, and confirmed that both the inferred velocity anisotropy and the DM density slope differ by much less than their respective 1-$\sigma$ uncertainties.\footnote{Precisely, we measure $\Delta \beta = 0.04$ and $\Delta \gamma = 0.06$, while the uncertainty on each parameter for the spherical case is of the order of $\sim0.15$ and $\sim1$, respectively.}

\subsection{Practical Quantities} \label{ssec: obs-param}

Given the choices of different DM parametrizations to compare with, and the fact that our data is not complete at all radii, we define here practical quantities to help us interpret our results. For example, the total DM mass (or $M_{-2}$ for NFW profiles) is not a well-constrained quantity, since our data does not really allow us to constrain in detail the shape of the DM density at large radii. Hence, a more suitable parameter to display and use for comparison purposes is the DM mass up to a fiducial radius. We do so, by displaying further in Tables~\ref{tab: mass-modeling-sph} and \ref{tab: mass-modeling-axi} the variable $M_{\rm dark}(r = R_{\max})$ -- i.e. the total mass of DM up to the maximum projected radius in our LOS$+$PM dataset, namely $R_{\rm max} = 900$~pc. To aid in the interpretation of this quantity we also compute, at this same radius, the circular velocity 
\begin{equation} \label{eq: vcirc}
    v_{\rm circ}(r) = \sqrt{\frac{G \, M(r)}{r}} \ ,
\end{equation}
which depends on the total\footnote{The total mass is a sum of the luminous and dark components.} cumulative mass up to a certain radius $r$, and on the gravitational constant $G$.

Similarly, due to the restricted spatial extent of our data, the inner dark matter slope parameter $\gamma_{\rm dark}$, or the respective Einasto index $n_{\rm dark}$, may not reflect accurately our fits and uncertainties of the DM slope where we are actually able to constrain it -- i.e. where we have both PM and LOS data. We therefore define an effective density slope parameter as
\begin{equation} \label{eq: gamma-practicle}
    \Gamma_{\rm dark} \equiv \displaystyle{\frac{\int_{r_{\rm min}}^{r_{\rm max}} \frac{{\rm d} \log{\rho}}{{\rm d} \log{r}} \, \rho(r) \, {\rm d}r}{\int_{r_{\rm min}}^{r_{\rm max}} \rho(r) \, {\rm d}r}} \ ,
\end{equation}
where $\rho(r)$ is the DM density. The $r_{\rm min}$ variable is defined as the minimum projected radius in the data where PM information is available. 
We define $r_{\rm max} = {\rm min}(R_{\rm max, PM}, r_{\Lambda{\rm CDM}}/3)$, where $R_{\rm max, PM}$ is the maximum projected radius in the data where PM information is available and $r_{\Lambda{\rm CDM}}$ is the scale radius of a NFW profile as expected for DM halos in $\Lambda$CDM simulations\footnote{We choose this value as reference because we wish to compare our observables to what is predicted from theory, while the $(1/3)$ factor is added with the intent of removing the part of the predicted density profile that has a cuspier drop due to the transition from the inner to the outer density profile.} assuming low-mass stellar components \citep{Read+17}.

In practice, the radial limits over which we average the logarithmic DM density slope are $r_{\rm min} = 42$~pc and $r_{\rm max} = 297$~pc, which translate to roughly $r_{\rm min} = 1.9$~arcmin and $r_{\rm max} = 22.5$~arcmin. If one considers the scale radius $r_{\Lambda{\rm CDM}}$, the conversion between $\Gamma_{\rm dark}$ and $\gamma_{\rm dark}$ in our PM radial range is such that cusp ($\gamma_{\rm dark} = -1$) and cored ($\gamma_{\rm dark} = 0$) values translate to $\Gamma_{\rm dark} = -1.2$ and $\Gamma_{\rm dark} = -0.38$, respectively, for a generalized NFW profile. For a generalized Plummer profile such as used in our axisymmetric fits, the respective numbers are $\Gamma_{\rm dark} = -1.07$ and $\Gamma_{\rm dark} = -0.14$ for cusp and cored models, thus providing a more subtle difference.

\newcommand\vmcite{{\citealt{Vitral&Mamon20}}}
\begin{deluxetable*}{rllllrrrllrrr}
\tablecaption{Main results of the \mpo\ spherical Jeans modeling.}
\label{tab: mass-modeling-sph}
\tablewidth{750pt}
\renewcommand{\arraystretch}{1.3}
\tabcolsep=3.2pt
\tabletypesize{\scriptsize}
\tablehead{
\colhead{ID} & 
\colhead{$\rho_{\rm dark}$} &
\colhead{Test} &
\colhead{$D$} &
\colhead{$\beta_{0}$} &
\colhead{$\beta_{\infty}$} &
\colhead{$r_{\star}$} &
\colhead{$M_{\star}$} &
\colhead{$r_{\rm dark}$} &
\colhead{$M_{\rm dark}^{R_{\rm max}}$} &
\colhead{$\gamma_{\rm dark}$ or $n$} &
\colhead{$\Gamma_{\rm dark}$} &
\colhead{$\Delta \rm AICc$} \\
\colhead{} &
\colhead{} &
\colhead{} &
\colhead{[kpc]} &
\colhead{} &
\colhead{} &
\colhead{$\left[10^{2}~\rm{pc}\right]$} & 
\colhead{$\left[10^{5}~\rm{M}_\odot\right]$} & 
\colhead{$\left[10^{2}~\rm{pc}\right]$} & 
\colhead{$\left[10^{8}~\rm{M}_\odot\right]$} & 
\colhead{} &
\colhead{} \\
\colhead{(1)} &
\colhead{(2)} & 
\colhead{(3)} & 
\colhead{(4)} & 
\colhead{(5)} & 
\colhead{(6)} & 
\colhead{(7)} & 
\colhead{(8)} &
\colhead{(9)} &
\colhead{(10)} &
\colhead{(11)} &
\colhead{(12)} &
\colhead{(13)}
}
\startdata
1 & GKAZ &  $\rho_{\rm dark}$  & $  75.76^{+ 3.23}_{- 4.22} $ & $  0.45^{+ 0.07}_{- 0.19} $ &  --  & $  1.74^{+ 0.10}_{- 0.11} $ & $  4.25^{+ 3.16}_{- 1.16} $ & $  1.93^{+ 12.17}_{- 0.27} $ & $  1.34^{+ 3.79}_{- 0.14} $ & $  0.76^{+ 0.32}_{- 1.43} $ & $ -0.11^{+ 0.28}_{- 0.79} $ & $  0.00 $ \\
2 & EIN &  $\rho_{\rm dark}$  & $  76.73^{+ 3.03}_{- 4.28} $ & $  0.40^{+ 0.06}_{- 0.23} $ &  --  & $  1.77^{+ 0.09}_{- 0.12} $ & $  5.05^{+ 2.24}_{- 1.97} $ & $  4.91^{+ 19.38}_{- 4.91} $ & $  1.45^{+ 4.49}_{- 1.45} $ & $  0.70^{+ 7.80}_{- 0.60} $ & $ -0.31^{+ 0.26}_{- 0.34} $ & $  0.11 $ \\
3 & GPLU &  $\rho_{\rm dark}$  & $  75.99^{+ 2.89}_{- 4.52} $ & $  0.39^{+ 0.13}_{- 0.14} $ &  --  & $  1.76^{+ 0.08}_{- 0.13} $ & $  5.07^{+ 2.06}_{- 2.07} $ & $  8.68^{+ 14.83}_{- 4.04} $ & $  1.41^{+ 0.68}_{- 0.43} $ & $ -0.24^{+ 1.25}_{- 0.55} $ & $ -0.41^{+ 0.88}_{- 0.42} $ & $  0.17 $ \\
4 & GNFW &  $\rho_{\rm dark}$  & $  74.74^{+ 4.76}_{- 2.65} $ & $  0.41^{+ 0.09}_{- 0.18} $ &  --  & $  1.75^{+ 0.11}_{- 0.09} $ & $  4.30^{+ 2.99}_{- 1.34} $ & $  2.47^{+ 15.46}_{- 0.06} $ & $  0.80^{+ 1.62}_{- 0.01} $ & $  1.44^{+ 0.01}_{- 2.00} $ & $ -0.24^{+ 0.01}_{- 0.72} $ & $  0.34 $ \\
\hline
5 & GPLU & $\beta_{\rm gOM}$ & $  74.85^{+ 4.61}_{- 3.10} $ & $  0.48^{+ 0.08}_{- 0.21} $ & $ -0.98^{+ 1.78}_{- 0.92} $ & $  1.74^{+ 0.11}_{- 0.10} $ & $  4.12^{+ 3.29}_{- 1.10} $ & $  4.94^{+ 16.52}_{- 0.88} $ & $  1.14^{+ 1.54}_{- 0.12} $ & $  0.43^{+ 0.75}_{- 1.22} $ & $ -0.17^{+ 0.45}_{- 0.69} $ & $  1.99 $ \\
6 & GPLU & $\beta_{\rm gTiret}$ & $  75.87^{+ 3.51}_{- 4.04} $ & $  0.52^{+ 0.19}_{- 0.23} $ & $ -0.21^{+ 0.95}_{- 1.26} $ & $  1.75^{+ 0.10}_{- 0.10} $ & $  4.70^{+ 2.69}_{- 1.67} $ & $  6.47^{+ 18.15}_{- 2.25} $ & $  1.25^{+ 1.05}_{- 0.32} $ & $ -0.03^{+ 1.10}_{- 0.81} $ & $ -0.35^{+ 0.71}_{- 0.54} $ & $  1.76 $ \\
7 & GPLU & Cusp & $  77.27^{+ 3.72}_{- 3.34} $ & $  0.25^{+ 0.11}_{- 0.17} $ &  --  & $  1.78^{+ 0.11}_{- 0.10} $ & $  5.04^{+ 2.10}_{- 2.06} $ & $  28.42^{+ 83.80}_{- 7.71} $ & $  1.26^{+ 0.12}_{- 0.10} $ & \darkg{$\gamma = -1 $} & $ -1.00^{+ 0.00}_{- 0.00} $ & $  1.52 $ \\
8 & GPLU & Core & $  74.57^{+ 3.97}_{- 3.20} $ & $  0.40^{+ 0.12}_{- 0.09} $ &  --  & $  1.72^{+ 0.11}_{- 0.09} $ & $  4.33^{+ 3.02}_{- 1.28} $ & $  6.92^{+ 1.93}_{- 0.82} $ & $  1.37^{+ 0.37}_{- 0.21} $ & \darkg{$\gamma = 0 $} & $ -0.29^{+ 0.10}_{- 0.07} $ & $ -2.16 $ \\
9 & GPLU & PM & $  75.68^{+ 4.93}_{- 5.65} $ & $  0.53^{+ 0.07}_{- 0.18} $ &  --  & $  1.73^{+ 0.14}_{- 0.13} $ & $  4.49^{+ 2.96}_{- 1.45} $ & $  6.04^{+ 47.73}_{- 2.31} $ & $  1.40^{+ 2.11}_{- 0.70} $ & $  0.34^{+ 0.70}_{- 1.13} $ & $ -0.09^{+ 0.27}_{- 0.81} $ &  --  \\
10 & GPLU & $\epsilon$ & $  73.59^{+ 5.20}_{- 2.62} $ & $  0.56^{+ 0.04}_{- 0.21} $ &  --  & $  1.69^{+ 0.14}_{- 0.07} $ & $  5.08^{+ 2.19}_{- 2.06} $ & $  5.79^{+ 8.53}_{- 1.30} $ & $  1.61^{+ 1.04}_{- 0.37} $ & $  0.70^{+ 0.64}_{- 1.24} $ & $  0.18^{+ 0.33}_{- 0.85} $ &  --  \\
\enddata
\begin{tablenotes}
\scriptsize
\item \textsc{Notes} --  
Columns are 
\textbf{(1)} Model ID; 
\textbf{(2)} Dark matter parametrization: ``GPLU'' for a generalized \cite{Plummer1911} model with free inner slope, ``GKAZ'' for a generalized \cite{Kazantzidis+04_densityprofs} model with free inner slope, ``GNFW'' for a generalized \cite*{Navarro+97} model with free inner slope, and ``EIN'' for the \cite{Einasto65} model;
\textbf{(3)} Test type: 
``$\rho_{\rm dark}$'' when testing different parametrizations for the DM density profile,
``$\beta_{\rm gOM}$'' for a generalized \citet{Osipkov79}--\citet{Merritt85} parametrization of the velocity anisotropy profile, 
``$\beta_{\rm gTiret}$'' for a generalized \cite{Tiret+07} parametrization of the velocity anisotropy profile,
``Cusp'' when forcing an inner density slope of $-1$ for the dark matter, 
``Core'' when forcing a cored model for the dark matter, 
``PM'' when only using PMs (no LOS data) and
``$\epsilon$'' when using a lower PM error threshold; 
\textbf{(4)} Heliocentric distance, in kpc;
\textbf{(5)} anisotropy value at $r=0$;
\textbf{(6)} anisotropy value at infinity (only for models with variable anisotropy);
\textbf{(7)} Plummer scale radius of the stellar component, in $10^{2}$~pc;
\textbf{(8)} Total mass of the stellar component, in $10^{5}$~M$_{\odot}$;
\textbf{(9)} Dark matter scale radius, in $10^{2}$~pc;
\textbf{(10)} Dark matter mass at the maximum projected data radius, in $10^{8}$~M$_{\odot}$;
\textbf{(11)} Dark matter asymptotic density slope or Einasto index $n$;
\textbf{(12)} Dark matter density slope averaged over the spatial range where PMs are available;
\textbf{(13)} Difference in AICc relative to model~1.
Listed uncertainties are based on the 16th and 84th percentiles of the marginal distributions, unless the maximum likelihood solution was outside that boundary, in which case the uncertainties are related to the minimum or maximum value of the MCMC chain.
We did not consider the AICc diagnostic when the data set was different from the respective standard model.
\end{tablenotes}
\end{deluxetable*}

\subsection{Statistical tools} \label{ssec: stats}

We employ Bayesian evidence to compare our different \mpo\ mass-anisotropy models and correct for over- and under-fitting. This model selection involves comparing the maximum log posteriors using Bayesian information criteria.
We use the corrected Akaike Information Criterion (derived by \citealt{Sugiyara78} and independently by \citealt{HurvichTsai89} who demonstrated its utility for a wide range of models)
\begin{equation}
      \mathrm{AICc} =\mathrm{AIC} + 2 \, \frac{N_{\mathrm{free}} \,  (1 + N_{\mathrm{free}})}{N_{\mathrm{data}} - N_{\mathrm{free}} - 1} \ ,
\end{equation}
where AIC is the original Akaike Information Criterion \citep{akaike1973information}
\begin{equation}
 \label{eq: AIC}
    \mathrm{AIC} = - 2 \, \ln \mathcal{L_{\mathrm{MLE}}} + 2 \, N_{\mathrm{free}} \ ,
\end{equation}
and where ${\cal L}_{\rm MLE}$ is the maximum likelihood estimate found when exploring the parameter space, $N_{\rm free}$ is the number of free parameters, and $N_{\rm data}$ the number of data points. 
We prefer AICc to the other popular simple Bayesian evidence model, the Bayes Information Criterion (BIC, \citealt{Schwarz78}), because  AICc is more robust for situations where the true model is not among the tested ones (for example our choice of a Plummer density profile for the stellar component is purely empirical and not theoretically motivated), in contrast with BIC
\citep{Burnham&Anderson02}.

The likelihood (given the data) of one model relative to a reference one is \citep{Akaike83}
\begin{equation}
\rm \exp\left(-\frac{AIC-AIC_{\rm ref}}{2}\right) \ ,
\label{eq: pAIC}
\end{equation}
and we use it to infer likelihood probabilities.
In general $\Delta {\rm AICc}$ differences $\gtrsim 4$ (i.e. a confidence level $\gtrsim 85\%$, according to eq.~[\ref{eq: pAIC}] above) are required to prefer one model over another.
It is also important to mention that such diagnostics are purely statistical, and do not account for intrinsic astrophysical phenomena that might favor or disfavor a particular model.

\section{Results} \label{sec: results}

\subsection{Spherical modeling} \label{ssec: sph-results}

We first present the results of spherical Jeans modeling with \mpo. Key outcomes are listed in Table~\ref{tab: mass-modeling-sph}. Our velocity dispersion goodness-of-fits for the spherical case were very similar to the ones presented in Figure~\ref{fig: disp-axi} for the case of axisymmetric models, so we do not show the spherical fits separately (caveat:  our spherical modeling neglects rotation, so the spherical model predictions in the lower left panel are zero). Below, we detail our results.

\subsubsection{Dark Matter density parametrization} \label{sssec: dark-matter-sph}

The first four lines of Table~\ref{tab: mass-modeling-sph} address the comparison between the four DM density parametrizations we use: \cite{Kazantzidis+04_densityprofs} listed as GKAZ, \cite{Einasto65} listed as EIN, generalized \cite{Plummer1911} with free inner slope listed as GPLU, and, finally, the generalized \cite{Navarro+97} with free inner slope listed as GNFW. We refer the reader back to Section~\ref{sssec: priors} for the motivations of each parametrization and now focus on the practical fitting results.

A comparison of these models' AICc yields a modest preference for the GKAZ profile, followed by EIN, GPLU and GNFW. However, the AICc differences reach, at most, $0.34$ between the GKAZ and NFW profiles, which translates to the GKAZ model being $\sim1.2$ times more likely than GNFW (from Eq.~[\ref{eq: pAIC}]). This is definitely not enough to robustly distinguish those models, meaning that they all fit the data equally well. This is true even though the GPLU yields a slightly more cuspy DM profile, but not significantly so given the high uncertainties associated with this parameter when using spherical models. Therefore, given the analytical simplicity of the GPLU profile and its physical meaning,\footnote{Models such as GNFW do not have a finite mass, while EIN models are not able to reproduce centrally decreasing density profiles.} we choose to use this model for further tests, as well as when modeling Draco with \jampy\ further on.

\subsubsection{Velocity anisotropy parametrization} \label{sssec: anis-result}

Models~5 and 6 from Table~\ref{tab: mass-modeling-sph} display our tests with different velocity anisotropy parametrizations, specifically the gOM and gTiret generalizations with free inner and outer anisotropy values. The results show that although the inner velocity anisotropy tends to agree with the constant anisotropy case (i.e. preferring radial anisotropy), the velocity anisotropy at infinity prefers a more tangential behavior. Nevertheless, the uncertainties associated with this parameter are extremely large and basically encompass very radially anisotropic cases as well. Since the anisotropy at large radii is not well constrained by the data, the improvement in the fit is not enough for us to actually prefer more general models such as this (i.e. AICc is higher). Hence we use the constant anisotropy case here and, when performing axisymmetric modeling, we use it as our standard model. More general anisotropy parameterizations could over-fit the data.

\subsubsection{Cusp vs. Core under the spherical assumption} \label{sssec: cusp-core-sph}

One of the main goals of our dynamical modeling is to constrain the DM slope of Draco. From our fits that leave this slope as a free variable, we observe a general preference midway between a cored and a cuspy profile where the PM data is present (i.e. column~12). A different test is then to directly compare two mass models, one with a fixed inner cusp (i.e. $\gamma = -1$) and another with a fixed core (i.e. $\gamma = 0$). We perform those runs and display them in Table~\ref{tab: mass-modeling-sph} as models~7 and 8.

The cusped and cored models prefer velocity anisotropies that are consistent with each other at the $1\sigma$ levels, with cuspy models preferring slightly lower values. The AICc comparison between these two models shows a significant preference for the cored case, which is $\sim 6$ times more likely than the cuspy counterpart, or equivalently, rules out the latter with nearly $85\%$ probability (from Eq.~[\ref{eq: pAIC}]). This is opposite to what was found by the spherical modeling from \cite{Massari+20}, although the much better completeness and accuracy of our dataset (cf. Figure~\ref{fig: draco-fields}) can easily account for such differences. In the next section, we move on to test this result under the more suitable geometric assumption of axisymmetry.

\begin{deluxetable*}{rrlrrrlrrrr}
\tablecaption{Main results of the \jampy\ axisymmetric Jeans modeling.}
\label{tab: mass-modeling-axi}
\tablewidth{750pt}
\renewcommand{\arraystretch}{1.3}
\tabcolsep=4pt
\tabletypesize{\scriptsize}
\tablehead{
\colhead{$i$} & 
\colhead{$\Omega_{0}$} &
\colhead{$\Omega_{\infty}$} &
\colhead{$D$} &
\colhead{$\beta_{\rm J}$} &
\colhead{$r_{\rm dark}$} &
\colhead{$M_{\rm dark}^{R_{\rm max}}$} &
\colhead{$\gamma_{\rm dark}$} &
\colhead{$\Gamma_{\rm dark}$} &
\colhead{$\overline{\beta_{\rm B}}$} &
\colhead{$v_{\rm circ}^{R_{\rm max}}$} \\
\colhead{[deg]} &
\colhead{} &
\colhead{} &
\colhead{[kpc]} &
\colhead{} &
\colhead{$\left[10^{2}~\rm{pc}\right]$} &
\colhead{$\left[10^{8}~\msun\right]$} &
\colhead{} &
\colhead{} &
\colhead{} &
\colhead{$\left[\rm{km \, s^{-1}}\right]$} \\
\colhead{(1)} &
\colhead{(2)} & 
\colhead{(3)} & 
\colhead{(4)} & 
\colhead{(5)} & 
\colhead{(6)} & 
\colhead{(7)} & 
\colhead{(8)} &
\colhead{(9)} &
\colhead{(10)} &
\colhead{(11)}
}
\startdata
$  43.0 $ & $  0.36^{+ 0.64}_{- 6.42} $ & $ -2.14^{+ 2.95}_{- 371.34} $ & $  81.32^{+ 4.34}_{- 4.63} $ & $  0.95^{+ 0.02}_{- 0.02} $ & $  9.43^{+ 102.82}_{- 3.64} $ & $  1.91^{+ 1.37}_{- 0.20} $ & $ -0.66^{+ 0.71}_{- 0.24} $ & $ -0.77^{+ 0.47}_{- 0.15} $ & $ -0.02^{+ 0.26}_{- 0.34} $ & $  29.21^{+ 9.14}_{- 1.34} $ \\
$  47.7 $ & $  0.46^{+ 0.02}_{- 7.22} $ & $ -1.95^{+ 2.55}_{- 1.45} $ & $  76.83^{+ 3.39}_{- 4.37} $ & $  0.78^{+ 0.03}_{- 0.18} $ & $  6.22^{+ 131.43}_{- 0.44} $ & $  1.41^{+ 0.81}_{- 0.18} $ & $ -0.24^{+ 0.11}_{- 0.80} $ & $ -0.51^{+ 0.10}_{- 0.55} $ & $ -0.01^{+ 0.09}_{- 0.67} $ & $  25.80^{+ 6.57}_{- 1.64} $ \\
$  52.4 $ & $  0.39^{+ 0.21}_{- 6.67} $ & $ -2.42^{+ 3.10}_{- 1.43} $ & $  73.33^{+ 5.16}_{- 1.85} $ & $  0.51^{+ 0.18}_{- 0.16} $ & $ 6.94^{+ 74.25}_{- 1.65} \ \, $ & $  1.06^{+ 0.72}_{- 0.02} $ & $ -0.63^{+ 0.72}_{- 0.51} $ & $ -0.79^{+ 0.46}_{- 0.35} $ & $ -0.34^{+ 0.43}_{- 0.42} $ & $  22.91^{+ 6.63}_{- 0.44} $ \\
$  57.1 $ & $  0.28^{+ 0.32}_{- 8.10} $ & $ -2.49^{+ 3.11}_{- 1.06} $ & $  74.98^{+ 3.95}_{- 3.21} $ & $  0.35^{+ 0.20}_{- 0.22} $ & $  7.01^{+ 295.92}_{- 4.56} $ & $  0.88^{+ 0.67}_{- 0.50} $ & $ -0.79^{+ 0.22}_{- 0.46} $ & $ -0.94^{+ 0.22}_{- 0.31} $ & $ -0.41^{+ 0.35}_{- 0.46} $ & $  20.72^{+ 6.77}_{- 7.22} $ \\
$  61.8 $ & $  0.45^{+ 0.54}_{- 8.20} $ & $ -1.59^{+ 2.33}_{- 0.91} $ & $  73.19^{+ 5.51}_{- 2.71} $ & $  0.45^{+ 0.04}_{- 0.33} $ & $  10.71^{+ 289.80}_{- 4.11} $ & $  1.16^{+ 0.31}_{- 0.28} $ & $ -0.78^{+ 0.36}_{- 0.45} $ & $ -0.84^{+ 0.25}_{- 0.39} $ & $ -0.11^{+ 0.06}_{- 0.59} $ & $  24.00^{+ 2.73}_{- 3.27} $ \\
$  66.5 $ & $  0.32^{+ 0.68}_{- 8.32} $ & $ -1.46^{+ 2.13}_{- 1.17} $ & $  73.42^{+ 4.59}_{- 3.45} $ & $  0.39^{+ 0.07}_{- 0.46} $ & $  10.16^{+ 276.71}_{- 2.40} $ & $  1.02^{+ 0.34}_{- 0.18} $ & $ -0.80^{+ 0.17}_{- 0.45} $ & $ -0.87^{+ 0.13}_{- 0.38} $ & $ -0.11^{+ 0.11}_{- 0.77} $ & $  22.50^{+ 3.50}_{- 2.17} $ \\
$  71.2 $ & $ -5.23^{+ 4.91}_{- 3.19} $ & $ \ \ \, 0.55^{+ 0.23}_{- 2.58} $ & $  75.99^{+ 2.30}_{- 5.91} $ & $  0.24^{+ 0.21}_{- 0.35} $ & $  70.10^{+ 209.14}_{- 64.08} $ & $  1.30^{+ 0.01}_{- 0.49} $ & $ -1.08^{+ 0.72}_{- 0.22} $ & $ -1.08^{+ 0.50}_{- 0.22} $ & $ -0.32^{+ 0.35}_{- 0.51} $ & $  24.90^{+ 0.40}_{- 5.07} $ \\
$  75.9 $ & $  0.19^{+ 0.81}_{- 6.54} $ & $ -1.66^{+ 2.32}_{- 0.49} $ & $  72.44^{+ 5.44}_{- 1.79} $ & $  0.24^{+ 0.20}_{- 0.29} $ & $  12.74^{+ 381.65}_{- 6.02} $ & $  0.94^{+ 0.34}_{- 0.17} $ & $ -0.98^{+ 0.53}_{- 0.31} $ & $ -1.03^{+ 0.37}_{- 0.28} $ & $ -0.26^{+ 0.29}_{- 0.42} $ & $  21.77^{+ 3.26}_{- 2.23} $ \\
$  80.6 $ & $ -4.87^{+ 4.59}_{- 2.66} $ & $  \ \ \, 0.39^{+ 0.32}_{- 2.49} $ & $  73.07^{+ 5.56}_{- 2.52} $ & $  0.17^{+ 0.23}_{- 0.34} $ & $  24.04^{+ 260.36}_{- 14.83} $ & $  1.13^{+ 0.13}_{- 0.37} $ & $ -1.04^{+ 0.32}_{- 0.32} $ & $ -1.05^{+ 0.24}_{- 0.32} $ & $ -0.34^{+ 0.34}_{- 0.48} $ & $  23.77^{+ 1.02}_{- 4.45} $ \\
$  85.3 $ & $ -5.70^{+ 5.34}_{- 2.09} $ & $  \ \ \, 0.81^{+ 0.19}_{- 2.92} $ & $  74.32^{+ 4.53}_{- 3.93} $ & $  0.19^{+ 0.19}_{- 0.38} $ & $  173.12^{+ 141.10}_{- 164.65} $ & $  1.29^{+ 0.63}_{- 0.50} $ & $ -1.03^{+ 0.35}_{- 0.28} $ & $ -1.03^{+ 0.24}_{- 0.29} $ & $ -0.31^{+ 0.30}_{- 0.53} $ & $  25.10^{+ 7.18}_{- 5.55} $ \\
$  90.0 $ & $ -0.05^{+ 0.06}_{- 6.60} $ & $ -2.18^{+ 2.93}_{- 318.09} $ & $  75.92^{+ 2.44}_{- 5.28} $ & $ -0.01^{+ 0.39}_{- 0.12} $ & $  16.82^{+ 294.31}_{- 7.06} $ & $  0.87^{+ 0.39}_{- 0.11} $ & $ -1.21^{+ 0.47}_{- 0.14} $ & $ -1.23^{+ 0.42}_{- 0.12} $ & $ -0.56^{+ 0.54}_{- 0.19} $ & $  20.48^{+ 4.43}_{- 1.13} $ \\
\rowcolor{lavender} $ \left< . \right>_{i} $ & $  0.15^{+ 0.24}_{- 7.24} $ & $ -1.14^{+ 1.85}_{- 1.81} $ & $  75.37^{+ 4.73}_{- 4.00} $ & $  0.56^{+ 0.25}_{- 0.42} $ & $  11.18^{+ 191.03}_{- 4.89} $ & $  1.20^{+ 0.79}_{- 0.28} $ & $ -0.71^{+ 0.44}_{- 0.49} $ & $ -0.83^{+ 0.32}_{- 0.37} $ & $ -0.20^{+ 0.28}_{- 0.53} $ & $  24.19^{+ 6.31}_{- 2.97} $ \\
\enddata
\begin{tablenotes}
\scriptsize
\item \textsc{Notes} --  
Columns are 
\textbf{(1)} Inclination, in degrees ($90\degree$ is edge-on); 
\textbf{(2)} Rotation parameter at $r=0$ (see Eq.~[\ref{eq: rotation-jampy}]);
\textbf{(3)} Rotation parameter at $r \rightarrow \infty$;
\textbf{(4)} Heliocentric distance, in kpc; 
\textbf{(5)} $\beta_{\rm J}$ velocity anisotropy parameter, as defined in Eq.~(\ref{eq: beta-jampy}); 
\textbf{(6)} Dark matter scale radius, in $10^{2}$~pc;
\textbf{(7)} Dark matter mass at maximum projected data radius, in $10^{8}$~M$_{\odot}$;
\textbf{(8)} Dark matter asymptotic density slope;
\textbf{(9)} Dark matter density slope averaged over the spatial range where PMs are available;
\textbf{(10)} Globally-averaged $\beta_{\rm B}$ velocity anisotropy, as defined in Eq.~(\ref{eq: binney-beta});
\textbf{(11)} Circular velocity at maximum projected data radius, in $\kms$;
The uncertainties are based on the 16th and 84th percentiles of the marginal distributions, unless the maximum likelihood solution was outside that boundary, in which case the uncertainties are based on the minimum or maximum value of the MCMC chain. 
The last row displays the integrated estimate for each parameter, averaged over the inclination probability distribution of Draco, as described in Section~\ref{ssec: axi-inc}. The baryonic mass and respective Plummer major axis are fixed in all cases to $4.7 \times 10^{5}~\msun$ and $9.1$~arcmin.
\end{tablenotes}
\end{deluxetable*}

\subsubsection{Velocity anisotropy under the spherical assumption} \label{sssec: anis-sph}

Along with the measurement of the DM slope, our study is the first to consistently constrain the velocity anisotropy of Draco. As discussed in Section~\ref{sssec: PMsigmaprofiles}, the projected anisotropy is directly measured by the data, independently of any model assumption. However, to recover the intrinsic 3D value of $\beta$, assumptions are required; here, we analyze this problem under the consideration of spherical symmetry.

Throughout all models in Table~\ref{tab: mass-modeling-sph}, the velocity anisotropy $\beta_{\rm B}$ remains within the 1-$\sigma$ range of $\sim 0.2 - 0.6$. This means that under spherical assumptions, our data consistently constrain the orbital shapes in Draco to be radially anisotropic. Previous PM studies of Draco \citep{Massari+20,delPino+22} hint towards a similar behavior, although their error bars are too high to discard tangential orbits at a 1-$\sigma$ level in the former work. Given the unprecedentedly low PM uncertainties of our dataset, we are able to rule out negative values of $\beta_{\rm B}$ in model~3 with $98.5\%$ confidence, if one assumes sphericity.

\subsection{Axisymmetric modeling} \label{ssec: axi-results}

Having explored spherical models, we now move on to more realistic axisymmetric models. Because \jampy\ is much slower than the equivalent \mpo\ MCMC routines, we are forced here to reduce our set of dynamical models. We thus use the priors and assumptions specified in Section~\ref{sssec: priors-jampy}. Specifically, informed by the results of the spherical models, we focus on axisymmetric luminous models with a Plummer density distribution and constant $\beta_{\rm J}$ as function of radius, embedded in a spherical dark halo with a generalized Plummer profile. 

\subsubsection{Inclination dependence} \label{ssec: axi-inc}

We ran \jampy\ over a set of eleven inclinations linearly spaced from $43\degree$ to $90\degree$ (edge-on), which encompasses Draco's inclination PDF computed in Section~\ref{sssec: inc}, and display the results in Table~\ref{tab: mass-modeling-axi}. In addition to the parameter $\beta_{\rm J}$ defined by Eq.~(\ref{eq: beta-jampy}), we also list the globally-averaged Binney anisotropy $\overline{\beta_{\rm B}}$. The latter was obtained by first calculating the mass-weighted second velocity moments averaged over the entire system, and then substituting those into Eq.~(\ref{eq: binney-beta}).  

Overall, the results for $\Omega_{\infty}$ and $r_{\rm dark}$ slightly increase with inclination, while other parameters tend to decrease with $i$. In particular, models that are closer to edge-on prefer more cuspy DM profiles, lower velocity anisotropies $\beta_{\rm J}$ (still positive, but closer to $\beta_{\rm J} = 0$) and lower dynamical distances. 

In spite of these correlations, we can provide a global estimate and uncertainty for every model parameter, integrated over the inclination distribution that we calculated for Draco in Section~\ref{sssec: inc}. That is, for a parameter $\Theta$, we apply
\begin{equation} \label{eq: bayes-inc}
    p(\Theta) = \int_{i_{\rm min}}^{i_{\rm max}} p(\Theta | i) \, p(i) \, {\rm d}i \ ,
\end{equation}
where $p(\Theta)$ is the final posterior distribution integrated over all inclinations, $p(\Theta | i)$ is the posterior distribution obtained from the MCMC chain, and $ p(i) {\rm d}i$ is the probability of falling into a certain inclination range, which we compute empirically from the distribution computed in Section~\ref{sssec: inc}. In practice, the integral above is a discrete sum over the inclination ranges probed by the eleven values we calculated. 
We use the same procedure to obtain the most probable value of every parameter, i.e. we average the results in Table~\ref{tab: mass-modeling-axi} weighted by the respective inclination probabilities. Finally, we define the confidence regions as the 16th and 84th percentiles of the numerically computed  $p(\Theta)$. 

\begin{figure}
\centering
\includegraphics[width=\hsize]{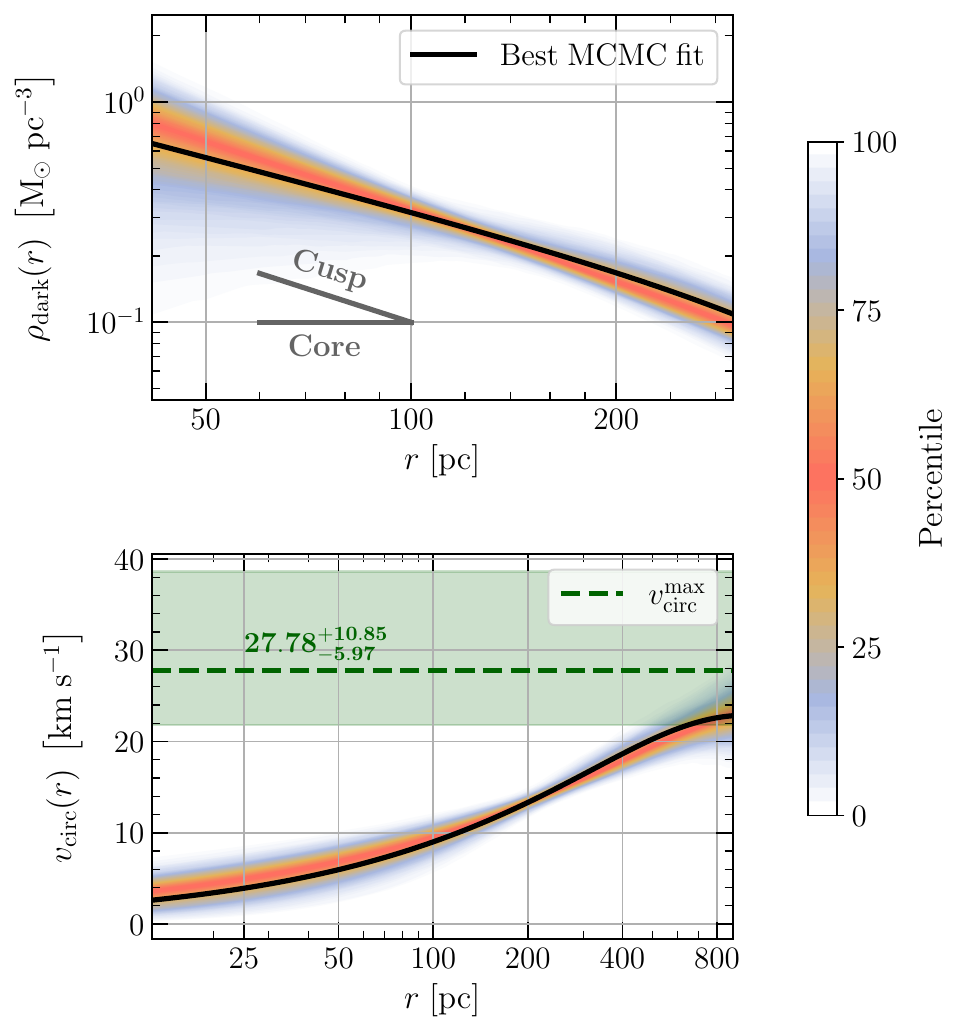}
\caption{\textit{Dark matter density and circular velocity:} The \textit{upper} panel shows the dark matter density profiles from our \jampy\ MCMC chains for $i = 57.1\degree$, color-coded according to their respective percentiles. We display the most probable solution (as defined in Section~\ref{ssec: jeans-jampy}) as a black line. The \textit{lower} panel shows the circular velocity curves, defined by Eq.~(\ref{eq: vcirc}), for the same models, as well as their maximum value in green. The radial extent of the upper panel covers the extent of our PM data (where $\Gamma_{\rm dark}$ was computed), while the lower panel covers the extent of our entire dataset.}
\label{fig: rhodm-axi}
\end{figure}

\subsubsection{Data-Model Comparison} \label{ssec: axi-good-fit}

The parameter estimates thus marginalized over all inclinations are listed in the last row of Table~\ref{tab: mass-modeling-axi}. They resemble the estimates for the cases $i = 52.4\degree$ and $i = 57.1\degree$. We choose the case of $i = 57.1\degree$ as a canonical model for display, as it represents a value between the median ($56.2\degree$) and the mean ($58.9\degree$) of the inclination PDF. In Figure~\ref{fig: disp-axi} we display, for this inclination, the data-model comparison for the three velocity dispersion components, and the LOS rotation velocity amplitude. Figure~\ref{fig: rhodm-axi} shows the dark matter density profile we estimate over the range where we calculate its slope, as well as the circular velocity over the entire data range. 

Figure~\ref{fig: disp-axi} shows a very satisfactory fit from \jampy, which helps to strengthen both our parametrization choices and our conclusions further on.
Figure~\ref{fig: rhodm-axi} shows that both the circular velocity and DM cusp (down to $\sim 100$~pc) are well constrained by our models.

We verified that our results do not depend sensitively on the adopted parameterization for $\rho_{\star}$. For this we ran a comparison $i = 57.1\degree$ model in which the stellar density was taken to follow a S\'ersic \citep{Sersic63,Sersic68} profile,\footnote{We assigned the structural parameters provided by \cite{Odenkirchen+01} for this specific parameterization.} instead of the canonical Plummer one. This yielded a considerably higher $\chi^{2}$ statistic, driven in part by a poorer fit to Draco's rotation profile. Nonetheless, the inferred results for, e.g. $\Gamma_{\rm dark}$ and $\overline{\beta_{\rm B}}$ (namely: $ -0.70^{+ 0.64}_{- 0.14} $ and $ -1.04^{+ 0.66}_{- 0.40} $, respectively), agree within 1-$\sigma$ to what is listed for $i = 57.1\degree$, in Table~\ref{tab: mass-modeling-axi}. 

\begin{figure}
\centering
\includegraphics[width=0.75\hsize]{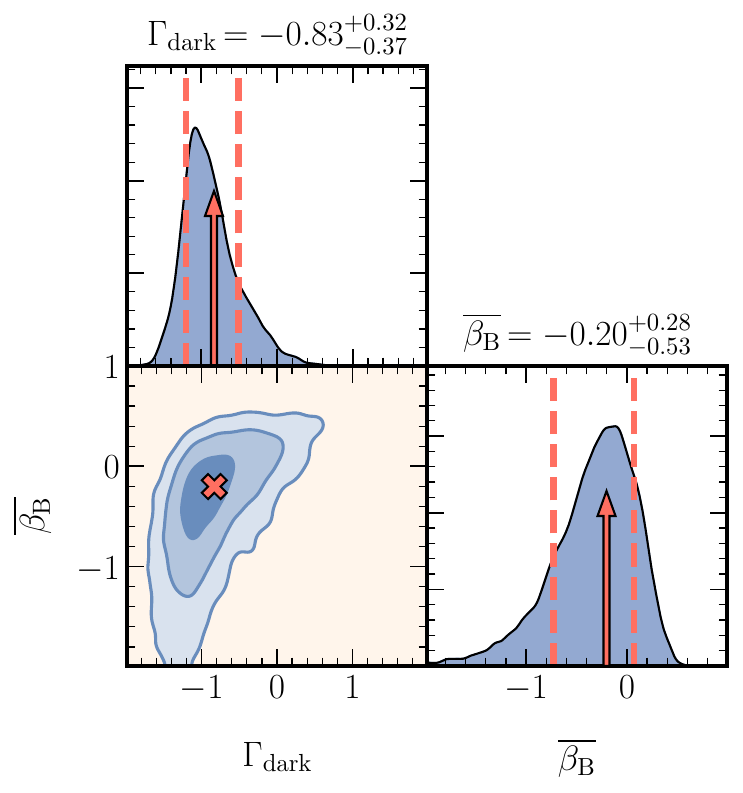}
\caption{\textit{Averaged dark matter slope and velocity anisotropy:} Final posterior probability distributions of Draco's dark matter density slope, $\Gamma_{\rm dark}$, averaged over the range where we have PM data, and the globally-averaged $\beta_{\rm B}$ velocity anisotropy (see Eq.~[\ref{eq: binney-beta}]). The distributions were marginalized over inclinations as described in Section~\ref{ssec: axi-inc}, and further smoothed for simple visualization purposes. The values highlighted correspond to the best likelihood estimates (title texts, arrows, and cross), and the respective 16th and 84th percentiles (dashed lines). The results imply a classic $\Lambda$CDM-like slope in the Draco Sph galaxy, and a marginal preference for overall tangential velocity anisotropy.}
\label{fig: gamma-dm-pdf}
\end{figure}

\begin{figure}
\centering
\includegraphics[width=0.75\hsize]{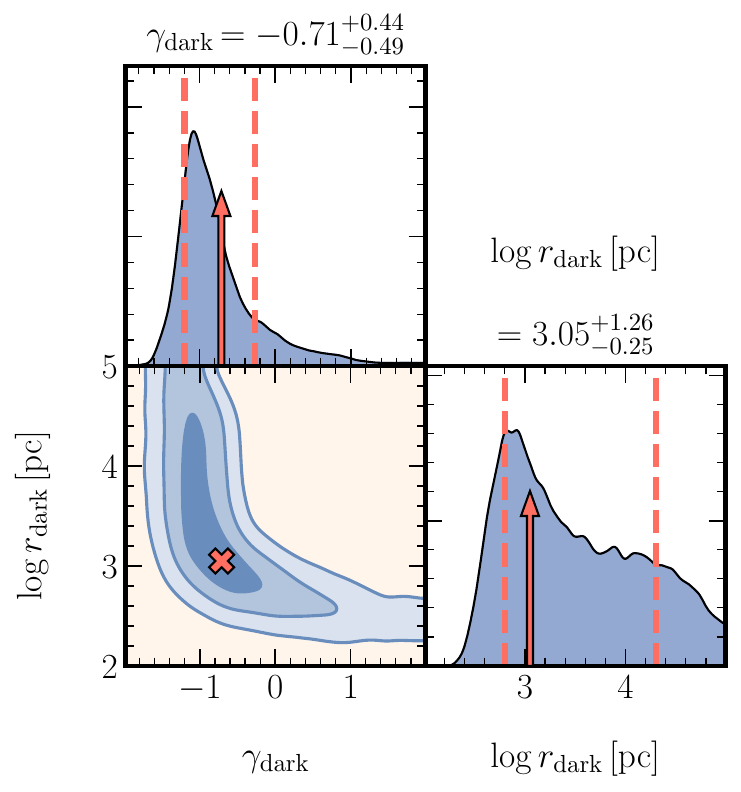}
\caption{\textit{Asymptotic dark matter slope and scale radius:} Final posterior probability distributions of Draco's asymptotic dark matter density slope, $\gamma_{\rm dark}$, and the logarithm of the dark matter scale radius, $\log r_{\rm dark}$, with $r_{\rm dark}$ in pc. The plotting details are similar to the ones in Figure~\ref{fig: gamma-dm-pdf}. The results rule out a core larger than 942~pc at 3-$\sigma$ confidence.}
\label{fig: gamma-rdark}
\end{figure}

\subsubsection{Inferred Quantities} \label{ssec: overall-results}

We display in Figure~\ref{fig: gamma-dm-pdf} the posterior distributions for and correlation between the parameters $\Theta = \{ \Gamma_{\rm dark}, \, \overline{\beta_{\rm B}} \}$, marginalized over all inclinations as described in Section~\ref{ssec: axi-inc}. The Figure shows that models with a core generally require a more radial velocity anisotropy to fit the data, consistent with what has been found in other contexts \citep[e.g. Figure~3 in][]{vanderMarel+00}. Our averaged slope estimate, $\Gamma_{\rm dark} = -0.83^{+ 0.32}_{- 0.37} $, is consistent with a classic $\Lambda$CDM slope, even though the posterior distribution we derive has a long tail towards larger (including even positive) $\Gamma_{\rm dark}$ values.
The globally-averaged Binney $\overline{\beta_{\rm B}}$ is well below the parameter $\beta_{\rm J}$, defined by Eq.~(\ref{eq: beta-jampy}) for all inclinations. This is because $\beta_{\rm B}$ depends on 
$\left< v_{\phi}^2 \right>$, while $\beta_{\rm J}$ does not. Axisymmetric models generally have $\left< v_{\phi}^2 \right>$ increasing from the symmetry axis towards the equatorial plane (see Appendix~\ref{sapp: xi-dependence-scf}). So while $\beta_{\rm B} = \beta_{\rm J}$ on the symmetry axis, instead $\beta_{\rm B} \leq \beta_{\rm J}$ in the equatorial plane. While Table~\ref{tab: mass-modeling-axi} shows that our best-fit models have $\beta_{\rm J}$ positive and increasing with decreasing inclination, the inferred value of $\overline{\beta_{\rm B}}$ depends less on inclination to within the statistical uncertainties.
The overall anisotropy marginalized over inclinations is $\overline{\beta_{\rm B}} = -0.20^{+ 0.28}_{- 0.53}$. So our best-fit models are radially anisotropic on the symmetry axis, and tangentially anisotropic in the equatorial plane. When integrated over the entire meridional $(R, z)$ plane they are tangentially anisotropic, but still statistically consistent with isotropy.

Similarly, we plot in Figure~\ref{fig: gamma-rdark} an equivalent case for the parameters $\Theta = \{ \gamma_{\rm dark}, \, \log r_{\rm dark} \}$, with $r_{\rm dark}$ in pc. As expected from our conversions between $\Gamma_{\rm dark}$ and $\gamma_{\rm dark}$ in Section~\ref{ssec: obs-param}, one has a remarkable agreement between the peak of $\gamma_{\rm dark}$'s PDF and a classic $\Lambda$CDM slope. Besides, our uncertainties on $\gamma_{\rm dark}$ are consistent with what is expected from PM datasets having a similar number of stars as ours, as argued in \cite{Guerra+23}. More importantly, this figure allows us to probe the core radius that our 3D data is able to constrain: while negative asymptotic slopes agree with a large set of DM scale radii, positive slopes require that the respective core (or even a drop in the density) be limited within $\lesssim 1$~kpc. Indeed, upon analyses of our MCMC chains, cores larger than 487~pc, 717~pc and 942~pc are ruled out at 1-, 2- and 3-$\sigma$ confidence,\footnote{These numbers are derived upon selecting the elements of the corner plot in Figure~\ref{fig: gamma-rdark} that correspond to $\gamma_{\rm dark} \geq 0$, and retrieving the values that encompasses 68\%, 95\% and 99.7\% of the respective $\log{r_{\rm dark}}$ distribution.} respectively. For reference, the scale radius predicted by $\Lambda$CDM, previously mentioned in Section~\ref{ssec: obs-param}, equals $1.06$~kpc.

The circular velocity at our outermost data point in our best-fit models is $v_{\rm circ}^{R_{\rm max}} = 24.19^{+ 6.31}_{- 2.97}~\kms$. Similarly, the maximum value of the circular velocity is $v_{\rm circ}^{\rm max}(r) = 27.78^{+ 10.85}_{- 5.97}~\kms$. This $v_{\rm circ}^{\rm max}(r)$ measurement is generally higher than most previous calculations, namely \cite[][$15-35~\kms$]{Strigari+07_vmax}, \cite[][$18.2^{+3.2}_{-1.6}~\kms$]{Martinez15} and \cite[][$10.2-17.0~\kms$]{Massari+20}. Correspondingly, the dark mass in our models is higher as well. 
It is difficult to precisely determine where such differences could come from, but  one could speculate that this relates to different completeness of the respective datasets used in each work. 
Indeed, the circular velocity values measured by the likewise axisymmetric modeling from \citet[][figure~9]{Hayashi+20} over similar radial ranges lie closer to ours (i.e. $\sim25 - 30~\kms$).

From Table~\ref{tab: mass-modeling-axi}, one sees that higher heliocentric distances of Draco are usually related to lower inclinations and more cored models, and vice-versa.
Our estimate of Draco's distance, $D = 75.37^{+ 4.73}_{- 4.00}$~kpc, provides the first dynamical distance for this dwarf. Comparatively, \cite{Bonanos+04} measured $75.8\pm5.44$~kpc using a set of 146 RR Lyrae stars, \cite{Aparicio+01} found $80\pm7$~kpc from analyses of the magnitude of the horizontal branch at the RR Lyrae instability strip, and \cite{Muraveva+20} reported $80.5\pm2.6$~kpc when using 285 RR Lyrae stars. Our measurement is thus comparable to and competitive with other literature results based on stellar population methods. Thus, high-quality astrometric data also provides a valuable validation of standard distance determination techniques.

\subsubsection{Spherical vs. axisymmetric models} \label{ssec: sph-vs-axi}

Until recently, there were no PM dispersion profiles available for internal mass modeling of dSphs. Hence, methods employed to analyze LOS velocities had to make substantial assumptions to remove degeneracies in the data. Among other things, models usually assumed spherical geometry 
\citep[e.g.][]{Wilkinson+02,Read+18,Massari+20}, with the important exception of \cite{Hayashi+20}. The velocity moments that are derived from the Jeans equations then depend only on the projected radius to the system's center. Instead, observed quantities in axisymmetric models depend on the position angle on the sky. We have found that for the new PM dataset presented here, axisymmetric models yield substantially different results from spherical models. This is true especially for the quantities most of interest, namely the dark matter cusp slope and the velocity anisotropy. Axisymmetric models imply lower anisotropy $\overline{\beta_{\rm B}}$  and higher cusp slope $\Gamma_{\rm dark}$. Hence, it is critically important to construct axisymmetric models that properly take position angle  dependencies into account. 

In Appendices~\ref{sapp: beta-dependence-scf} and~\ref{sapp: xi-dependence-scf}, we again use the scale-free dynamical models of the type discussed in \cite{deBruijne+96} to explain this result. Figure~\ref{fig: b3-app} in the appendix shows that there is a tight monotonic relation between the PM anisotropy $\sigma_{\rm POSt} / \sigma_{\rm POSr}$ integrated over the sky, and the globally-integrated intrinsic $\overline{\beta_{\rm B}}$. This relation is very similar for spherical and axisymmetric models. However, for Draco we have not measured $\sigma_{\rm POSt} / \sigma_{\rm POSr}$ over the entire projected image of the galaxy, but only for two fields along the major axis (see Figure~\ref{fig: observed-fields-dss}). Figure~\ref{fig: axi-scf} in the Appendix shows that in axisymmetric geometry, $\sigma_{\rm POSt} / \sigma_{\rm POSr}$ is {\it not} constant with position angle on the sky. Instead, it is much lower on the major axis than on the minor axis. Hence, a spherical model that assumes that $\sigma_{\rm POSt} / \sigma_{\rm POSr}$ is the same everywhere as measured on the major axis will {\it overestimate} the radial anisotropy $\beta_{\rm B}$. So while our axisymmetric models imply that  
$\overline{\beta_{\rm B}} = -0.20^{+ 0.28}_{- 0.53}$, our best-fit canonical spherical model (model~3 in Table~\ref{tab: mass-modeling-sph}) has instead $\beta_{\rm B} = 0.39^{+0.13}_{-0.14}$. 
As mentioned in Section~\ref{sssec: anis-sph}, the latter is consistent with previous studies that assumed sphericity \citep[e.g.][]{Massari+20,delPino+22}. The higher $\beta_{\rm B}$ in spherical models translates to a shallower DM slope, given the correlation in Figure~\ref{fig: gamma-dm-pdf}.

Systematic biases of this type are lessened, but not necessarily erased, when using more spatially complete datasets. This is something to keep in mind as we enter a new era of galactic PMs: although spherical models are less costly and thus helpful to understand general choices of parametrization and priors, robust results and respective conclusions can only be obtained when considering more complex models that take into account the real shape of the galaxy \citep[this has also been argued previously by][]{Genina+18}. 

Even in our case, there are still degeneracies that are not fully taken into account. Specifically, as highlighted in \cite{Cappellari20}, the deprojection of the surface brightness to obtain the intrinsic luminosity density is not unique unless the axisymmetric galaxy is seen edge-on \citep{Rybicki87,Kochanek&Rybicki96}. The degeneracy increases considerably when the galaxy is seen at low inclinations \citep{Romanowsky&Kochanek97, vandenBosch97, Magorrian99}. Our results for $\Gamma_{\rm dark}$ do not depend strongly on the assumed inclination (cd.~Table~\ref{tab: mass-modeling-axi}). But it should be kept in mind that at low inclination other deprojections of the luminous density may be possible that differ from the Plummer models assumed here.

\section{Robustness of Modeling Results} \label{sec: robust}

\subsection{The effect of binaries} \label{ssec: binaries}

As explained in Section~\ref{sssec: binaries-los}, recent works have proposed that the LOS velocity dispersion from dwarf galaxies could be inflated due to the presence of unresolved binaries from single-epoch exposures. In particular, such inflation could translate to mass overestimation, and could eventually bias our DM measurements. Beyond the tests performed in Section~\ref{sssec: binaries-los}, which have shown that the LOS velocity dispersion remains consistent when using data with different number of epochs, another independent test is to compare mass-modeling with either LOS$+$PM data (case i) or with PM data alone (case ii). Due to the lack of high quality PM data, such a test has never been performed before. Our new state-of-the-art PM catalog can thus shed light onto this question.

Model~9 in Table~\ref{tab: mass-modeling-sph} depicts the results from spherical Jeans modeling with PM data alone, but with the same priors as in our preferred model 3 (LOS$+$PM). Despite the slight increase of Poisson uncertainties, due to the fact that the PM subset alone has about half as many stars as the LOS$+$PM subset, the overall predictions of model~9 agree within 1-$\sigma$ to all the predictions from model 3. Therefore, we conclude from both the tests performed in the current Section and those from Section~\ref{sssec: binaries-los}, that our dynamical modeling results for Draco do not have significant biases due to the presence of unresolved binaries. Although we can only constrain this for the specific galaxy analyzed here, this is an important result that sets the stage for future interpretations on the impact of unresolved binaries on the velocity dispersion profiles of other dwarf galaxies.

\subsection{The effect of tides} \label{ssec: tides}

To derive mass estimates and density profile shapes, our analyses assume that Draco is in dynamical equilibrium. However, a number of recent studies have suggested that the excessive mass-to-light ratios measured in dSphs could be due to out-of-equilibrium dynamics, which in turn inflate the velocity dispersion \citep[e.g.][]{Klessen&Kroupa&Kroupa98,Hammer+18}. Thus, it is of interest to gauge the possibility that our conclusions could be biased by MW tidal effects.

On this subject, the past literature strongly supports that Draco is a galaxy with no clear signs of tidal disruption: Both \cite{Odenkirchen+01} and \cite{Segall+07} have found no evident stellar streams, asymmetric disturbances or density breaks that are characteristic of a tidally-perturbed system. 
Orbital analyses from \cite{Sohn+17} support that the last pericenter passage of Draco, when accounting for a massive Large Magellanic Cloud (see their table~5), happens $\sim 4$~Gyr ago, at an average closest MW distance of $\sim 100$~kpc, while \cite{Pace+22} report a pericenter of $r_{\rm peri} = 58.0^{+11.4}_{-9.5}$~kpc and \cite{Battaglia+22} report $r_{\rm peri} = 51.7^{+4.1}_{-6.1}$~kpc ($37.6^{+4.2}_{-4.4}$~kpc) for a lighter (heavier) MW, mostly from \gaia\ DR3 data.

For the lower pericenter distances found by \cite{Battaglia+22} and \cite{Pace+22}, one could envision the gas-expansion scenario proposed by \cite{Hammer+24} and \cite{Wang+24}, which could produce higher velocity dispersion profiles than the stellar component alone (although still with typically $\lesssim$ half of the values observed in Figure~\ref{fig: disp-axi}).  
However, \cite{Sohn+17} also provide arguments (see their figure~4) against a mean radial expansion of the stellar component.
Hence, there is no indication that strong tidal effects could be biasing our results, thus making
Draco an ideal galaxy for our equilibrium-based dynamical modeling.

\subsection{Error threshold} \label{ssec: error-limit}

As discussed in Section~\ref{sssec: under-err}, dynamical modeling should be careful when including data points with PM uncertainties higher than the system's intrinsic velocity dispersion. The larger the PM uncertainties, the more important it is that they are known very accurately. This is difficult to guarantee, since not all sources of observational uncertainty are always known or easily quantified. If the PM uncertainties are underestimated, then the galaxy PM dispersion will be overestimated, which biases the dynamical modeling results. Hence, in our analysis we have only included stars for which the observational PM uncertainty is smaller than $0.024~\masyr$, similar to the intrinsic PM dispersion of the galaxy (Section~\ref{sssec: under-err}). Also, 
previous works with globular cluster PM data have often considered only low PM-error stars when deriving PM dispersion profiles from \gaia\ \citep[][]{Baumgardt+19,Vitral+22} and \hst\ \citep[][]{Bellini+14,Watkins+15-disp}. 

To further test our choice of a PM error threshold, we also performed, for comparison, a \mpo\ analysis with only the subset of stars meeting a lower threshold of $0.022~\masyr$. We display this run as model~10 in Table~\ref{tab: mass-modeling-sph}, which should be compared to its counterpart with the higher error threshold, model~3. Model~10 has 239 stars with measured PMs, compared to 364 in model~3. This implies a decrease of 35\% in statistical completeness, and results in somewhat larger uncertainties for all inferred model parameters. Within 1-$\sigma$ all parameters agree between these two models. This further supports that our results are robust to our choice of PM error threshold, and that there is no indication of biases due to underestimated errors in our standard dataset.

\subsection{Higher-Order Velocity Moments} \label{ssec: higherorder}

The shape of the LOS velocity distribution can in principle be used to obtain a constraint on the velocity anisotropy that is independent of our Jeans results \citep{vanderMarel&Franx93}. However, detailed modeling of this shape requires more complicated modeling techniques than those presented here \citep*[e.g.][]{Chaname+08}, which is outside the scope of the present paper. Nonetheless, we describe in Appendix~\ref{sapp: GaussHermite} that approximate modeling is again possible with the scale-free dynamical models of the type discussed in \cite{deBruijne+96}. Figure~\ref{fig: b5-app} in the Appendix shows that plausible axisymmetric models exist that both: (a) have values of $\overline{\beta_{\rm B}}$ consistent with those derived from our Jeans models; and (b) predict a kurtosis for the LOS velocity distribution that is consistent with the observed value. 
Hence, we would not expect that future detailed modeling of higher-order moments would change our conclusions about Draco's velocity anisotropy, and hence its mass distribution. However, per Figure~\ref{fig: b5-app}, it might be helpful to constrain Draco's viewing inclination, as well as details of its phase-space distribution function.

\section{Cosmological Implications and Future Work} \label{sec: discussion}
 
Our newly measured asymptotic DM slope for Draco is $\gamma_{\rm dark} = -0.71^{+ 0.44}_{- 0.49}$, consistent with the behavior expected in $\Lambda$CDM \citep{Navarro+97}, especially when comparing the most likely results from our MCMC chains in Figure~\ref{fig: gamma-rdark} (in other words, the peak of the respective posterior PDF).
Instead, various early studies of dSph galaxies found that observations favored shallow inner DM density profile slopes, consistent with a constant density `core' at the center \citep[e.g.][]{Battaglia+08,Walker&Penarrubia&Penarrubia11,Amorisco&Evans12,Brownsberger&Randall&Randall21},\footnote{Notice that such analyses considered the Sculptor and Fornax dwarfs, which have a higher stellar content, from whence it is not completely excluded that a cuspy DM density profile could have been lowered by baryonic effects.} and inconsistent with predictions of standard $\Lambda$CDM cosmology, from DM-only simulations.
So if future PM studies of other dSph galaxies were to support our findings, then this supports the standard cosmological hypothesis that the DM in the Universe behaves as a cold particle. The cold aspect of the DM particles would cause them to clump towards the deeper regions of the potential well, and form diverging cusps such as the one we measure.

Early suggestions of DM cores in dwarf galaxies inspired some studies to propose fundamental changes in the nature of DM, such as warm DM (WDM), e.g. sterile neutrinos and gravitinos, that predict lower central DM densities and cored profiles \citep{Dalcanton&Hogan01}, or SIDM for which DM particles in the central region thermalize via collisions and thereby form a cored profile \citep[e.g.][]{Sameie+20}. The fact that our study can rule out cored profiles larger than $942$~pc at a 3-$\sigma$ level thus imposes useful constraints on the SIDM cross-section.\footnote{Specific constraints on how the core radius relates to the DM cross-section and particle mass can be found in \cite{Read+18} and \cite[][see \citeyear{Maccio+13} erratum]{Maccio+12}, respectively.}

Independent evidence for the presence of DM cores in gas-rich star-forming dwarf galaxies exists on the basis of HI rotation curve modeling \citep[e.g.][]{Moore94,Flores&Primack94,Burkert95}. A likely explanation for this in the context of standard $\Lambda$CDM cosmology is the impact of baryons on the DM density profile. This may transform DM cusps into cores by transferring energy and mass to the outer parts of the halos, e.g. via supernova feedback \citep{Read&Gilmore05,Pontzen&Governato12,Brooks&Zolotov&Zolotov14}, or star formation events \citep{Read+18}. Our findings in the present paper suggest that these processes have not been important for Draco. However, that is not unexpected. 
Draco is below the limit where stellar feedback as implemented in current cosmological simulations should still produce a small core \citep{Fitts+17}. Moreover, its star formation shut down long ago ($\sim$10 Gyr; \citealt{Aparicio+01}). Hence, our work does not provide insight into how the DM profiles of higher-mass or star-forming galaxies may be impacted by baryonic physics.

To further probe how the variety of proposed theoretical mechanisms to form a core compare to observations, it will be essential to measure and model the 3D dynamics of dwarfs with different characteristics than Draco. For example, dSphs having more troubled dynamical pasts, having suffered close encounters with other satellites or the Milky Way itself, or instead dSphs with star formation histories that shut down more recently. Efforts in these directions are already underway -- we have acquired multi-epoch datasets of the Sculptor dSph galaxy from the same \hst\ programs described here \citep{2021hst..prop16737S}, and are in the process of also acquiring analogous \hst\ data for the Ursa Minor dSph galaxy (\citealt{Vitral+23-UMi}). Moreover, we have ongoing programs on \jwst\ to further extend our time baselines for both Draco and Sculptor (\citealt{2023jwst.prop.4513V}). The \textit{Nancy Grace Roman Space Telescope} will provide the opportunity to further extend such studies over large fields of view \citep{Han+23}. This will help lift various degeneracies that have so far complicated a full exploration of existing discrepancies between observations and theoretical predictions for dwarf galaxies. 

\section{Conclusions} \label{sec: conclusion}

For many decades, 3D velocity datasets of the internal kinematics of dwarf galaxies were only conceivable through numerical simulations. Thanks to \hst's record time in operation, and its exquisite astrometric capabilities, we can now study other galaxies using plane-of-sky velocities that are not simulated, but observed. With four epochs of \hst\ observations of the Draco dwarf spheroidal galaxy spanning an 18 year temporal baseline, we measure precise proper motions for hundreds of stars, with uncertainties below the intrinsic velocity dispersion of the galaxy. This provides the most precise PM dataset of Draco to date. We make this dataset publicly available as online material,\footnote{The dataset will be available at a TBD \textsc{zenodo} link upon paper acceptance.} so that it can also be used for other studies than those described here. 

By combining the PMs with existing line-of-sight velocities, we derive for the first time radially-resolved 3D velocity dispersion profiles for any dwarf galaxy. With suitable modeling, these directly constrain the intrinsic velocity anisotropy of the galaxy, and resolve the mass-anisotropy degeneracy that often plagues dynamical modeling. 

To fit the measurements and infer the radial mass profile, we solve the Jeans equations in both spherical and axisymmetric geometries. The latter provides the first axisymmetric modeling of any pressure-supported external galaxy that is observationally constrained by all three orthogonal components of the stellar velocity field. 

The viewing inclination of the galaxy is not constrained by the data. 
So we marginalize our modeling results over all possible inclinations, informed by the overall distribution of projected galaxy shapes for elliptical galaxies in the nearby Universe. None of our conclusions depend sensitively on the actual inclination. 

Below, we summarize our main findings for the Draco dSph galaxy, as well as the cosmological implications of our study.

\begin{itemize}

    \item We provide new estimates of the galaxy center, based on elliptical Bayesian fits to the \gaia\ EDR3 stellar counts (see Table~\ref{tab: overview}).
    
    \item We determine the line-of-sight rotation curve  of the galaxy (see Figure~\ref{fig: vlos-rotation}) and find that it has measurable rotation. Over the radial region covered by the LOS dataset, $\left< v / \sigma \right> = 0.22 \pm 0.09$.

    \item We show that the impact of unresolved binaries on the LOS data is negligible, and does not significantly alter the dynamical modeling results. 

    \item For the PMs in our \hst\ fields along the projected major axis, we measure averaged observed velocity dispersion ratios of  $\left< \sigma_{\rm POSt} \right> / \left< \sigma_{\rm POSr} \right> = 0.80 \pm 0.08$ and $\left< \sigma_{\rm LOS} \right> / \left< \sigma_{\rm POS} \right> = 1.08 \pm 0.09$, where $\sigma_{\rm POS}$ represents an average over both PM directions. The first ratio is independent of galaxy distance, while the second is inversely proportional to it.\footnote{We assume $D = 75.37^{+ 4.73}_{- 4.00}$~kpc.} The tight observational constraints on ratios like these enable dynamical models to strongly constrain the structure of Draco.
    
    \item Axisymmetric models imply that the velocity dispersion tensor of the galaxy is radially anisotropic along the symmetry axis, with $\beta_{\rm J} = 0.56^{+ 0.25}_{- 0.42}$. This is similar to the anisotropy everywhere in our best-fit spherical model ($\beta_{\rm B} = 0.39^{+0.13}_{-0.14}$). However, the best-fit axisymmetric models are tangentially anisotropic in the equatorial plane, as required to maintain hydrostatic equilibrium in an oblate system. Their globally averaged anisotropy is $\overline{\beta_{\rm B}} = -0.20^{+ 0.28}_{- 0.53}$, so tangentially anisotropic, but still statistically consistent with isotropy.

    \item Construction of axisymmetric models is essential for flattened galaxies. This is particularly important for PM datasets such as those presented here, which do not cover all position angles. Spherical models then yield biased estimates for both the velocity anisotropy and the inferred cusp slope.

    \item We infer the DM density slope averaged over the spatial range for which we have PM measurements, $\Gamma_{\rm dark}$, to be $-0.83^{+ 0.32}_{- 0.37}$, and an asymptotic DM density slope of $\gamma_{\rm dark} = -0.71^{+ 0.44}_{- 0.49}$. Cores larger than 487~pc, 717~pc and 942~pc are ruled out at 1-, 2- and 3-$\sigma$ confidence, respectively. The data do not have the constraining power to distinguish between different plausible parametrizations for the cusped DM density profile (see Table~\ref{tab: mass-modeling-sph}).
    
    \item The measured slope is in good agreement with $\Lambda$CDM predictions,\footnote{For reference, a traditional NFW profile has an asymptotic cusp slope of $-1$.} given that our measurements fall well within the break radius of the DM density profile predicted by cosmological simulations. Our best likelihood results corroborate the idea that DM is formed by some sort of cold particle.
    An asymptotic core is marginally inconsistent with the data at 89.5\% confidence, when marginalized over all other quantities, arguing against modified DM scenarios such as warm DM or SIDM. Nonetheless, a small asymptotic core cannot be effectively ruled out.
    
    \item The measured cusp slope provides no evidence that it has been lowered through baryonic feedback processes, although this cannot be ruled out either. However, this is not unexpected given Draco's low mass and ancient star formation history. This provides no insight into how the DM profiles of higher-mass or star-forming galaxies may be impacted by baryonic physics.

    \item We measure Draco's DM halo to have a mass of $1.20^{+ 0.79}_{- 0.28} \times 10^{8}~\msun$ at the outermost data point ($R_{\rm max} = 900$~pc), where the circular velocity reaches $v_{\rm circ} = 24.19^{+ 6.31}_{- 2.97}~\kms$. The maximum circular velocity in our best-fit models is $v_{\rm circ}^{\rm max}(r) = 27.78^{+ 10.85}_{- 5.97}~\kms$.
    
    \item We infer a dynamical distance of $75.37^{+ 4.73}_{- 4.00}$~kpc. 
    This is consistent with estimates obtained in the literature using other methods. Our $\sim 6$\% distance uncertainty is competitive with the uncertainties inherent to galaxy distances based on stellar evolution and the cosmic distance ladder. This is similar to the situation for globular clusters \citep{Watkins+15}. 
\end{itemize}

We have obtained here one of the most reliable constraints to date on the DM density profiles of dwarf galaxies. The results lessen the tension around the `cusp-core' problem, and give further credence to standard $\Lambda$CDM cosmology. Our study is just a first step into the realm of 3D axisymmetric dynamics of dSphs. The methods we used can be largely generalized to other systems, and further studies on both Draco and other galaxies are already underway. Hence, many substantial advances are likely to be made in this area in the coming years.

\section{Acknowledgments}

Support for this work was provided by NASA through grants for programs GO-12966 and GO-16737 from the Space Telescope Science Institute (STScI), which is operated by the Association of Universities for Research in Astronomy (AURA), Inc., under NASA contract NAS5-26555,
and through funding to the \jwst\ Telescope Scientist Team (PI: M. Mountain) through grant 80NSSC20K0586.
M.G.W. acknowledges support from National Science Foundation (NSF) grants AST-1909584 and AST-2206046.  We thank Jorge Peñarrubia for thoughtful suggestions concerning our models, and Justin Read for sharing his \textsc{Python} algorithm to compute $M_{\star} - M_{200}$ relations as published in \cite{Read+17}. \\
The Digitized Sky Surveys were produced at the Space Telescope Science Institute under U.S. Government grant NAG W-2166. The images of these surveys are based on photographic data obtained using the Oschin Schmidt Telescope on Palomar Mountain and the UK Schmidt Telescope. 
The plates were processed into the present compressed digital form with the permission of these institutions. \\
This work has made use of data from the European Space Agency (ESA) mission \gaia\ (\url{https://www.cosmos.esa.int/gaia}), processed by the \gaia\ Data Processing and Analysis Consortium (DPAC, \url{https://www.cosmos.esa.int/web/gaia/dpac/consortium}). Funding for the DPAC has been provided by national institutions, in particular the institutions participating in the \gaia\ Multilateral Agreement.

This project is part of the HSTPROMO (High-resolution Space Telescope PROper MOtion) Collaboration (\url{https://www.stsci.edu/~marel/hstpromo.html}), a set of projects aimed at improving our dynamical understanding of stars, clusters, and galaxies in the nearby Universe through measurement and interpretation of PMs from \hst, \gaia, and other space observatories. We thank the collaboration members for the sharing of their ideas and software.

\vspace{3mm}
\textit{Data availability:} The data that we used to construct proper motions were obtained from the Mikulski Archive for Space Telescopes (MAST) at the Space Telescope Science Institute. They relate to \hst\ observations from programs GO-10229, GO-10812, GO-12966 and GO-16737, which can be accessed via \dataset[DOI: 10.17909/f6wq-1098]{https://doi.org/10.17909/f6wq-1098}. We also share our final data products on \textsc{Zenodo} at \dataset[DOI: 10.5281/zenodo.11111113]{https://zenodo.org/records/11111113}, which consist of proper motions, 3D velocity dispersion profiles, rotation profile and the catalog ID of the stars from \cite{Walker+23} used in this work.

%

\vspace{3mm}
\facilities{
HST; Gaia; MMT.
}


\software{
{\sc Python} \citep{VanRossum09}, 
{\sc BALRoGO} \citep{Vitral21},
\jampy\ \citep{Cappellari20},
\mpo\ (Mamon \& Vitral, in prep.),
{\sc emcee} \citep{emcee},
{\sc Scipy} \citep{Jones+01},
{\sc Numpy} \citep{vanderWalt11},
{\sc Matplotlib} \citep{Hunter07},
{\sc aplpy} \citep{aplpy2012},
\scf\ \citep{deBruijne+96},
{\sc GeoGebra} \citep{Hohenwarter02}, as well as software from Wolfram Research, Inc., Mathematica, Version 12.0, Champaign, IL (2019).
}

\bibliography{src}{}

\begin{thebibliography}{}
\expandafter\ifx\csname natexlab\endcsname\relax\def\natexlab#1{#1}\fi
\providecommand{\url}[1]{\href{#1}{#1}}
\providecommand{\dodoi}[1]{doi:~\href{http://doi.org/#1}{\nolinkurl{#1}}}
\providecommand{\doeprint}[1]{\href{http://ascl.net/#1}{\nolinkurl{http://ascl.net/#1}}}
\providecommand{\doarXiv}[1]{\href{https://arxiv.org/abs/#1}{\nolinkurl{https://arxiv.org/abs/#1}}}

\bibitem[{Akaike(1973)}]{akaike1973information}
Akaike, H. 1973, Information Theory and an Extension of the Maximum Likelihood Principle (New York, NY: Springer New York), 199--213

\bibitem[{{Akaike}(1983)}]{Akaike83}
{Akaike}, H. 1983, Internaltional Statistical Institute, 44, 277

\bibitem[{{Amorisco} \& {Evans}(2012)}]{Amorisco&Evans12}
{Amorisco}, N.~C., \& {Evans}, N.~W. 2012, \mnras, 419, 184, \dodoi{10.1111/j.1365-2966.2011.19684.x}

\bibitem[{{Anderson}(2022)}]{Anderson2022}
{Anderson}, J. 2022, {One-Pass HST Photometry with hst1pass}, Instrument Science Report ACS 2022-02

\bibitem[{{Anderson} {et~al.}(2006){Anderson}, {Bedin}, {Piotto}, {Yadav}, \& {Bellini}}]{Anderson+06}
{Anderson}, J., {Bedin}, L.~R., {Piotto}, G., {Yadav}, R.~S., \& {Bellini}, A. 2006, \aap, 454, 1029, \dodoi{10.1051/0004-6361:20065004}

\bibitem[{{Aparicio} {et~al.}(2001){Aparicio}, {Carrera}, \& {Mart{\'\i}nez-Delgado}}]{Aparicio+01}
{Aparicio}, A., {Carrera}, R., \& {Mart{\'\i}nez-Delgado}, D. 2001, \aj, 122, 2524, \dodoi{10.1086/323535}

\bibitem[{{Armandroff} {et~al.}(1995){Armandroff}, {Olszewski}, \& {Pryor}}]{Armandroff+95}
{Armandroff}, T.~E., {Olszewski}, E.~W., \& {Pryor}, C. 1995, \aj, 110, 2131, \dodoi{10.1086/117675}

\bibitem[{{Bacon} {et~al.}(1983){Bacon}, {Simien}, \& {Monnet}}]{Bacon+83}
{Bacon}, R., {Simien}, F., \& {Monnet}, G. 1983, \aap, 128, 405

\bibitem[{{Battaglia} {et~al.}(2008){Battaglia}, {Helmi}, {Tolstoy}, {Irwin}, {Hill}, \& {Jablonka}}]{Battaglia+08}
{Battaglia}, G., {Helmi}, A., {Tolstoy}, E., {et~al.} 2008, \apjl, 681, L13, \dodoi{10.1086/590179}

\bibitem[{{Battaglia} {et~al.}(2022){Battaglia}, {Taibi}, {Thomas}, \& {Fritz}}]{Battaglia+22}
{Battaglia}, G., {Taibi}, S., {Thomas}, G.~F., \& {Fritz}, T.~K. 2022, \aap, 657, A54, \dodoi{10.1051/0004-6361/202141528}

\bibitem[{{Baumgardt} {et~al.}(2019){Baumgardt}, {Hilker}, {Sollima}, \& {Bellini}}]{Baumgardt+19}
{Baumgardt}, H., {Hilker}, M., {Sollima}, A., \& {Bellini}, A. 2019, \mnras, 482, 5138, \dodoi{10.1093/mnras/sty2997}

\bibitem[{{Bellini} {et~al.}(2014){Bellini}, {Anderson}, {van der Marel}, {Watkins}, {King}, {Bianchini}, {Chanam{\'e}}, {Chandar}, {Cool}, {Ferraro}, {Ford}, \& {Massari}}]{Bellini+14}
{Bellini}, A., {Anderson}, J., {van der Marel}, R.~P., {et~al.} 2014, \apj, 797, 115, \dodoi{10.1088/0004-637X/797/2/115}

\bibitem[{{Bellini} {et~al.}(2018){Bellini}, {Libralato}, {Bedin}, {Milone}, {van der Marel}, {Anderson}, {Apai}, {Burgasser}, {Marino}, \& {Rees}}]{Bellini+2018}
{Bellini}, A., {Libralato}, M., {Bedin}, L.~R., {et~al.} 2018, \apj, 853, 86, \dodoi{10.3847/1538-4357/aaa3ec}

\bibitem[{{Bianchini} {et~al.}(2016){Bianchini}, {Norris}, {van de Ven}, {Schinnerer}, {Bellini}, {van der Marel}, {Watkins}, \& {Anderson}}]{Bianchini+16}
{Bianchini}, P., {Norris}, M.~A., {van de Ven}, G., {et~al.} 2016, \apjl, 820, L22, \dodoi{10.3847/2041-8205/820/1/L22}

\bibitem[{{Binney}(1980)}]{Binney80}
{Binney}, J. 1980, \mnras, 190, 873

\bibitem[{{Binney} \& {Mamon}(1982)}]{Binney&Mamon82}
{Binney}, J., \& {Mamon}, G.~A. 1982, \mnras, 200, 361, \dodoi{10.1093/mnras/200.2.361}

\bibitem[{{Binney} \& {Tremaine}(1987)}]{Binney&Tremaine87}
{Binney}, J., \& {Tremaine}, S. 1987, Galactic dynamics (Princeton, NJ: Princeton University Press)

\bibitem[{{Bonanos} {et~al.}(2004){Bonanos}, {Stanek}, {Szentgyorgyi}, {Sasselov}, \& {Bakos}}]{Bonanos+04}
{Bonanos}, A.~Z., {Stanek}, K.~Z., {Szentgyorgyi}, A.~H., {Sasselov}, D.~D., \& {Bakos}, G.~{\'A}. 2004, \aj, 127, 861, \dodoi{10.1086/381073}

\bibitem[{{Breddels} {et~al.}(2013){Breddels}, {Helmi}, {van den Bosch}, {van de Ven}, \& {Battaglia}}]{Breddels+13}
{Breddels}, M.~A., {Helmi}, A., {van den Bosch}, R.~C.~E., {van de Ven}, G., \& {Battaglia}, G. 2013, \mnras, 433, 3173, \dodoi{10.1093/mnras/stt956}

\bibitem[{{Brooks} \& {Zolotov}(2014)}]{Brooks&Zolotov&Zolotov14}
{Brooks}, A.~M., \& {Zolotov}, A. 2014, \apj, 786, 87, \dodoi{10.1088/0004-637X/786/2/87}

\bibitem[{{Brownsberger} \& {Randall}(2021)}]{Brownsberger&Randall&Randall21}
{Brownsberger}, S.~R., \& {Randall}, L. 2021, \mnras, 501, 2332, \dodoi{10.1093/mnras/staa3719}

\bibitem[{{Burkert}(1995)}]{Burkert95}
{Burkert}, A. 1995, \apjl, 447, L25

\bibitem[{{Burnham} \& {Anderson}(2002)}]{Burnham&Anderson02}
{Burnham}, K.~P., \& {Anderson}, D.~R. 2002, A Practival Information-Theoretic Approach, 2nd edn. (New York: Springer)

\bibitem[{{Cappellari}(2008)}]{Cappellari08}
{Cappellari}, M. 2008, \mnras, 390, 71, \dodoi{10.1111/j.1365-2966.2008.13754.x}

\bibitem[{{Cappellari}(2020)}]{Cappellari20}
---. 2020, \mnras, 494, 4819, \dodoi{10.1093/mnras/staa959}

\bibitem[{{Chanam{\'e}} {et~al.}(2008){Chanam{\'e}}, {Kleyna}, \& {van der Marel}}]{Chaname+08}
{Chanam{\'e}}, J., {Kleyna}, J., \& {van der Marel}, R. 2008, \apj, 682, 841, \dodoi{10.1086/589429}

\bibitem[{{Dalcanton} \& {Hogan}(2001)}]{Dalcanton&Hogan01}
{Dalcanton}, J.~J., \& {Hogan}, C.~J. 2001, \apj, 561, 35, \dodoi{10.1086/323207}

\bibitem[{{de Bruijne} {et~al.}(1996){de Bruijne}, {van der Marel}, \& {de Zeeuw}}]{deBruijne+96}
{de Bruijne}, J. H.~J., {van der Marel}, R.~P., \& {de Zeeuw}, P.~T. 1996, \mnras, 282, 909, \dodoi{10.1093/mnras/282.3.909}

\bibitem[{{del Pino} {et~al.}(2022){del Pino}, {Libralato}, {van der Marel}, {Bennet}, {Fardal}, {Anderson}, {Bellini}, {Tony Sohn}, \& {Watkins}}]{delPino+22}
{del Pino}, A., {Libralato}, M., {van der Marel}, R.~P., {et~al.} 2022, \apj, 933, 76, \dodoi{10.3847/1538-4357/ac70cf}

\bibitem[{{Dutton} \& {Macci{\`o}}(2014)}]{Dutton&Maccio14}
{Dutton}, A.~A., \& {Macci{\`o}}, A.~V. 2014, \mnras, 441, 3359, \dodoi{10.1093/mnras/stu742}

\bibitem[{{Einasto}(1965)}]{Einasto65}
{Einasto}, J. 1965, {Trudy Inst. Astroz. Alma-Ata}, 51, 87

\bibitem[{{Evans} \& {de Zeeuw}(1994)}]{Evans&deZeeuw94}
{Evans}, N.~W., \& {de Zeeuw}, P.~T. 1994, \mnras, 271, 202, \dodoi{10.1093/mnras/271.1.202}

\bibitem[{{Fardal} {et~al.}(2021){Fardal}, {van der Marel}, {del Pino}, \& {Sohn}}]{Fardal+21}
{Fardal}, M.~A., {van der Marel}, R., {del Pino}, A., \& {Sohn}, S.~T. 2021, \aj, 161, 58, \dodoi{10.3847/1538-3881/abcccf}

\bibitem[{{Fitts} {et~al.}(2017){Fitts}, {Boylan-Kolchin}, {Elbert}, {Bullock}, {Hopkins}, {O{\~n}orbe}, {Wetzel}, {Wheeler}, {Faucher-Gigu{\`e}re}, {Kere{\v{s}}}, {Skillman}, \& {Weisz}}]{Fitts+17}
{Fitts}, A., {Boylan-Kolchin}, M., {Elbert}, O.~D., {et~al.} 2017, \mnras, 471, 3547, \dodoi{10.1093/mnras/stx1757}

\bibitem[{{Flores} \& {Primack}(1994)}]{Flores&Primack94}
{Flores}, R.~A., \& {Primack}, J.~R. 1994, \apjl, 427, L1, \dodoi{10.1086/187350}

\bibitem[{{Foreman-Mackey} {et~al.}(2013){Foreman-Mackey}, {Hogg}, {Lang}, \& {Goodman}}]{emcee}
{Foreman-Mackey}, D., {Hogg}, D.~W., {Lang}, D., \& {Goodman}, J. 2013, PASP, 125, 306, \dodoi{10.1086/670067}

\bibitem[{{Genina} {et~al.}(2018){Genina}, {Ben{\'\i}tez-Llambay}, {Frenk}, {Cole}, {Fattahi}, {Navarro}, {Oman}, {Sawala}, \& {Theuns}}]{Genina+18}
{Genina}, A., {Ben{\'\i}tez-Llambay}, A., {Frenk}, C.~S., {et~al.} 2018, \mnras, 474, 1398, \dodoi{10.1093/mnras/stx2855}

\bibitem[{{Genina} {et~al.}(2020){Genina}, {Read}, {Frenk}, {Cole}, {Ben{\'\i}tez-Llambay}, {Ludlow}, {Navarro}, {Oman}, \& {Robertson}}]{Genina+20}
{Genina}, A., {Read}, J.~I., {Frenk}, C.~S., {et~al.} 2020, \mnras, 498, 144, \dodoi{10.1093/mnras/staa2352}

\bibitem[{{Gilmore} {et~al.}(2022){Gilmore}, {Randich}, {Worley}, {Hourihane}, {Gonneau}, {Sacco}, {Lewis}, {Magrini}, {Fran{\c{c}}ois}, {Jeffries}, {Koposov}, {Bragaglia}, {Alfaro}, {Allende Prieto}, {Blomme}, {Korn}, {Lanzafame}, {Pancino}, {Recio-Blanco}, {Smiljanic}, {Van Eck}, {Zwitter}, {Bensby}, {Flaccomio}, {Irwin}, {Franciosini}, {Morbidelli}, {Damiani}, {Bonito}, {Friel}, {Vink}, {Prisinzano}, {Abbas}, {Hatzidimitriou}, {Held}, {Jordi}, {Paunzen}, {Spagna}, {Jackson}, {Ma{\'\i}z Apell{\'a}niz}, {Asplund}, {Bonifacio}, {Feltzing}, {Binney}, {Drew}, {Ferguson}, {Micela}, {Negueruela}, {Prusti}, {Rix}, {Vallenari}, {Bergemann}, {Casey}, {de Laverny}, {Frasca}, {Hill}, {Lind}, {Sbordone}, {Sousa}, {Adibekyan}, {Caffau}, {Daflon}, {Feuillet}, {Gebran}, {Gonzalez Hernandez}, {Guiglion}, {Herrero}, {Lobel}, {Merle}, {Mikolaitis}, {Montes}, {Morel}, {Ruchti}, {Soubiran}, {Tabernero}, {Tautvai{\v{s}}ien{\.{e}}}, {Traven}, {Valentini}, {Van der Swaelmen}, {Villanova}, {Viscasillas V{\'a}zquez}, {Bayo},
  {Biazzo}, {Carraro}, {Edvardsson}, {Heiter}, {Jofr{\'e}}, {Marconi}, {Martayan}, {Masseron}, {Monaco}, {Walton}, {Zaggia}, {Aguirre B{\o}rsen-Koch}, {Alves}, {Balaguer-Nunez}, {Barklem}, {Barrado}, {Bellazzini}, {Berlanas}, {Binks}, {Bressan}, {Capuzzo-Dolcetta}, {Casagrande}, {Casamiquela}, {Collins}, {D'Orazi}, {Dantas}, {Debattista}, {Delgado-Mena}, {Di Marcantonio}, {Drazdauskas}, {Evans}, {Famaey}, {Franchini}, {Fr{\'e}mat}, {Fu}, {Geisler}, {Gerhard}, {Gonz{\'a}lez Solares}, {Grebel}, {Guti{\'e}rrez Albarr{\'a}n}, {Jim{\'e}nez-Esteban}, {J{\"o}nsson}, {Khachaturyants}, {Kordopatis}, {Kos}, {Lagarde}, {Ludwig}, {Mahy}, {Mapelli}, {Marfil}, {Martell}, {Messina}, {Miglio}, {Minchev}, {Moitinho}, {Montalban}, {Monteiro}, {Morossi}, {Mowlavi}, {Mucciarelli}, {Murphy}, {Nardetto}, {Ortolani}, {Paletou}, {Palou{\v{s}}}, {Pickering}, {Quirrenbach}, {Re Fiorentin}, {Read}, {Romano}, {Ryde}, {Sanna}, {Santos}, {Seabroke}, {Spina}, {Steinmetz}, {Stonkut{\'e}}, {Sutorius}, {Th{\'e}venin}, {Tosi}, {Tsantaki},
  {Wright}, {Wyse}, {Zoccali}, {Zorec}, \& {Zucker}}]{Gilmore+22}
{Gilmore}, G., {Randich}, S., {Worley}, C.~C., {et~al.} 2022, \aap, 666, A120, \dodoi{10.1051/0004-6361/202243134}

\bibitem[{{Graham} \& {Driver}(2005)}]{Graham&Driver05}
{Graham}, A.~W., \& {Driver}, S.~P. 2005, \pasa, 22, 118, \dodoi{10.1071/AS05001}

\bibitem[{{Guerra} {et~al.}(2023){Guerra}, {Geha}, \& {Strigari}}]{Guerra+23}
{Guerra}, J., {Geha}, M., \& {Strigari}, L.~E. 2023, \apj, 943, 121, \dodoi{10.3847/1538-4357/aca8a5}

\bibitem[{{Hammer} {et~al.}(2018){Hammer}, {Yang}, {Arenou}, {Babusiaux}, {Wang}, {Puech}, \& {Flores}}]{Hammer+18}
{Hammer}, F., {Yang}, Y., {Arenou}, F., {et~al.} 2018, \apj, 860, 76, \dodoi{10.3847/1538-4357/aac3da}

\bibitem[{{Hammer} {et~al.}(2024){Hammer}, {Wang}, {Mamon}, {Pawlowski}, {Yang}, {Jiao}, {Li}, {Bonifacio}, {Caffau}, \& {Wang}}]{Hammer+24}
{Hammer}, F., {Wang}, J., {Mamon}, G.~A., {et~al.} 2024, \mnras, 527, 2718, \dodoi{10.1093/mnras/stad2922}

\bibitem[{{Han} {et~al.}(2023){Han}, {Dey}, {Price-Whelan}, {Najita}, {Schlafly}, {Saydjari}, {Wechsler}, {Bonaca}, {Schlegel}, {Conroy}, {Raichoor}, {Drlica-Wagner}, {Kollmeier}, {Koposov}, {Besla}, {Rix}, {Goodman}, {Finkbeiner}, {Anand}, {Ashby}, {Bahr-Kalus}, {Beaton}, {Behera}, {Bell}, {Bellm}, {BenZvi}, {Beraldo e Silva}, {Birrer}, {Blanton}, {Bock}, {Broekgaarden}, {Brout}, {Brown}, {Brown}, {Bulbul}, {Calderon}, {Carlin}, {Carrillo}, {Castander}, {Chakraborty}, {Chandra}, {Chiang}, {Choi}, {Clark}, {Clarkson}, {Cooper}, {Crill}, {Cunha}, {Cunningham}, {Dalcanton}, {Danieli}, {Daylan}, {de Jong}, {DeRose}, {Dey}, {Dickinson}, {Dominguez}, {Dong}, {Eifler}, {El-Badry}, {Erkal}, {Escala}, {Fazio}, {Ferguson}, {Ferraro}, {Filion}, {Forero-Romero}, {Fu}, {Galbany}, {Garavito-Camargo}, {Gawiser}, {Geha}, {Gnedin}, {Gomez}, {Greene}, {Guy}, {Hadzhiyska}, {Hawkins}, {Heinrich}, {Hernquist}, {Hirata}, {Hora}, {Horowitz}, {Horta}, {Huang}, {Huang}, {Huanyuan}, {Hunt}, {Ibata}, {Jannuzi}, {Johnston}, {Jones},
  {Juneau}, {Kado-Fong}, {Kalari}, {Kallivayalil}, {Karim}, {Keeley}, {Khoperskov}, {Kim}, {Kov{\'a}cs}, {Krause}, {Kremer}, {Kremin}, {Krolewski}, {Kulkarni}, {Kuna}, {L'Huillier}, {Lacy}, {Lan}, {Lang}, {Leahy}, {Li}, {Lim}, {L{\'o}pez-Morales}, {Macri}, {Marc}, {Mau}, {McCarthy}, {McDonald}, {McQuinn}, {Meisner}, {Melnick}, {Merloni}, {Millard}, {Millon}, {Minchev}, {Montero-Camacho}, {Morales-Gutierrez}, {Morrell}, {Moustakas}, {Moustakas}, {Murray}, {Mutlu-Pakdil}, {Myeong}, {Myers}, {Nadler}, {Navarete}, {Ness}, {Nidever}, {Nikutta}, {Nushkia}, {Olsen}, {Pace}, {Pacucci}, {Padmanabhan}, {Parkinson}, {Pearson}, {Peng}, {Petric}, {Petric}, {Ratcliffe}, {Razieh}, {Reiprich}, {Rezaie}, {Ricci}, {Rich}, {Richstein}, {Riley}, {Rockosi}, {Rossi}, {Salvato}, {Samushia}, {Sanchez}, {Sand}, {E Sanderson}, {{\v{S}}ar{\v{c}}evi{\'c}}, {Sarkar}, {Savino}, {Schweizer}, {Shafieloo}, {Shengqi}, {Shields}, {Shipp}, {Simon}, {Siudek}, {Siwei}, {Slepian}, {Smith}, {Sobeck}, {Sohn}, {Som}, {Speagle}, {Spergel}, {Szabo},
  {Tan}, {Theissen}, {Tollerud}, {Tolls}, {Tran}, {Tsiane}, {Vacca}, {Valluri}, {Verberi}, {Warfield}, {Weaverdyck}, {Weiner}, {Weisz}, {Wetzel}, {White}, {Williams}, {Wolk}, {Wu}, {Wyse}, {Yang}, {Zaritsky}, {Zelko}, {Zhimin}, \& {Zucker}}]{Han+23}
{Han}, J.~J., {Dey}, A., {Price-Whelan}, A.~M., {et~al.} 2023, arXiv e-prints, arXiv:2306.11784, \dodoi{10.48550/arXiv.2306.11784}

\bibitem[{{Hargreaves} {et~al.}(1996){Hargreaves}, {Gilmore}, {Irwin}, \& {Carter}}]{Hargreaves+96}
{Hargreaves}, J.~C., {Gilmore}, G., {Irwin}, M.~J., \& {Carter}, D. 1996, \mnras, 282, 305, \dodoi{10.1093/mnras/282.2.305}

\bibitem[{{Hattori} {et~al.}(2021){Hattori}, {Valluri}, \& {Vasiliev}}]{Hattori+21}
{Hattori}, K., {Valluri}, M., \& {Vasiliev}, E. 2021, \mnras, 508, 5468, \dodoi{10.1093/mnras/stab2898}

\bibitem[{{Hayashi} {et~al.}(2020){Hayashi}, {Chiba}, \& {Ishiyama}}]{Hayashi+20}
{Hayashi}, K., {Chiba}, M., \& {Ishiyama}, T. 2020, \apj, 904, 45, \dodoi{10.3847/1538-4357/abbe0a}

\bibitem[{Hohenwarter(2002)}]{Hohenwarter02}
Hohenwarter, M. 2002, Master's thesis, Paris Lodron University, Salzburg, Austria

\bibitem[{{Hunter}(2007)}]{Hunter07}
{Hunter}, J.~D. 2007, Computing in Science \& Engineering, 9, 90, \dodoi{10.1109/MCSE.2007.55}

\bibitem[{{Hurvich} \& {Tsai}(1989)}]{HurvichTsai89}
{Hurvich}, C.~M., \& {Tsai}, C.-L. 1989, Biometrika, 76, 297

\bibitem[{{Jiao} {et~al.}(2023){Jiao}, {Hammer}, {Wang}, {Wang}, {Amram}, {Chemin}, \& {Yang}}]{Jiao+23}
{Jiao}, Y., {Hammer}, F., {Wang}, H., {et~al.} 2023, \aap, 678, A208, \dodoi{10.1051/0004-6361/202347513}

\bibitem[{{Jing} \& {Suto}(2000)}]{JS00}
{Jing}, Y.~P., \& {Suto}, Y. 2000, \apjl, 529, L69

\bibitem[{Jones {et~al.}(2001--)Jones, Oliphant, Peterson, {et~al.}}]{Jones+01}
Jones, E., Oliphant, T., Peterson, P., {et~al.} 2001--, {SciPy}: Open source scientific tools for {Python}.
\newblock \url{http://www.scipy.org/}

\bibitem[{{Kazantzidis} {et~al.}(2004{\natexlab{a}}){Kazantzidis}, {Kravtsov}, {Zentner}, {Allgood}, {Nagai}, \& {Moore}}]{Kazantzidis+04_DMshape}
{Kazantzidis}, S., {Kravtsov}, A.~V., {Zentner}, A.~R., {et~al.} 2004{\natexlab{a}}, \apjl, 611, L73, \dodoi{10.1086/423992}

\bibitem[{{Kazantzidis} {et~al.}(2004{\natexlab{b}}){Kazantzidis}, {Mayer}, {Mastropietro}, {Diemand}, {Stadel}, \& {Moore}}]{Kazantzidis+04_densityprofs}
{Kazantzidis}, S., {Mayer}, L., {Mastropietro}, C., {et~al.} 2004{\natexlab{b}}, \apj, 608, 663, \dodoi{10.1086/420840}

\bibitem[{Kenney \& Keeping(1951)}]{Kenney&KeepingPT2}
Kenney, J., \& Keeping, E. 1951, Mathematics of Statistics No. pt. 2 (Van Nostrand).
\newblock \url{https://books.google.com/books?id=AH4GxAEACAAJ}

\bibitem[{Kenney \& Keeping(1963)}]{Kenney&KeepingPT1}
---. 1963, Mathematics of Statistics No. pt. 1 (D. Van Nostrand).
\newblock \url{https://books.google.com/books?id=kROSAQAACAAJ}

\bibitem[{{Klessen} \& {Kroupa}(1998)}]{Klessen&Kroupa&Kroupa98}
{Klessen}, R.~S., \& {Kroupa}, P. 1998, \apj, 498, 143, \dodoi{10.1086/305540}

\bibitem[{{Kleyna} {et~al.}(2002){Kleyna}, {Wilkinson}, {Evans}, {Gilmore}, \& {Frayn}}]{Kleyna+02}
{Kleyna}, J., {Wilkinson}, M.~I., {Evans}, N.~W., {Gilmore}, G., \& {Frayn}, C. 2002, \mnras, 330, 792, \dodoi{10.1046/j.1365-8711.2002.05155.x}

\bibitem[{{Kochanek} \& {Rybicki}(1996)}]{Kochanek&Rybicki96}
{Kochanek}, C.~S., \& {Rybicki}, G.~B. 1996, \mnras, 280, 1257, \dodoi{10.1093/mnras/280.4.1257}

\bibitem[{{Kowalczyk} {et~al.}(2019){Kowalczyk}, {del Pino}, {{\L}okas}, \& {Valluri}}]{Kowalczyk+19}
{Kowalczyk}, K., {del Pino}, A., {{\L}okas}, E.~L., \& {Valluri}, M. 2019, \mnras, 482, 5241, \dodoi{10.1093/mnras/sty3100}

\bibitem[{{Kozhurina-Platais} {et~al.}(2015){Kozhurina-Platais}, {Borncamp}, {Anderson}, {Grogin}, \& {Hack}}]{Kozhurina-Platais+2015}
{Kozhurina-Platais}, V., {Borncamp}, D., {Anderson}, J., {Grogin}, N., \& {Hack}, M. 2015, {ACS/WFC Revised Geometric Distortion for DrizzlePac}, Instrument Science Report ACS/WFC 2015-06, 47 pages

\bibitem[{{Kroupa} {et~al.}(1993){Kroupa}, {Tout}, \& {Gilmore}}]{Kroupa+93}
{Kroupa}, P., {Tout}, C.~A., \& {Gilmore}, G. 1993, \mnras, 262, 545, \dodoi{10.1093/mnras/262.3.545}

\bibitem[{{Lambas} {et~al.}(1992){Lambas}, {Maddox}, \& {Loveday}}]{Lambas+92}
{Lambas}, D.~G., {Maddox}, S.~J., \& {Loveday}, J. 1992, \mnras, 258, 404, \dodoi{10.1093/mnras/258.2.404}

\bibitem[{{Lazar} \& {Bullock}(2020)}]{Lazar&Bullock20}
{Lazar}, A., \& {Bullock}, J.~S. 2020, \mnras, 493, 5825, \dodoi{10.1093/mnras/staa692}

\bibitem[{{Leonard} \& {Merritt}(1989)}]{Leonard&Merritt89}
{Leonard}, P. J.~T., \& {Merritt}, D. 1989, \apj, 339, 195, \dodoi{10.1086/167287}

\bibitem[{{Lewis} \& {Bridle}(2002)}]{Lewis&Bridle02}
{Lewis}, A., \& {Bridle}, S. 2002, \prd, 66, 103511, \dodoi{10.1103/PhysRevD.66.103511}

\bibitem[{{Li} {et~al.}(2021){Li}, {Koposov}, {Erkal}, {Ji}, {Shipp}, {Pace}, {Hilmi}, {Kuehn}, {Lewis}, {Mackey}, {Simpson}, {Wan}, {Zucker}, {Bland-Hawthorn}, {Cullinane}, {Da Costa}, {Drlica-Wagner}, {Hattori}, {Martell}, {Sharma}, \& {S5 Collaboration}}]{Li+21}
{Li}, T.~S., {Koposov}, S.~E., {Erkal}, D., {et~al.} 2021, \apj, 911, 149, \dodoi{10.3847/1538-4357/abeb18}

\bibitem[{{Libralato} {et~al.}(2014){Libralato}, {Bellini}, {Bedin}, {Piotto}, {Platais}, {Kissler-Patig}, \& {Milone}}]{Libralato+14}
{Libralato}, M., {Bellini}, A., {Bedin}, L.~R., {et~al.} 2014, \aap, 563, A80, \dodoi{10.1051/0004-6361/201322059}

\bibitem[{{Libralato} {et~al.}(2018){Libralato}, {Bellini}, {van der Marel}, {Anderson}, {Watkins}, {Piotto}, {Ferraro}, {Nardiello}, \& {Vesperini}}]{Libralato+2018}
{Libralato}, M., {Bellini}, A., {van der Marel}, R.~P., {et~al.} 2018, \apj, 861, 99, \dodoi{10.3847/1538-4357/aac6c0}

\bibitem[{{Libralato} {et~al.}(2022){Libralato}, {Bellini}, {Vesperini}, {Piotto}, {Milone}, {van der Marel}, {Anderson}, {Aparicio}, {Barbuy}, {Bedin}, {Borsato}, {Cassisi}, {Dalessandro}, {Ferraro}, {King}, {Lanzoni}, {Nardiello}, {Ortolani}, {Sarajedini}, \& {Sohn}}]{Libralato+22}
{Libralato}, M., {Bellini}, A., {Vesperini}, E., {et~al.} 2022, \apj, 934, 150, \dodoi{10.3847/1538-4357/ac7727}

\bibitem[{{Lima Neto} {et~al.}(1999){Lima Neto}, {Gerbal}, \& {M{\'a}rquez}}]{LimaNeto+99}
{Lima Neto}, G.~B., {Gerbal}, D., \& {M{\'a}rquez}, I. 1999, \mnras, 309, 481, \dodoi{10.1046/j.1365-8711.1999.02849.x}

\bibitem[{{Lindegren} {et~al.}(2018){Lindegren}, {Hern{\'a}ndez}, {Bombrun}, {Klioner}, {Bastian}, {Ramos-Lerate}, {de Torres}, {Steidelm{\"u}ller}, {Stephenson}, {Hobbs}, {Lammers}, {Biermann}, {Geyer}, {Hilger}, {Michalik}, {Stampa}, {McMillan}, {Casta{\~n}eda}, {Clotet}, {Comoretto}, {Davidson}, {Fabricius}, {Gracia}, {Hambly}, {Hutton}, {Mora}, {Portell}, {van Leeuwen}, {Abbas}, {Abreu}, {Altmann}, {Andrei}, {Anglada}, {Balaguer-N{\'u}{\~n}ez}, {Barache}, {Becciani}, {Bertone}, {Bianchi}, {Bouquillon}, {Bourda}, {Br{\"u}semeister}, {Bucciarelli}, {Busonero}, {Buzzi}, {Cancelliere}, {Carlucci}, {Charlot}, {Cheek}, {Crosta}, {Crowley}, {de Bruijne}, {de Felice}, {Drimmel}, {Esquej}, {Fienga}, {Fraile}, {Gai}, {Garralda}, {Gonz{\'a}lez-Vidal}, {Guerra}, {Hauser}, {Hofmann}, {Holl}, {Jordan}, {Lattanzi}, {Lenhardt}, {Liao}, {Licata}, {Lister}, {L{\"o}ffler}, {Marchant}, {Martin-Fleitas}, {Messineo}, {Mignard}, {Morbidelli}, {Poggio}, {Riva}, {Rowell}, {Salguero}, {Sarasso}, {Sciacca}, {Siddiqui}, {Smart},
  {Spagna}, {Steele}, {Taris}, {Torra}, {van Elteren}, {van Reeven}, \& {Vecchiato}}]{Lindegren+18}
{Lindegren}, L., {Hern{\'a}ndez}, J., {Bombrun}, A., {et~al.} 2018, \aap, 616, A2, \dodoi{10.1051/0004-6361/201832727}

\bibitem[{{Macci{\`o}} {et~al.}(2012){Macci{\`o}}, {Paduroiu}, {Anderhalden}, {Schneider}, \& {Moore}}]{Maccio+12}
{Macci{\`o}}, A.~V., {Paduroiu}, S., {Anderhalden}, D., {Schneider}, A., \& {Moore}, B. 2012, \mnras, 424, 1105, \dodoi{10.1111/j.1365-2966.2012.21284.x}

\bibitem[{{Macci{\`o}} {et~al.}(2013){Macci{\`o}}, {Paduroiu}, {Anderhalden}, {Schneider}, \& {Moore}}]{Maccio+13}
---. 2013, \mnras, 428, 3715, \dodoi{10.1093/mnras/sts251}

\bibitem[{{Magorrian}(1999)}]{Magorrian99}
{Magorrian}, J. 1999, \mnras, 302, 530, \dodoi{10.1046/j.1365-8711.1999.02135.x}

\bibitem[{{Mamon} {et~al.}(2013){Mamon}, {Biviano}, \& {Bou{\'e}}}]{Mamon+13}
{Mamon}, G.~A., {Biviano}, A., \& {Bou{\'e}}, G. 2013, \mnras, 429, 3079, \dodoi{10.1093/mnras/sts565}

\bibitem[{{Martin} {et~al.}(2008){Martin}, {de Jong}, \& {Rix}}]{Martin+08}
{Martin}, N.~F., {de Jong}, J. T.~A., \& {Rix}, H.-W. 2008, \apj, 684, 1075, \dodoi{10.1086/590336}

\bibitem[{{Martinez}(2015)}]{Martinez15}
{Martinez}, G.~D. 2015, \mnras, 451, 2524, \dodoi{10.1093/mnras/stv942}

\bibitem[{{Mart{\'\i}nez-Garc{\'\i}a} {et~al.}(2021){Mart{\'\i}nez-Garc{\'\i}a}, {del Pino}, {Aparicio}, {van der Marel}, \& {Watkins}}]{MartinezGarcia+21}
{Mart{\'\i}nez-Garc{\'\i}a}, A.~M., {del Pino}, A., {Aparicio}, A., {van der Marel}, R.~P., \& {Watkins}, L.~L. 2021, \mnras, 505, 5884, \dodoi{10.1093/mnras/stab1568}

\bibitem[{{Mart{\'\i}nez-Garc{\'\i}a} {et~al.}(2023){Mart{\'\i}nez-Garc{\'\i}a}, {del Pino}, {{\L}okas}, {van der Marel}, \& {Aparicio}}]{MartinezGarcia+23}
{Mart{\'\i}nez-Garc{\'\i}a}, A.~M., {del Pino}, A., {{\L}okas}, E.~L., {van der Marel}, R.~P., \& {Aparicio}, A. 2023, \mnras, 526, 3589, \dodoi{10.1093/mnras/stad2941}

\bibitem[{{Massari} {et~al.}(2020){Massari}, {Helmi}, {Mucciarelli}, {Sales}, {Spina}, \& {Tolstoy}}]{Massari+20}
{Massari}, D., {Helmi}, A., {Mucciarelli}, A., {et~al.} 2020, \aap, 633, A36, \dodoi{10.1051/0004-6361/201935613}

\bibitem[{{Massari} {et~al.}(2017){Massari}, {Posti}, {Helmi}, {Fiorentino}, \& {Tolstoy}}]{Massari+17}
{Massari}, D., {Posti}, L., {Helmi}, A., {Fiorentino}, G., \& {Tolstoy}, E. 2017, \aap, 598, L9, \dodoi{10.1051/0004-6361/201630174}

\bibitem[{{McConnachie}(2012)}]{McConnachie12}
{McConnachie}, A.~W. 2012, \aj, 144, 4, \dodoi{10.1088/0004-6256/144/1/4}

\bibitem[{{McGaugh} \& {Wolf}(2010)}]{McGaugh&Wolf&Wolf10}
{McGaugh}, S.~S., \& {Wolf}, J. 2010, \apj, 722, 248, \dodoi{10.1088/0004-637X/722/1/248}

\bibitem[{{Merritt}(1985)}]{Merritt85}
{Merritt}, D. 1985, \apj, 289, 18

\bibitem[{{Moore}(1994)}]{Moore94}
{Moore}, B. 1994, \nat, 370, 629+

\bibitem[{{Muraveva} {et~al.}(2020){Muraveva}, {Clementini}, {Garofalo}, \& {Cusano}}]{Muraveva+20}
{Muraveva}, T., {Clementini}, G., {Garofalo}, A., \& {Cusano}, F. 2020, \mnras, 499, 4040, \dodoi{10.1093/mnras/staa2984}

\bibitem[{{Navarro} {et~al.}(1997){Navarro}, {Frenk}, \& {White}}]{Navarro+97}
{Navarro}, J.~F., {Frenk}, C.~S., \& {White}, S. D.~M. 1997, \apj, 490, 493, \dodoi{10.1086/304888}

\bibitem[{{Odenkirchen} {et~al.}(2001){Odenkirchen}, {Grebel}, {Harbeck}, {Dehnen}, {Rix}, {Newberg}, {Yanny}, {Holtzman}, {Brinkmann}, {Chen}, {Csabai}, {Hayes}, {Hennessy}, {Hindsley}, {Ivezi{\'c}}, {Kinney}, {Kleinman}, {Long}, {Lupton}, {Neilsen}, {Nitta}, {Snedden}, \& {York}}]{Odenkirchen+01}
{Odenkirchen}, M., {Grebel}, E.~K., {Harbeck}, D., {et~al.} 2001, \aj, 122, 2538, \dodoi{10.1086/323715}

\bibitem[{{Osipkov}(1979)}]{Osipkov79}
{Osipkov}, L.~P. 1979, Soviet Astronomy Letters, 5, 42

\bibitem[{{Pace} {et~al.}(2022){Pace}, {Erkal}, \& {Li}}]{Pace+22}
{Pace}, A.~B., {Erkal}, D., \& {Li}, T.~S. 2022, \apj, 940, 136, \dodoi{10.3847/1538-4357/ac997b}

\bibitem[{{Pianta} {et~al.}(2022){Pianta}, {Capuzzo-Dolcetta}, \& {Carraro}}]{Pianta+22}
{Pianta}, C., {Capuzzo-Dolcetta}, R., \& {Carraro}, G. 2022, \apj, 939, 3, \dodoi{10.3847/1538-4357/ac9303}

\bibitem[{{Plummer}(1911)}]{Plummer1911}
{Plummer}, H.~C. 1911, \mnras, 71, 460

\bibitem[{{Pontzen} \& {Governato}(2012)}]{Pontzen&Governato12}
{Pontzen}, A., \& {Governato}, F. 2012, \mnras, 421, 3464, \dodoi{10.1111/j.1365-2966.2012.20571.x}

\bibitem[{{Pryor} \& {Kormendy}(1990)}]{Pryor&Kormendy&Kormendy90}
{Pryor}, C., \& {Kormendy}, J. 1990, \aj, 100, 127, \dodoi{10.1086/115496}

\bibitem[{{Read} \& {Gilmore}(2005)}]{Read&Gilmore05}
{Read}, J.~I., \& {Gilmore}, G. 2005, \mnras, 356, 107, \dodoi{10.1111/j.1365-2966.2004.08424.x}

\bibitem[{{Read} {et~al.}(2017){Read}, {Iorio}, {Agertz}, \& {Fraternali}}]{Read+17}
{Read}, J.~I., {Iorio}, G., {Agertz}, O., \& {Fraternali}, F. 2017, \mnras, 467, 2019, \dodoi{10.1093/mnras/stx147}

\bibitem[{{Read} \& {Steger}(2017)}]{Read&Steger17}
{Read}, J.~I., \& {Steger}, P. 2017, \mnras, 471, 4541, \dodoi{10.1093/mnras/stx1798}

\bibitem[{{Read} {et~al.}(2018){Read}, {Walker}, \& {Steger}}]{Read+18}
{Read}, J.~I., {Walker}, M.~G., \& {Steger}, P. 2018, \mnras, 481, 860, \dodoi{10.1093/mnras/sty2286}

\bibitem[{{Read} {et~al.}(2021){Read}, {Mamon}, {Vasiliev}, {Watkins}, {Walker}, {Pe{\~n}arrubia}, {Wilkinson}, {Dehnen}, \& {Das}}]{Read+21}
{Read}, J.~I., {Mamon}, G.~A., {Vasiliev}, E., {et~al.} 2021, \mnras, 501, 978, \dodoi{10.1093/mnras/staa3663}

\bibitem[{{Robitaille} \& {Bressert}(2012)}]{aplpy2012}
{Robitaille}, T., \& {Bressert}, E. 2012, {APLpy: Astronomical Plotting Library in Python}, Astrophysics Source Code Library.
\newblock \doeprint{1208.017}

\bibitem[{{Romanowsky} \& {Kochanek}(1997)}]{Romanowsky&Kochanek97}
{Romanowsky}, A.~J., \& {Kochanek}, C.~S. 1997, \mnras, 287, 35, \dodoi{10.1093/mnras/287.1.35}

\bibitem[{{Rybicki}(1987)}]{Rybicki87}
{Rybicki}, G.~B. 1987, in Structure and Dynamics of Elliptical Galaxies, ed. P.~T. {de Zeeuw}, Vol. 127, 397, \dodoi{10.1007/978-94-009-3971-4_41}

\bibitem[{{Salpeter}(1955)}]{Salpeter55}
{Salpeter}, E.~E. 1955, \apj, 121, 161

\bibitem[{{Sameie} {et~al.}(2020){Sameie}, {Yu}, {Sales}, {Vogelsberger}, \& {Zavala}}]{Sameie+20}
{Sameie}, O., {Yu}, H.-B., {Sales}, L.~V., {Vogelsberger}, M., \& {Zavala}, J. 2020, \prl, 124, 141102, \dodoi{10.1103/PhysRevLett.124.141102}

\bibitem[{{Schwarz}(1978)}]{Schwarz78}
{Schwarz}, G. 1978, Ann. Statist., 6, 461

\bibitem[{{Sedain} \& {Kacharov}(2023)}]{Sedain&Kacharov23}
{Sedain}, A., \& {Kacharov}, N. 2023, arXiv e-prints, arXiv:2305.11256, \dodoi{10.48550/arXiv.2305.11256}

\bibitem[{{S{\'e}gall} {et~al.}(2007){S{\'e}gall}, {Ibata}, {Irwin}, {Martin}, \& {Chapman}}]{Segall+07}
{S{\'e}gall}, M., {Ibata}, R.~A., {Irwin}, M.~J., {Martin}, N.~F., \& {Chapman}, S. 2007, \mnras, 375, 831, \dodoi{10.1111/j.1365-2966.2006.11356.x}

\bibitem[{{S\'ersic}(1963)}]{Sersic63}
{S\'ersic}, J.~L. 1963, Bull. Assoc. Argentina de Astron., 6, 41

\bibitem[{{S\'ersic}(1968)}]{Sersic68}
---. 1968, Atlas de galaxias australes (Cordoba, Argentina: Observatorio Astronomico)

\bibitem[{{Simonneau} \& {Prada}(2004)}]{Simonneau&Prada04}
{Simonneau}, E., \& {Prada}, F. 2004, \rmxaa, 40, 69

\bibitem[{{Sohn} {et~al.}(2021){Sohn}, {del Pino Molina}, {Besla}, {Libralato}, {Patel}, {Pawlowski}, {Watkins}, \& {van der Marel}}]{2021hst..prop16737S}
{Sohn}, S.~T., {del Pino Molina}, A., {Besla}, G., {et~al.} 2021, {Internal Proper Motion Kinematics of Dwarf Spheroidal Galaxies: Constraining the Density and Properties of Dark Matter}, HST Proposal. Cycle 29, ID. \#16737

\bibitem[{{Sohn} {et~al.}(2017){Sohn}, {Patel}, {Besla}, {van der Marel}, {Bullock}, {Strigari}, {van de Ven}, {Walker}, \& {Bellini}}]{Sohn+17}
{Sohn}, S.~T., {Patel}, E., {Besla}, G., {et~al.} 2017, \apj, 849, 93, \dodoi{10.3847/1538-4357/aa917b}

\bibitem[{{Spencer} {et~al.}(2018){Spencer}, {Mateo}, {Olszewski}, {Walker}, {McConnachie}, \& {Kirby}}]{Spencer+18}
{Spencer}, M.~E., {Mateo}, M., {Olszewski}, E.~W., {et~al.} 2018, \aj, 156, 257, \dodoi{10.3847/1538-3881/aae3e4}

\bibitem[{{Strigari} {et~al.}(2018){Strigari}, {Frenk}, \& {White}}]{Strigari+18}
{Strigari}, L.~E., {Frenk}, C.~S., \& {White}, S. D.~M. 2018, \apj, 860, 56, \dodoi{10.3847/1538-4357/aac2d3}

\bibitem[{{Strigari} {et~al.}(2007){Strigari}, {Koushiappas}, {Bullock}, \& {Kaplinghat}}]{Strigari+07_vmax}
{Strigari}, L.~E., {Koushiappas}, S.~M., {Bullock}, J.~S., \& {Kaplinghat}, M. 2007, \prd, 75, 083526, \dodoi{10.1103/PhysRevD.75.083526}

\bibitem[{{Sugiura}(1978)}]{Sugiyara78}
{Sugiura}, N. 1978, Communications in Statistics - Theory and Methods, 7, 13, \dodoi{10.1080/03610927808827599}

\bibitem[{{Tiret} {et~al.}(2007){Tiret}, {Combes}, {Angus}, {Famaey}, \& {Zhao}}]{Tiret+07}
{Tiret}, O., {Combes}, F., {Angus}, G.~W., {Famaey}, B., \& {Zhao}, H.~S. 2007, \aap, 476, L1

\bibitem[{{Tolstoy} {et~al.}(2004){Tolstoy}, {Irwin}, {Helmi}, {Battaglia}, {Jablonka}, {Hill}, {Venn}, {Shetrone}, {Letarte}, {Cole}, {Primas}, {Francois}, {Arimoto}, {Sadakane}, {Kaufer}, {Szeifert}, \& {Abel}}]{Tolstoy+04}
{Tolstoy}, E., {Irwin}, M.~J., {Helmi}, A., {et~al.} 2004, \apjl, 617, L119, \dodoi{10.1086/427388}

\bibitem[{{van den Bosch}(1997)}]{vandenBosch97}
{van den Bosch}, F.~C. 1997, \mnras, 287, 543, \dodoi{10.1093/mnras/287.3.543}

\bibitem[{{van der Marel}(1991)}]{vanderMarel91}
{van der Marel}, R.~P. 1991, \mnras, 253, 710, \dodoi{10.1093/mnras/253.4.710}

\bibitem[{{van der Marel} {et~al.}(2002){van der Marel}, {Alves}, {Hardy}, \& {Suntzeff}}]{vanderMarel+02}
{van der Marel}, R.~P., {Alves}, D.~R., {Hardy}, E., \& {Suntzeff}, N.~B. 2002, \aj, 124, 2639, \dodoi{10.1086/343775}

\bibitem[{{van der Marel} \& {Anderson}(2010)}]{vanderMarel&Anderson&Anderson10}
{van der Marel}, R.~P., \& {Anderson}, J. 2010, \apj, 710, 1063, \dodoi{10.1088/0004-637X/710/2/1063}

\bibitem[{{van der Marel} {et~al.}(2023){van der Marel}, {Anderson}, {Bellini}, {Libralato}, {Sohn}, \& {Watkins}}]{2023jwst.prop.4513V}
{van der Marel}, R.~P., {Anderson}, J., {Bellini}, A., {et~al.} 2023, {Internal Dynamics of Milky Way Dwarf Spheroidal Galaxies}, JWST Proposal. Cycle 3, ID. \#4513

\bibitem[{{van der Marel} \& {Franx}(1993)}]{vanderMarel&Franx93}
{van der Marel}, R.~P., \& {Franx}, M. 1993, \apj, 407, 525, \dodoi{10.1086/172534}

\bibitem[{{van der Marel} {et~al.}(2000){van der Marel}, {Magorrian}, {Carlberg}, {Yee}, \& {Ellingson}}]{vanderMarel+00}
{van der Marel}, R.~P., {Magorrian}, J., {Carlberg}, R.~G., {Yee}, H.~K.~C., \& {Ellingson}, E. 2000, \aj, 119, 2038

\bibitem[{{van der Walt} {et~al.}(2011){van der Walt}, {Colbert}, \& {Varoquaux}}]{vanderWalt11}
{van der Walt}, S., {Colbert}, S.~C., \& {Varoquaux}, G. 2011, Computing in Science Engineering, 13, 22

\bibitem[{Van~Rossum \& Drake(2009)}]{VanRossum09}
Van~Rossum, G., \& Drake, F.~L. 2009, Python 3 Reference Manual (Scotts Valley, CA: CreateSpace)

\bibitem[{{Vasiliev}(2019)}]{Vasiliev19agama}
{Vasiliev}, E. 2019, \mnras, 482, 1525, \dodoi{10.1093/mnras/sty2672}

\bibitem[{{Vasiliev} \& {Baumgardt}(2021)}]{Vasiliev&Baumgardt&Baumgardt21}
{Vasiliev}, E., \& {Baumgardt}, H. 2021, \mnras, 505, 5978, \dodoi{10.1093/mnras/stab1475}

\bibitem[{{Vera-Ciro} {et~al.}(2014){Vera-Ciro}, {Sales}, {Helmi}, \& {Navarro}}]{Vera-Ciro+14}
{Vera-Ciro}, C.~A., {Sales}, L.~V., {Helmi}, A., \& {Navarro}, J.~F. 2014, \mnras, 439, 2863, \dodoi{10.1093/mnras/stu153}

\bibitem[{{Vitral}(2021)}]{Vitral21}
{Vitral}, E. 2021, \mnras, 504, 1355, \dodoi{10.1093/mnras/stab947}

\bibitem[{{Vitral} {et~al.}(2022){Vitral}, {Kremer}, {Libralato}, {Mamon}, \& {Bellini}}]{Vitral+22}
{Vitral}, E., {Kremer}, K., {Libralato}, M., {Mamon}, G.~A., \& {Bellini}, A. 2022, \mnras, 514, 806, \dodoi{10.1093/mnras/stac1337}

\bibitem[{{Vitral} {et~al.}(2023{\natexlab{a}}){Vitral}, Libralato, Kremer, Mamon, Bellini, Bedin, \& Anderson}]{Vitral+23}
{Vitral}, E., Libralato, M., Kremer, K., {et~al.} 2023{\natexlab{a}}, \mnras, 522, 5740, \dodoi{10.1093/mnras/stad1068}

\bibitem[{{Vitral} \& {Mamon}(2020)}]{Vitral&Mamon20}
{Vitral}, E., \& {Mamon}, G.~A. 2020, \aap, 635, A20, \dodoi{10.1051/0004-6361/201937202}

\bibitem[{{Vitral} \& {Mamon}(2021)}]{VitralMamon21}
---. 2021, \aap, 646, A63, \dodoi{10.1051/0004-6361/202039650}

\bibitem[{{Vitral} {et~al.}(2023{\natexlab{b}}){Vitral}, {Sohn}, {Bellini}, {Bennet}, {Chaname}, {Libralato}, {Patel}, {Watkins}, {del Pino Molina}, \& {van der Marel}}]{Vitral+23-UMi}
{Vitral}, E., {Sohn}, S.~T., {Bellini}, A., {et~al.} 2023{\natexlab{b}}, {Shedding Light on Dark Matter: Internal Proper Motions in Ursa Minor}, HST Proposal. Cycle 31, ID. \#17434

\bibitem[{{Walker} {et~al.}(2023){Walker}, {Caldwell}, {Mateo}, {Olszewski}, {Pace}, {Bailey}, {Koposov}, \& {Roederer}}]{Walker+23}
{Walker}, M.~G., {Caldwell}, N., {Mateo}, M., {et~al.} 2023, \apjs, 268, 19, \dodoi{10.3847/1538-4365/acdd79}

\bibitem[{{Walker} {et~al.}(2007){Walker}, {Mateo}, {Olszewski}, {Bernstein}, {Sen}, \& {Woodroofe}}]{Walker+07}
{Walker}, M.~G., {Mateo}, M., {Olszewski}, E.~W., {et~al.} 2007, \apjs, 171, 389, \dodoi{10.1086/517886}

\bibitem[{{Walker} {et~al.}(2009){Walker}, {Mateo}, {Olszewski}, {Pe{\~n}arrubia}, {Evans}, \& {Gilmore}}]{Walker+09}
---. 2009, \apj, 704, 1274, \dodoi{10.1088/0004-637X/704/2/1274}

\bibitem[{{Walker} {et~al.}(2015){Walker}, {Olszewski}, \& {Mateo}}]{Walker+15}
{Walker}, M.~G., {Olszewski}, E.~W., \& {Mateo}, M. 2015, \mnras, 448, 2717, \dodoi{10.1093/mnras/stv099}

\bibitem[{{Walker} \& {Pe{\~n}arrubia}(2011)}]{Walker&Penarrubia&Penarrubia11}
{Walker}, M.~G., \& {Pe{\~n}arrubia}, J. 2011, \apj, 742, 20, \dodoi{10.1088/0004-637X/742/1/20}

\bibitem[{{Wang} {et~al.}(2024){Wang}, {Hammer}, {Yang}, {Pawlowski}, {Mamon}, \& {Wang}}]{Wang+24}
{Wang}, J., {Hammer}, F., {Yang}, Y., {et~al.} 2024, \mnras, 527, 7144, \dodoi{10.1093/mnras/stad3651}

\bibitem[{{Wang} {et~al.}(2023){Wang}, {Zhu}, {Jing}, {Grand}, {Li}, {Fu}, {Li}, {Han}, {Li}, {Feng}, \& {Frenk}}]{Wang+23}
{Wang}, W., {Zhu}, L., {Jing}, Y., {et~al.} 2023, \apj, 956, 91, \dodoi{10.3847/1538-4357/acf314}

\bibitem[{{Watkins} {et~al.}(2015{\natexlab{a}}){Watkins}, {van der Marel}, {Bellini}, \& {Anderson}}]{Watkins+15}
{Watkins}, L.~L., {van der Marel}, R.~P., {Bellini}, A., \& {Anderson}, J. 2015{\natexlab{a}}, \apj, 812, 149, \dodoi{10.1088/0004-637X/812/2/149}

\bibitem[{{Watkins} {et~al.}(2015{\natexlab{b}}){Watkins}, {van der Marel}, {Bellini}, \& {Anderson}}]{Watkins+15-disp}
---. 2015{\natexlab{b}}, \apj, 803, 29, \dodoi{10.1088/0004-637X/803/1/29}

\bibitem[{{Wegg} {et~al.}(2019){Wegg}, {Gerhard}, \& {Bieth}}]{Wegg+19}
{Wegg}, C., {Gerhard}, O., \& {Bieth}, M. 2019, \mnras, 485, 3296, \dodoi{10.1093/mnras/stz572}

\bibitem[{{Wilkinson} {et~al.}(2002){Wilkinson}, {Kleyna}, {Evans}, \& {Gilmore}}]{Wilkinson+02}
{Wilkinson}, M.~I., {Kleyna}, J., {Evans}, N.~W., \& {Gilmore}, G. 2002, \mnras, 330, 778, \dodoi{10.1046/j.1365-8711.2002.05154.x}

\bibitem[{{Wilkinson} {et~al.}(2004){Wilkinson}, {Kleyna}, {Evans}, {Gilmore}, {Irwin}, \& {Grebel}}]{Wilkinson+04}
{Wilkinson}, M.~I., {Kleyna}, J.~T., {Evans}, N.~W., {et~al.} 2004, \apjl, 611, L21, \dodoi{10.1086/423619}

\bibitem[{{Wilson}(1955)}]{Wilson55}
{Wilson}, A.~G. 1955, \pasp, 67, 27, \dodoi{10.1086/126754}

\bibitem[{{Zhao}(1996)}]{Zhao96}
{Zhao}, H. 1996, \mnras, 278, 488

\bibitem[{{Zhu} {et~al.}(2024){Zhu}, {Lu}, {Cappellari}, {Li}, {Mao}, {Gao}, \& {Ge}}]{Zhu+24}
{Zhu}, K., {Lu}, S., {Cappellari}, M., {et~al.} 2024, \mnras, 527, 706, \dodoi{10.1093/mnras/stad3213}

\bibitem[{{Zhu} {et~al.}(2016){Zhu}, {van de Ven}, {Watkins}, \& {Posti}}]{Zhu+16}
{Zhu}, L., {van de Ven}, G., {Watkins}, L.~L., \& {Posti}, L. 2016, \mnras, 463, 1117, \dodoi{10.1093/mnras/stw2081}

\end{thebibliography}
\bibliographystyle{aasjournal}



\appendix

\section{Density profiles} \label{app: dens-prof}

We present here the analytical forms of the luminous and DM density profiles considered throughout our work. Those are, respectively, the \cite{Plummer1911} density profile,
\begin{equation} \label{eq: plummer}
    \rho_{\rm PLU}(r) = \frac{3 M_{\infty}}{4 \pi a^{3}} \left(1 + \frac{r^{2}}{a^{2}} \right)^{-\frac{5}{2}} \ ,
\end{equation}
where $M_{\infty}$ is the total tracer mass at infinity, and $a$ is a scale radius (the same one depicted in Table~\ref{tab: mass-modeling-sph}).
%
In projected geometry, we have the Plummer axisymmetric surface density profile, which we use in Section~\ref{sssec: center}:
\begin{equation} \label{eq: axi-sd-plummer}
\begin{split}
    \Sigma_{\rm PLU, axi}(R, \xi) =& \frac{M_{\infty}}{\pi a^{2} (1 - \epsilon)} \\
    \times& \left[1 + \frac{R^{2}}{a^{2}} \left(\cos^{2}\xi + \frac{\sin^{2}\xi}{(1-\epsilon)^{2}}\right) \right]^{-2} \ .
\end{split}
\end{equation}
where $\xi$ is the projected angle on the sky ($\xi = 0$ on the projected major axis) and $\epsilon$ is defined as $\epsilon \equiv 1 - b/a$ ($b$ and $a$ are minor and major axes, respectively).

Back to the spherical profiles used for Jeans modeling, there is the generalized Plummer model,
\begin{equation} \label{eq: gplummer}
    \rho_{\rm GPLU}(r) = (3+\gamma) \frac{M_{\infty}}{4 \pi a^{3}} \left(\frac{r}{a}\right)^{\gamma} \left(1 + \frac{r^{2}}{a^{2}} \right)^{-\frac{(\gamma + 5)}{2}} \ ,
\end{equation}
where the new parameter $\gamma < 3$ measures the inner density slope. The \cite{Kazantzidis+04_densityprofs} profile,
\begin{equation} \label{eq: gkaz}
    \rho_{\rm GKAZ}(r) = \frac{1}{\Gamma(3+\gamma)} \frac{M_{\infty}}{4 \pi a^{3}} \left(\frac{r}{a}\right)^{\gamma} \mathrm{exp}\left(-\frac{r}{a} \right) \ .
\end{equation}

The generalized \citet[][NFW]{Navarro+97} profile,
\begin{equation} \label{eq: gNFW}
\begin{split}
    \rho_{\rm GNFW}(r) =& \frac{M_{-2}/(4 \pi a^{3})}{_{2}F_{1}(1, 1; 4+\gamma; -2-\gamma)} \\ 
    \times& \left(\frac{3+\gamma}{2+\gamma}\right)^{3+\gamma} \left(\frac{r}{a}\right)^{\gamma} \left(1 + \frac{r}{a} \right)^{-(3+\gamma)}  \ ,
\end{split}
\end{equation}
where $M_{-2}$ stands for the cumulative mass within the radius where the condition ${\rm d} \log{\rho} / {\rm d} \log{r} = -2$ is satisfied. The \cite{Einasto65} profile,
\begin{equation} \label{eq: einasto}
    \rho_{\rm EIN}(r) = \frac{1}{n \, \Gamma(3 n)} \frac{M_{\infty}}{4 \pi a^{3}} \, \mathrm{exp}\left[-\left(\frac{r}{a} \right)^{\frac{1}{n}}\right] \ ,
\end{equation}
where $n$ is the Einasto index. We also use the \cite{Sersic63,Sersic68} model,
\begin{equation}  \label{eq: Sersic}
    \Sigma_{\rm SER}(R) = \frac{M_{\infty} \, b_n^{2n}}{2\pi R_{\rm e}^{2} \, n \, \Gamma(2n)}\,\exp \left [-b_n\, \left(\frac{R}{R_{\rm e}}\right)^{\frac{1}{n}}\right] \ ,
\end{equation}
where $R$ is the projected distance to the source center, $R_{\mathrm{e}}$ is the effective radius containing half of the projected luminosity and $n$ is the S\'ersic index. The term $b_{n}$ is a function of $n$, obtained by solving the equation:
\begin{equation} \label{eq: solveb}
    \Gamma(2n)/2 = \gamma(2n,b_{n}) \ ,
\end{equation}
where $\gamma(a,x) = \int_{0}^{x} t^{a-1} e^{-t} \mathrm{d} t$ is the lower incomplete gamma function. This model does not have an exact analytical deprojection into volume density (see \citealt{Graham&Driver05} for a review). Instead, we use a combination of the most precise analytical approximations for different domains of S\'ersic indexes and radii. This combination is described thoroughly in appendix~A from \cite{VitralMamon21}, and essentially uses the methods from \citet{LimaNeto+99} plus \citet{Simonneau&Prada04}, and \cite{Vitral&Mamon20}.

\section{Scale-free Reference Models} \label{app: scalefree-vpos}

\subsection{Model Parameters and Distribution Functions}
\label{app: scalefree-intro}

The Jeans models in Section~\ref{sec: methods} are largely based on numerical solution of ordinary differential equations. This means that it is somewhat complicated to quickly or approximately explore how the intrinsic properties or predicted observables of the models depend on its key parameters. For the latter purpose we have found it more convenient to resort to a simpler class of models, namely the scale-free models introduced by \cite*{deBruijne+96}.

These models consider the case of an axisymmetric luminous density with axial ratio $q$ and radial dependence $\rho_{\star} \propto r^{-\kappa}$,\footnote{We relabel the parameter $\gamma$ used in \cite*{deBruijne+96} here as $\kappa$, to avoid confusion with the central cusp slope of the DM density defined in Appendix~\ref{app: dens-prof}.} embedded in a spherical gravitational potential with circular velocity $v_{\rm circ} \propto r^{-\delta/2}$. The case of a logarithmic potential corresponds to $\delta=0$, while a Kepler potential corresponds to $\delta=1$.

For these potentials, there are different classes of phase-space distribution functions (DFs) that can yield hydrostatic equilibrium in the same geometry and with the same globally-averaged anisotropy, but with a different variation of the anisotropy in the meridional $(R, z$) plane. Specifically, \cite{deBruijne+96} defined distinct Case~I and Case~II DFs. In the Case~I DFs (for which $f(E,L_z)$ models form a subset) the quantity $\beta_{\rm B}$, as defined in Eq.~(\ref{eq: binney-beta}), is more tangential in the equatorial plane than on the symmetry axis. In the Case~II DFs, instead, $\beta_{\rm B}$ is constant throughout the system. The intrinsic and projected quantities of interest for these DFs can be expressed semi-analytically, so that they can be computed quickly.

A software package called \scf\ originally written by one of us (RvdM) for the \cite{deBruijne+96} paper is available for this purpose. The scale-free models have the additional advantage that the full DF is known, so that higher-order moments can be calculated in addition to the second-order moments constrained by the Jeans equations (see Appendix~\ref{sapp: GaussHermite}). 

The scale-free models exactly reproduce the asymptotic large-radii limit of our Jeans models in Section~\ref{sec: methods}, at intrinsic and projected radii well in excess of the scale radii $r_{\star}$ and $r_{\rm dark}$ of the luminous and dark matter (see e.g. Table~\ref{tab: mass-modeling-sph}). But they also provide a reasonable approximation at the intermediate radii where we have PM measurements, if the power-law slopes $\kappa$ and $\delta$ are set to reproduce the average slopes of the luminous density and circular velocity over those radii.

From exploration of the known properties of our best-fit Jeans models in Section~\ref{sec: methods}, we found that the combination of $\kappa=2$ and $\delta=0$ (i.e. an axisymmetric isothermal luminous density in a spherical isothermal potential) is reasonable over the relevant radial range in Draco. We show below some predictions of such models as a function of the geometry (spherical, or axisymmetric of a given axial ratio and inclination), of the type of DF, and of the globally-averaged $\overline{\beta_{\rm B}}$. The latter can be calculated numerically for given $\beta_{\rm p}$, where $\beta_{\rm p}$ is a parameter that enters into the analytical DFs and which controls the amount of anisotropy in the models.\footnote{We relabel the parameter $\beta$ in \cite{deBruijne+96} here as $\beta_{\rm p}$, to set it apart from the distinct quantities $\overline{\beta_{\rm B}}$ and $\beta_{\rm J}$ already defined by equations~(\ref{eq: binney-beta}) and~(\ref{eq: beta-jampy}).} 

\begin{figure}
\centering
\includegraphics[width=\hsize]{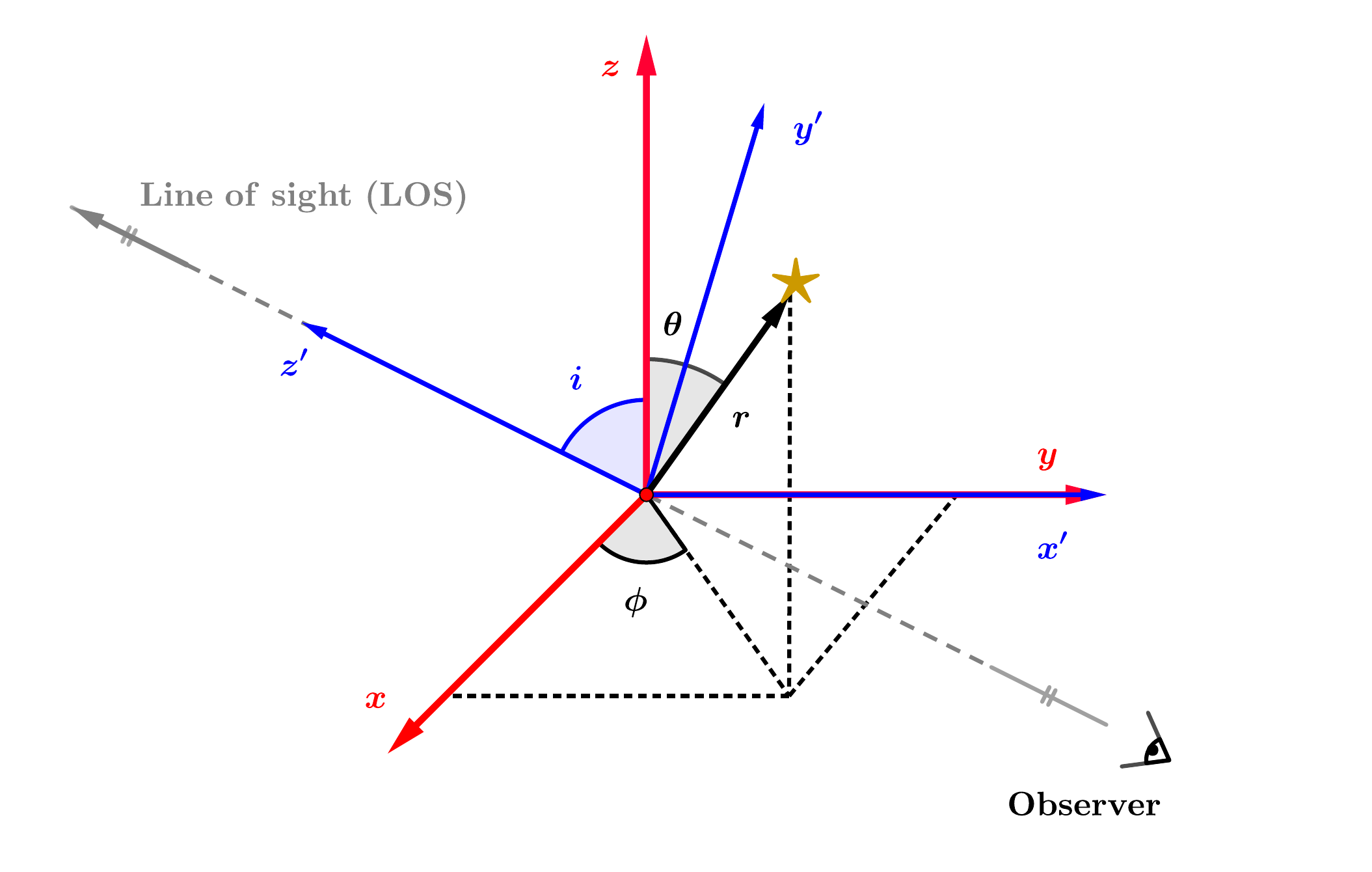}
\caption{\textit{Geometry of the system}: Set of coordinate definitions. The astronomical source is inclined by an angle $i$ and the line-of-sight direction is the same than the $z'$ direction depicted above.
}
\label{fig: geom-frame}
\end{figure}

\subsection{Evaluation of Proper Motions} \label{sapp: formalism-scf}

The \cite{deBruijne+96} paper addressed the evaluation of projected line-of-sight velocities of the scale-free models, but not the calculation of PMs, which were not observationally accessible at the time. We therefore extended their formalism to plane-of-sky (POS) velocities.\footnote{The corresponding software can be found at the link \\ \url{https://gitlab.com/eduardo-vitral/scalefree}.} 

Figure~\ref{fig: geom-frame} shows the general geometry of the problem, with the same conventions as in 
\cite{Evans&deZeeuw94}. The Cartesian frame $(x, y, z)$ is aligned with the axisymmetric luminous density, so that the $z$-axis is its symmetry axis. The $(r, \theta, \phi)$ are spherical coordinates in this system. A second Cartesian frame $(x', y', z')$ is introduced so that the $(x',y')$ frame corresponds to the plane of the sky, with $x' \equiv y$ along the projected major axis of the luminous density, and $z'$ along the LOS direction. The viewing inclination $i$ is the angle between the $z$ and $z'$ axes. The coordinates are related through 
\begin{subequations}
\begin{flalign}
    x' &= y \ , \\
    y' &= -x \cos{i} + z \sin{i} \ , \\
    z' &= x \sin{i} + z \cos{i} \ .
\end{flalign}
\end{subequations}

The velocity along any coordinate direction in the $(x', y', z')$ frame can be expressed as a combination of radial, tangential and azimuthal components,
\begin{equation}
    v_{\rm dir} = C_{r} v_{r} + C_{\theta} v_{\theta} + C_{\phi} v_{\phi} \ .
\end{equation}
For the LOS direction dir$=$$z'$, the $C_i$ terms are given by eqs~(49) in \cite{deBruijne+96} 
\begin{subequations}
\begin{flalign}
    C_{r}& = \sin {\theta} \sin {i} \cos {\phi}+\cos {\theta} \cos {i} \ , \\
    C_{\theta} &=\cos {\theta} \sin {i} \cos {\phi}-\sin {\theta} \cos {i} \ , \\
    C_{\phi} &= -\sin {i} \sin {\phi} \ .
\end{flalign}
\end{subequations}
The velocities in the radial and tangential directions on the plane of the sky can be written as
\begin{subequations}
\begin{flalign}
    v_{\rm POSr} &= \displaystyle{\frac{x' v_{x'} + y' v_{y'}}{\sqrt{x'^{2} + y'^{2}}}} \ , \\
    v_{\rm POSt} &= \displaystyle{\frac{y' v_{x'} - x' v_{y'}}{\sqrt{x'^{2} + y'^{2}}}} \ .
\end{flalign}
\end{subequations}
From these relations it follows\footnote{We computed the relations in eqs.~(\ref{eq: vposr-sf})--(\ref{eq: cposr_r2}) using the \textsc{Mathematica} software.} that the $C_{\rm i}$ terms for the POSr and POSt directions are, respectively:
\begin{itemize}[leftmargin=*]
    \item {\bf $v_{\bf POSr}$}:
    \begin{subequations} \label{eq: vposr-sf}
    \begin{flalign}
        C_{r} &= \sqrt{\Delta^{2}} \ , \\
        C_{\theta} &= [\sin{(2\theta)} (\cos ^2{i}  - \sin ^2{i} \cos ^2{\phi}) \\
        &-\cos(2 \theta ) \sin{(2 i)} \cos {\phi}] /
        (2 \, \sqrt{\Delta^{2}}) \ , \nonumber \\
        C_{\phi} &= \sin {\phi} \sin {i} \left[\sin {\theta}\sin{i}\cos {\phi} \right. \\
        &+\cos {\theta} \cos {i}\left. \right] / {\sqrt{\Delta^{2}}} \ . \nonumber
    \end{flalign}
    \end{subequations}
    \item  {\bf $v_{\bf POSt}$}:
    \begin{subequations} \label{eq: vpost-sf}
    \begin{flalign}
        C_{r} &= 0 \ , \\
        C_{\theta} &= \sin {i} \sin {\phi} / \sqrt{\Delta^{2}} \ , \\
        C_{\phi} &= \left[\cos {\theta} \sin {i} \cos {\phi}-\sin {\theta} \cos {i}\right] / \sqrt{\Delta^{2}} \ .
    \end{flalign}
    \end{subequations}
\end{itemize}
Above, the variable $\Delta^{2}$ is defined as:
\begin{equation} \label{eq: cposr_r2}
\begin{split}
    \Delta^{2} &\equiv \sin ^2{\theta} \sin ^2{\phi} \\
    &+(\cos {\theta} \sin {i} - \sin {\theta} \cos {i} \cos {\phi})^2 \ .
\end{split}
\end{equation}
With these equations one can calculate the $n$-th order instrinsic and projected moments for dir $=$ (LOS, POSr, POSt) starting from eq.~(50) in \cite{deBruijne+96},
\begin{equation} \label{eq: db+96-moments}
\begin{split}
    \rho\left<v_{\mathrm{dim}}^{n}\right> &= \sum_{j = 0}^{n} \sum_{k = 0}^{n - j} \left({n \atop j}\right) \left({n - j \atop k}\right) \\
    &\times C_{r}^{j} C_{\theta}^{k} C_{\phi}^{n-j-k} \rho\left<v_{r}^{j} v_{\theta}^{k} v_{\phi}^{n-j-k}\right> \ ,
\end{split}
\end{equation}
as described in their paper. 
All the quantities $\rho\left<v_{r}^{j} v_{\theta}^{k} v_{\phi}^{n-j-k}\right>$ are uniquely determined by the DF and the model parameters. 

\begin{figure}
\centering
\includegraphics[width=\hsize]{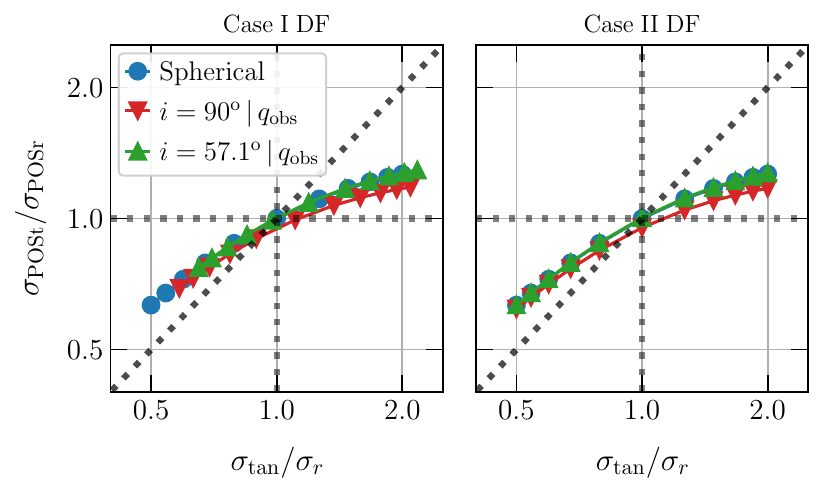}
\caption{\textit{Anisotropy dependence of kinematics:} projected anisotropy
$\sigma_{\rm POSt} / \sigma_{\rm POSr}$ as function of intrinsic three-dimensional velocity dispersion anisotropy $\sigma_{\rm tan}/\sigma_r$. Here $\sigma$ is used as shorthand for $\left< v^2 \right>^{1/2}$. Mass-weighted averaged predictions over all position angles are shown for scale-free models with $\kappa=2$ and $\delta=0$, and for Case~I (left) and Case~II (right) DFs, respectively. Each panel shows predictions for three different geometries, namely: (a) spherical; (b) axisymmetric and edge-on; and (c) axisymmetric and at viewing inclination $i=57.1\degree$. The axisymmetric models all have the same projected axial ratio as Draco. The projected anisotropy in the plane of the sky is tightly related to the intrinsic anisotropy.}
\label{fig: b3-app}
\end{figure}

\subsection{Anisotropy Dependence of observables} \label{sapp: beta-dependence-scf}

\citet{Leonard&Merritt89} and \citet{vanderMarel&Anderson&Anderson10} both showed that in spherical geometry there is a direct relation between the projected anisotropy $\sigma_{\rm POSt} / \sigma_{\rm POSr}$ and the intrinsic three-dimensional velocity dispersion anisotropy $\sigma_{\rm tan}/\sigma_r$, where $\sigma_{\rm tan} \equiv [(\sigma_{\theta}^2 + \sigma_{\phi}^2)/2]^{1/2}$. The scale-free models discussed above allow us to demonstrate that a similar relation holds in axisymmetric geometry as well. 

Figure~\ref{fig: b3-app} shows the relation for three different geometries, namely: (a) spherical; 
(b) axisymmetric and edge-on, with the same axial ratio as observed in projection of the sky (listed in Table~\ref{tab: overview}); and (c) axisymmetric and at viewing inclination $i=57.1\degree$, as in Section~\ref{ssec: axi-good-fit}, which is close to the median expected inclination given Draco's projected axial ratio. The intrinsic axial of the latter model is $q=0.608$. The figure shows ratios of mass-weighted averages over all position angles (either on the plane of the sky, or in the meridional $(R, z)$ plane, respectively), given that in axisymmetric models all quantities generally vary as a function of position angle (see Appendix~\ref{sapp: xi-dependence-scf} below). The two different panels show the predictions for Case~I and Case~II DFs, respectively. In the figures, $\sigma$ is used as shorthand for $\left< v^2 \right>^{1/2}$; these quantities are independent of whether or not the system has mean rotation, but the quantities are equal to the dispersion only if there is no rotation.

In all cases shown, the projected $\sigma_{\rm POSt} / \sigma_{\rm POSr}$ is simply a diluted measure (i.e. brought closer to unity due to projection effects) of the intrinsic 3D ratio $\sigma_{\rm tan} / \sigma_r$, with only subtle variations depending on the assumed geometry and the type of DF. The slope of the relation does vary with the radial slope $\kappa$ of the luminous tracer density distribution \citep[not shown here, but see][] {vanderMarel&Anderson&Anderson10},
but the relation remains monotonic and one-to-one. This explains why with suitable modeling, our new PM measurements have the constraining power to determine the intrinsic velocity dispersion anisotropy of Draco, even when not using LOS data. 

\begin{figure*}
\centering
\includegraphics[width=0.9\hsize]{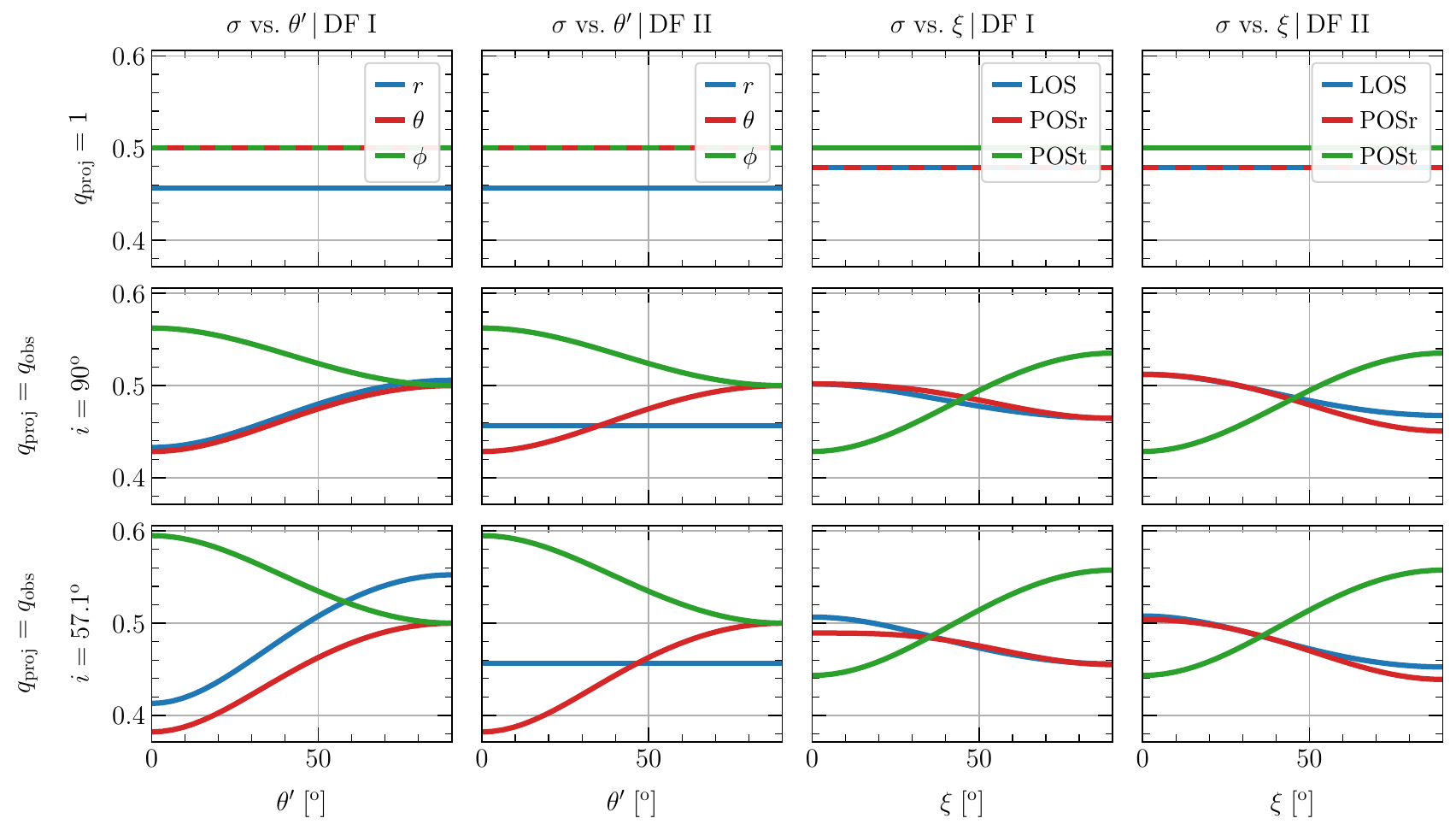}
\caption{\textit{Position-angle dependence of kinematics:} Rows show the predictions of scale-free models for the same three geometries as in Figure~\ref{fig: b3-app}, in each case chosen to have  globally-averaged anisotropy $\overline{\beta_{\rm B}}=-0.20$. The left two panels show the variations of $\sigma_r$, $\sigma_{\theta}$, and $\sigma_{\phi}$ with angle $\theta'$ in the meridional $(R, z)$ plane ($\theta' \equiv 90^{\circ} - \theta = 0$ in the equatorial plane). The right two panels show the variations of the projected $\sigma_{\rm LOS}$, $\sigma_{\rm POSr}$, and $\sigma_{\rm POSt}$ with the angle $\xi$ in the plane of the sky ($\xi = 0$ on the projected major axis). Here $\sigma$ is used as shorthand for $\left< v^2 \right>^{1/2}$. Velocities are expressed in the dimensionless units defined in section~2.1 of \citet{deBruijne+96}. The first and third columns show Case~I DFs, while the second and fourth columns show Case~II DFs. In axisymmetric geometry there are important variations of the kinematical observables with position angle on the sky, which are ignored when observations are interpreted with spherical models. This can introduce biases in model results.}
\label{fig: axi-scf}
\end{figure*}

\subsection{Position-Angle Dependence} \label{sapp: xi-dependence-scf}

For this paper, we have obtained new \hst\ PM data primarily along the projected major axis. So while Figure~\ref{fig: b3-app} shows the quantity $\sigma_{\rm POSt} / \sigma_{\rm POSr}$ integrated over all position angles, it is important to assess also how the kinematics vary with position angle in the models. The three rows of Figure~\ref{fig: axi-scf} show \scf\ model predictions for the same three geometries as in Figure~\ref{fig: b3-app}. For each of the geometries we show models with two different types of DF (Case~I and Case~II), but always with $\overline{\beta_{\rm B}}=-0.20$, consistent with the best fit from our axisymmetric Jeans models in Table~\ref{tab: mass-modeling-axi}. The left two panels show the variations of the internal kinematics in the 
meridional $(R, z)$ plane, while the right two panels show the variations of the projected kinematics in the plane of the sky. 

The main difference in the internal kinematics predictions for the axisymmetric models between the two types of DF is in the variation of $\sigma_r$ with $\theta'$ ($\equiv 90^{\circ} - \theta$). For either DF, $\sigma_{\phi}$ increases from the symmetry axis to the major axis (as before, we use in this discussion $\sigma$ as shorthand for $\left< v^2 \right>^{1/2}$). This is a direct consequence of the tensor virial theorem, which requires that overall the system has more dynamical pressure parallel to the equatorial plane than perpendicular to it, so as to support its flattened shape. The right two panels show that in projection, $\sigma_{\rm POSr}$ decreases from the major to the minor axis, while $\sigma_{\rm POSt}$ increases from the major to the minor axis. These behaviors are nearly independent of inclination (at fixed projected axial ratio) and of the specific type of DF. These variations are ignored when a spherical model is constructed, because the projected quantities are then independent of position angle. This can introduce important biases, as discussed in Section~\ref{ssec: sph-vs-axi}. 

While  Figure~\ref{fig: axi-scf} pertains to a specific set of parameter combinations, we verified that different parameter combinations yield qualitatively similar conclusions. For example, when considering flatter models with lower axial ratio (either intrinsically or in projection), the variations from major to minor axis become more pronounced than in Figure~\ref{fig: axi-scf}, but otherwise remain qualitatively similar.

\subsection{Higher-order moments} \label{sapp: GaussHermite}

Our analysis in this paper has been based on the second (and first) velocity moments that enter into the Jeans equations of hydrostatic equilibrium. However, for the observed LOS velocities we have enough measurements, with small enough error bars, to make it possible to also determine higher-order moments. Specifically, we calculated both the fourth-order Gauss-Hermite moment $h_4$ \citep[as in][]{vanderMarel&Franx93} and the kurtosis $\overline{K_{\rm LOS}}$ of the overall LOS velocity distribution. The kurtosis is defined here as 
\begin{equation}
  \overline{K_{\rm LOS}} = 
  \left[ \left< v_{\rm LOS}^4 \right> /
  \left< v_{\rm LOS}^2 \right>^2 \right] -3  ,
\end{equation}
where as before, $\left < v^k \right>$ indicates a $k$-th order  
velocity moment. We use all position angles on the sky and center the LOS velocity distribution on the bulk systemic velocity listed in Table~\ref{tab: overview}. Hence, 
odd moments are generally statistically consistent with zero. The observations then imply that $h_4 = 0.065 \pm 0.036$ and $\overline{K_{\rm LOS}} = 0.21 \pm 0.25$. Both of these quantities are positive, indicating that the LOS velocity distribution is more centrally peaked than a Gaussian. 

\begin{figure}
\centering
\includegraphics[width=\hsize]{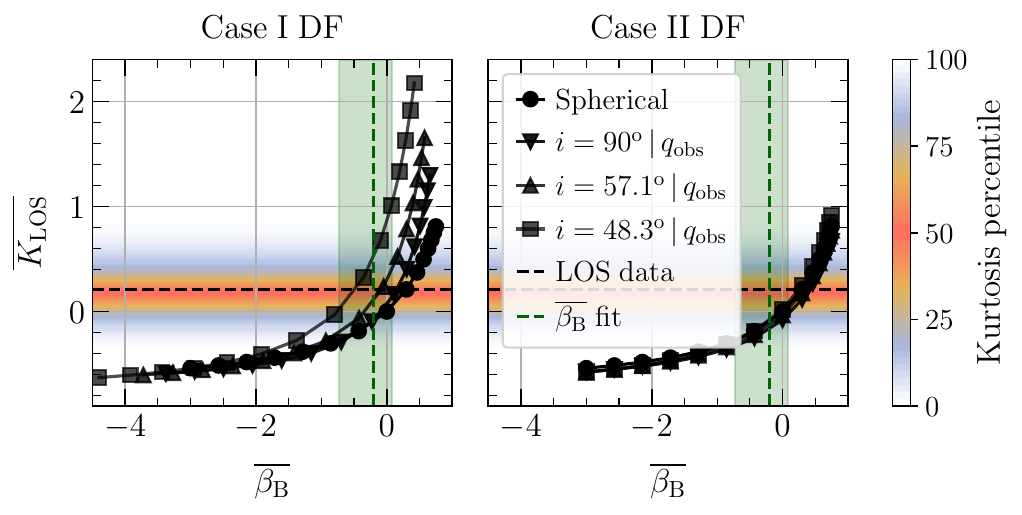}
\caption{\textit{Kurtosis:} Kurtosis of the grand-total LOS velocity distribution as a function of the globally averaged velocity anisotropy $\overline{\beta_{\rm B}}$. As in Figure~\ref{fig: b3-app}, different curves show predictions of scale-free models for different geometries, for Case~I (left panel) and Case~II (right panel) DFs, respectively. We now also include a flatter axisymmetric model with true axial ratio $q=0.45$, which at viewing inclination $48.3^{\circ}$ projects to the observed axial ratio for Draco. We show in \textit{dashed black} the observed kurtosis along with the percentiles of its distribution, color-coded as indicated on the right. The $\overline{\beta_{\rm B}}$ inferred from our axisymmetric Jeans models, with its 68\% confidence region, is depicted in \textit{green}. Plausible models exist that match all constraints (i.e. have predictions within the intersection of the orange and green bands).}
\label{fig: b5-app}
\end{figure}

The second-order Jeans equations cannot be used to interpret these measurements. However, the scale-free DF models predict all higher-order moments $\left < v^k \right>$ for any give set of model parameters. Reconstructing the full velocity distribution and Gauss-Hermite moments from the higher-order moments is not a straightfoward inversion problem 
\citep{deBruijne+96}, so we restrict data-model comparisons here to the kurtosis. Figure~\ref{fig: b5-app} shows the predicted kurtosis as function of the globally averaged Binney $\overline{\beta_{\rm B}}$, defined by Eq.~(\ref{eq: binney-beta}). As in Figure~\ref{fig: b3-app}, we show predictions for different geometries, with Case~I DFs in the left panel and Case~II DFs in the right panel. We now also include a flatter axisymmetric model with true axial ratio $q=0.45$, which at viewing inclination $48.3^{\circ}$ projects to the observed axial ratio for Draco. The horizontal orange-grey band shows the observed kurtosis with its uncertainty. For reference, the vertical green band shows the 68\% confidence band on $\overline{\beta_{\rm B}}$ for Draco from our axisymmetric Jeans models (bottom row of Table~\ref{tab: mass-modeling-axi}).

The predictions in the Figure show that more radially anisotropic models predict more peaked LOS velocity distributions (higher kurtosis), as has been previously established \citep[e.g.][]{vanderMarel&Franx93}. For the Case~II DFs, the predictions are more-or-less independent of geometry. But for the Case~I DFs, flatter axisymmetric models (seen at lower inclinations) predict higher kurtosis values. Hence, higher-order moments can help both to determine the viewing inclination and the full structure of the DF. More sophisticated dynamical modeling is required to fully exploit this information (e.g. \citealt{Chaname+08}). Nonetheless, we note here that an axisymmetric Case~I scale-free model at the median inclination of $57.1^{\circ}$ predicts the observed $\overline{K_{\rm LOS}} = 0.21$ when $\overline{\beta_{\rm B}} = -0.07$. The latter is fully consistent with the constraints that we have obtained from our axisymmetric Jeans models. Hence, there is no reason to expect that detailed modeling of higher-order moments would alter the conslusions about Draco's velocity anisotropy, and hence its mass distribution, that we have drawn in this paper.

\end{document}